\documentclass[preprint,12pt,authoryear]{elsarticle}
\usepackage{amsmath}
\usepackage{graphicx}
\usepackage[table]{xcolor}
\usepackage{listings}
\usepackage{enumerate}
\usepackage{natbib}
\usepackage[a4paper]{geometry}
\usepackage{graphicx}
\usepackage[textwidth=8em,textsize=small]{todonotes}
\usepackage{amsmath}
\usepackage{natbib}
\usepackage{array}
\usepackage[utf8]{inputenc}
\usepackage{graphicx}
\usepackage{longtable}
\graphicspath{ {./images/} }
\usepackage{multirow}
\usepackage{float}
 \usepackage{xspace}
 \usepackage{appendix}
 \usepackage{array}
\usepackage{amssymb}
\usepackage{dirtytalk}
\usepackage{etoolbox}
\usepackage{url}
\newcolumntype{C}[1]{>{\centering\arraybackslash}p{#1}}
\renewcommand{\arraystretch}{1.3}
\usepackage{booktabs}

\usepackage{url} 
\date{January 3, 2023}
\journal{ARXIV}

\begin{document}

\begin{frontmatter}

\title{Age-Gender-Country-Specific Death Rates Modelling and Forecasting: A Linear Mixed-Effects Model}
\author{Reza Dastranj\corref{cor1}}

\ead{dastranj@math.muni.cz}

\author{Martin Kol\'a\v r}
\ead{mkolar@math.muni.cz}

\address{Department of Mathematics and Statistics, Masaryk University, Kotlářská 2, 611 37 Brno, Czech Republic}

\cortext[cor1]{Corresponding Author}

\begin{abstract}
                                      
A linear mixed-effects (LME) model is proposed for modelling and forecasting single and multi-population age-specific death rates (ASDRs). The innovative approach that we take in this study treats age, the interaction between gender and age, their interactions with predictors, and cohort as fixed effects. Furthermore, we incorporate additional random effects to account for variations in the intercept, predictor coefficients, and cohort effects among different age groups of females and males across various countries. In the single-population case, we will see how the random effects of intercept and slope change over different age groups. We will show that the LME model is identifiable. Using simulating parameter uncertainty in the LME model, we will calculate $95\%$ uncertainty intervals for death rate forecasts. We will use data from the Human Mortality Database (HMD) to illustrate the procedure. We assess the predictive performance of the LME model in comparison to the Lee-Carter (LC) models fitted to individual populations. Additionally, we evaluate the predictive accuracy of the LME model relative to the Li-Lee (LL) model. Our results indicate that the LME model provides a more precise representation of observed mortality rates within the HMD, demonstrates robustness in calibration rate selection, and exhibits superior performance when contrasted with the LC and LL models.
\end{abstract}


\end{frontmatter}

\section{Introduction}\label{sec:intro}

Over the years, statisticians, actuaries, and demographers have been thinking about designing a suitable mortality projection model that can be used in the life insurance business, the pension funds industry, social security, epidemiology, and population studies. In the twentieth century, human mortality rates have declined remarkably around the world, and life expectancy has increased steadily. There is no clear indication that this trend of increasing life expectancy has reached its maximum \citep{burger2012human}. The continuing increase in life expectancy has brought additional stress in support systems for the elderly, such as governments, health care systems, the life insurance business, and the pension funds industry. For the pension funds industry and the insurance business, the pricing and reserving of annuities depend on future ASDRs. The continuous upward trajectory of life expectancy and its profound impact across various sectors, particularly the life insurance industry, highlights the criticality of developing robust models for projecting mortality rates.

Studying actuarial mortality tables, the British actuary Benjamin Gompertz observed that a law of geometrical progression pervades large portions of different mortality tables for humans \citep{gompertz1825xxiv}. He discovered that the probability of dying among individuals aged 30-80 years increases exponentially with age, and he proposed to represent the hazard rates as
\begin{equation}\label{Eq1}
    h(x)=\alpha \exp(\beta x),
\end{equation}
where $\alpha, \beta > 0 $, see \citet{macdonald2018modelling}. 

Lee and Carter are in that group of demographers who tackled this problem. The LC model defines the log ASDRs as
\begin{equation}\label{Eq2}
    \log (m_{x,t})=a_x+b_x k_t+\epsilon_{xt}, 
\end{equation}
in which the time-dependent component $ k_t$ is then forecast as a random walk with drift  \citep{lee1992modeling}. This model cannot be fitted as a regular regression because it does not include covariates. The LC model has become the ground for many extensions to forecasting single, and multiple-population mortality rates. Embedding the LC model in a Poisson regression model, \citet{brouhns2002poisson} proposed an approach for age–gender-specific mortality rates. The method of P-splines was applied to smoothing and forecasting mortality tables by \citet{currie2004smoothing}. Including a cohort effect, \citet{renshaw2006cohort} proposed an extension to the LC model. Another well-known technique is the Cairns-Blake-Dowd \citep{cairns2006two} model (specifically designed for pensioner ages), which introduces a two-factor model for the development of the mortality curve through time.  \citet{hyndman2007robust} used functional data analysis, nonparametric smoothing techniques, and robust statistics to model mortality rates (functional demographic model). A comparative evaluation by \citet{d2011mortality} established an overall comparison between the LC and the functional demographic model to the Italian mortality data. These methods focus originally on mortality forecasting of a single population.

If scientists have more samples to test in their experiments, it will increase the certainty of their estimates. In the case of multi-population mortality forecasting, exploring additional common effects from other populations, parameter estimation is more reliable. As a result, it improves the forecast accuracy for ASDRs. Modifying the original LC model, \citet{li2005coherent} have generated forecasts for multi-population death rates. The augmented common factor
(ACF) model assumes that 
\begin{equation}\label{Eq3}
y_{ixt}=\alpha_{ix}+\beta_{x}k_t+\beta_{ix} k_{it}+\epsilon_{ixt}, \qquad \epsilon_{ixt}\overset{\mathrm{iid}}{\sim}N(0\ ,\ \sigma_{i,\epsilon} ^2),
\end{equation}
for populations $i = 1, \cdots , M$, age groups $x = 0,1, \cdots , \omega$, and 
 time periods $t = 0,1, \cdots , T$.
\citet{kleinow2015common} has introduced a common age effect model to multiple populations. For constructing a two-population mortality model \citet{li2015step} have introduced a systematic process. \citet{richman2021neural} have proposed a neural network extension of the LC model. It is a fully connected neural network with multiple hidden layers. One of the advantages of the neural network extension is that the forecasts are not derived using time series methods. But, they are obtained implicitly once the neural network has been trained on a training dataset.
 For a review of the mortality modelling and forecasting methods, see \citet{booth2008mortality}.
 
Linear mixed-effects models play an increasingly important role in applied statistics \citep[see][for an overview of the theoretical
and computational aspects of LME models]{pinheiro2006mixed}. An age-specific death rate is a mortality rate in a particular age group. For each population, ASDRs are collected sequentially in different years. Mortality rates from the same age group are usually more similar than rates from other age groups. That means rates from the same age group are related to one another. Therefore, ASDRs form a longitudinal dataset (i.e. repeated measurements over time \citep{frees2004longitudinal, antonio2007actuarial} where observations within each age group are correlated. 
LME models offer a flexible and powerful approach for analyzing longitudinal data, such as ASDRs collected over time in different populations. In the context of mortality forecasting, LME models provide several advantages and valuable insights. At their core, LME models combine fixed effects and random effects to capture both population-level trends and individual-level variations within the data. Fixed effects represent the overall relationships between predictors (such as age, gender, cohort, and mortality covariates) and the response variable (log ASDRs). These fixed effects help uncover general patterns and associations between mortality rates and various factors of interest. For example, they can reveal how mortality rates change with age or differ between genders. However, mortality data often exhibit correlations and dependencies within specific groups. For instance, mortality rates within the same age group are typically more similar to each other than to rates from other age groups. LME models address this by incorporating random effects, which account for the correlated nature of observations within each age group. By modelling the individual-level variations within age groups, genders, and countries, LME models provide a more nuanced understanding of the mortality trends. The use of LME models in mortality forecasting allows for improved predictions and insights. By considering both fixed and random effects, these models provide a comprehensive understanding of the underlying factors influencing mortality rates. They can capture not only the overall population-level trends but also the specific variations within different subgroups. Furthermore, LME models account for the longitudinal nature of the data, where observations are collected sequentially over time. This is crucial in mortality forecasting because it allows for the incorporation of temporal dependencies and the exploration of trends and changes in mortality rates over different time periods. Therefore, LME models offer a valuable framework for mortality forecasting by integrating fixed effects and random effects to account for both population-level trends and individual-level variations within age groups, genders, and countries. By capturing correlations within the data and accounting for the longitudinal nature of the observations, LME models provide a more robust and accurate analysis of mortality trends, leading to improved predictions and valuable insights for understanding and forecasting mortality patterns.

\citet{tsutakawa1988mixed} proposes a mixed model for analyzing geographic variability in mortality rates, specifically focusing on lung cancer in Missouri. 
In an actuarial context, 
\citet{antonio2006lognormal} develop a flexible model for loss reserving by employing the theory of linear mixed models. Unlike traditional claims-reserving techniques, the mixed model allows for individual claim predictions with a sound statistical basis, including confidence regions. \citet{antonio2007actuarial} explore the use of generalized linear mixed models (GLMMs) for analyzing repeated measurements and longitudinal data in actuarial statistics. \citet{hughes2013dimension} propose a novel parameterization of the spatial generalized linear mixed model to address spatial dependence in non-Gaussian spatial data. Their method effectively tackles spatial confounding and improves computational efficiency by reducing the dimensionality of spatial random effects.
\citet{lima2021global} employ a linear mixed effects model to investigate age-specific trends in breast cancer (BC) incidence and mortality rates worldwide from 1990 to 2017. Their objective is to determine whether the increases in BC rates are evident across all age groups and regions and whether these changes can primarily be attributed to decreasing fertility rates. For mortality improvement rates with random effects, \citet{renshaw2021modelling} apply regression approaches within the framework of linear mixed-effects models.
\citet{verbeeck2023linear} employ linear mixed effects models for the estimation of excess mortality during the COVID-19 pandemic.

In this paper, we concentrate on mixed-effects linear regression. Utilizing classical regression models, such as linear or nonlinear regression, is not appropriate for this longitudinal data, where the subjects are age groups. This is due to the presence of repeated measurements within age groups and potential correlations over time. To adequately account for these dependencies and obtain accurate estimates of variance and p-values, it is necessary to employ a specialized statistical approach like linear mixed effects models. To develop a linear mixed-effects model for mortality rate forecasting, we will utilize two covariates: $k_{ct}$ and $k_t$. These covariates play distinct roles in capturing the overall mortality trend movement within our study. Specifically, $k_{ct}$ captures the overall mortality trend across two age groups: ages $0$ to $40$ and ages $41$ to $110$, for both females and males within each country at time $t$. In contrast, $k_t$ represents the general mortality trend across all age groups for females and males across all countries at time $t$. $k_{ct}$ is calculated by averaging log mortality rates across these two distinct age ranges: ages $0$ to $40$ and ages $41$ to $110$. We choose this division to emphasize the specific influence of external factors on mortality rates among different age groups. Utilizing the linear mixed-effects model, we can effectively predict and elucidate the variation in the logarithm of the ASDRs variable, $y_{cgxt}$, based on the covariates $k_{ct}$ and $k_t$.
The variance in random effect tells us how much variability there is among individuals across ages, and the variance in the residual term describes the amount of variance among observations within each age group. We will see that the sum of the variances of the random effects, $\sigma_{\eta}^2$, is higher than the residual variance $\sigma^2$. It confirms that observations from the same age group are generally more similar than observations from different age groups. The extent of dependence within the data can be quantified by employing the intra-class correlation coefficient (ICC)
\begin{equation}\label{Eq4}
  {\rm ICC} = \sigma_{\eta}^2 / (\sigma_{\eta}^2 + \sigma^2),
\end{equation}
see \citet{hox2017multilevel}.

Parameter estimation for the variance components $\sigma_{\eta}^2$ and $\sigma^2$ is typically done in two steps using a technique called restricted maximum likelihood (REML)\citep{Kenward1997restricted,pinheiro2006mixed,jiang2021linear}.

Model identifiability in the context of mortality models is a common issue. The LC model is underdetermined, which means that the parametrization in \eqref{Eq2} is not unique. To make sure the solutions in \eqref{Eq2} are unique, Lee and Carter introduced two constraints

\begin{equation}\label{Eq5}
    \sum _x b_x = 1, \qquad \sum _t k_t =0.
\end{equation}

A complete picture of the identifiability issues in mortality models is provided by \citet{hunt2020identifiability}. \citet{wang2013identifiability} studies the identifiability of linear mixed-effects models, and gives necessary and sufficient conditions of identifiability. One of the advantages of the LME model is that it is identifiable. We will discuss the identifiabilty of the LME in Section \ref{sec2}.

The paper is arranged as follows. Section \ref{sec2} defines the linear mixed-effects model of log ASDRs. In Section \ref{sec3}, we apply this LME model to a dataset spanning from 1961 to 2010, comprising twelve populations. We employ a stepwise model selection approach to identify the most suitable model. Subsequently, we extend our analysis to project mortality rates for these twelve populations up to 2019, assessing forecast accuracy by comparing the LME model's predictions to those of LC models fitted individually to each population. Additionally, we explore life expectancies for these populations, deriving values from the fitted LME model and contrasting them with those derived from LC models. Additionally, we employ both LME and LL models to analyze mortality rates across these populations within age groups spanning from 0 to 100, organized into 5-year age intervals. Furthermore, in Section \ref{sec3}, we delve into an exploration of how we can incorporate GDP per capita in the LME model, recognizing the profound influence of economic factors on mortality patterns. In Section \ref{sec4}, we fit the LME models to fifty-eight single population datasets in the HMD \citep{HMD} for the observations occurring in the years before 2000, and we forecast mortality rates from the year 2000 to 2019. In Section \ref{sec5}, we employ both the LME and LC model to analyze the present value and solvency capital requirement of a hypothetical portfolio. Section \ref{sec6} concludes with a critical discussion of the LME model.

\section{The Linear Mixed-Effects ASDRs Model}\label{sec2}

\begin{figure}[H]
    \centering
    \includegraphics[width=\textwidth]{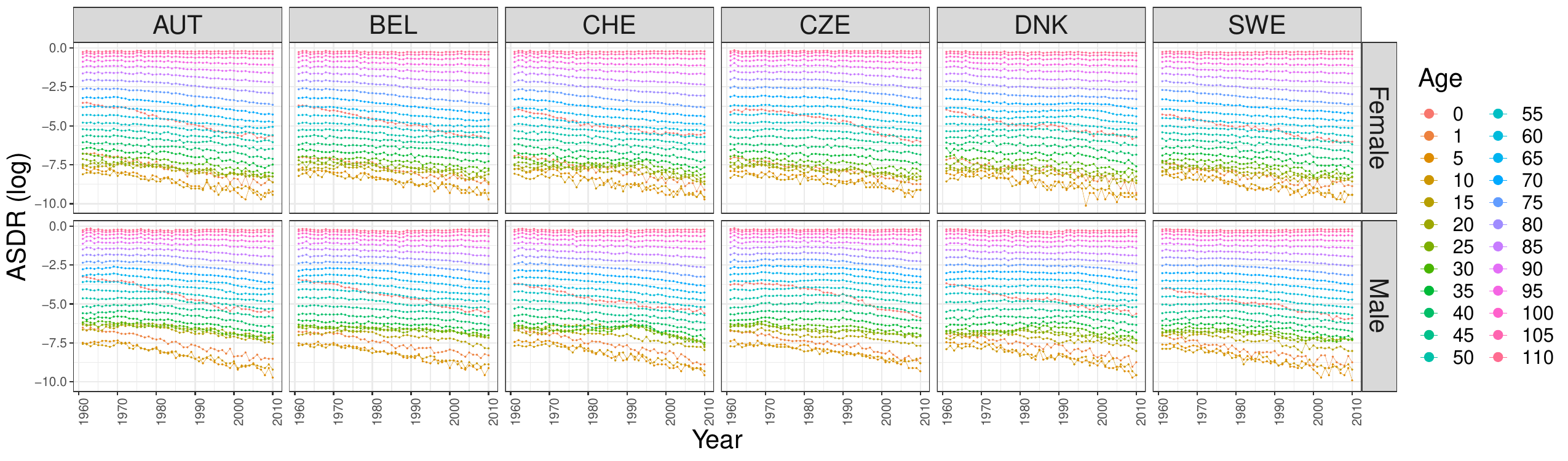}
    \caption{ASDRs (log) for twelve populations $1961-2010$.}
    \label{fig1}
\end{figure}

Projections of future mortality rates and life expectancy are needed to formulate plans for health and social services and economic development. We want to construct models not only for estimating the characteristics of the ASDRs but also for predicting how the mortality rates are most likely to evolve. To do so, we need critical parameters in the model capturing trends and evolution in mortality rates of different age groups over time. The age-specific mortality pattern reflects the underlying death mechanisms and epidemiological characteristics of the age group. Data visualization can often reveal patterns hidden from pure numerical analysis. It is essential to look at some mortality data to identify patterns in the age structure of a population. Figure \ref{fig1} shows the log ASDRs of twelve populations in $1961-2010$. Each curve is an age-specific death rate over 50 years on a natural logarithm scale. We see a clear evolution of mortality rates going down, which means, on average, people tend to live longer. We also see that the changes over time in mortality rates have been, for most log ASDRs, very regular and relatively slow and stable. We observe that the rates of the same age group in different populations follow a similar pattern. Death rates at some ages change much more rapidly than at other ages. The red curves in the middle of plots in Figure \ref{fig1} show the most substantial effects, infant mortality rates. Finding the regularity in both age patterns and trends over time, we conclude that the population age structure has a consistent and robust internal pattern.

\begin{figure}[H]
    \centering
    \includegraphics[width=\textwidth]{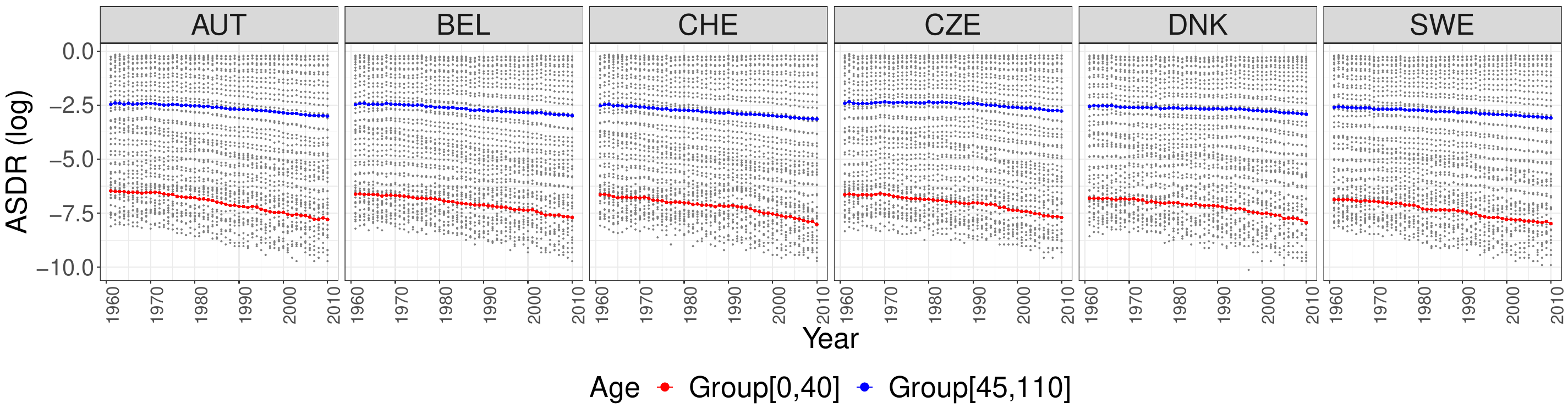}
    \caption{Plot of $k_{ct}$ for each country in our study ($1961-2010$).}
    \label{fig2}
\end{figure}

In Figures \ref{fig2}, the thick blue and red curves represent $k_{ct}$, which is calculated by averaging log mortality rates across two distinct age ranges: ages $0$ to $40$ and ages $41$ to $110$. We adopt this division to highlight the specific impact of external factors on mortality rates among different age groups. Historical data underscores a striking pattern: mortality rates for individuals in the younger age groups, typically up to $40$ or $50$ years old, exhibit substantial fluctuations during various time periods. This volatility is primarily attributed to external factors and societal shifts, with certain influences, such as the prevalence of AIDS, substance abuse, and increased violence, having a notably more pronounced effect on the younger demographic. These factors have played a significant role in shaping mortality trends \citep{plat2009stochastic}. Consequently, the introduction of the factor $k_{ct}$ into the LME model is aimed at providing a more comprehensive account of these dynamic influences, particularly their amplified impact on younger age groups.

\begin{figure}[H]
    \centering
    \includegraphics[width=\textwidth]{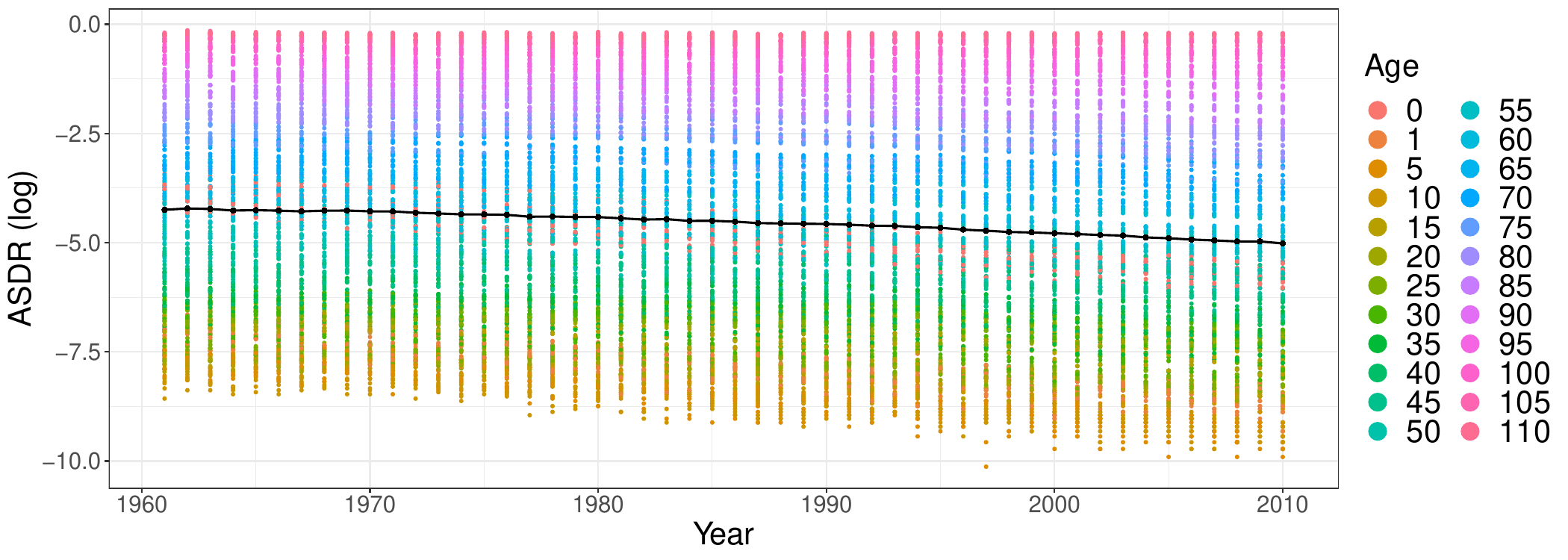}
    \caption{Plot of $k_t$: Average of log ASDRs across 12 populations and all age groups ($1961-2010$).}
    \label{fig3}
\end{figure}

In Figures \ref{fig3}, the thick black curve represents $k_t$, obtained by taking the average of log mortality rates across ages. For example, $k_{1961}$ is the 1961 average of log ASDRs. $k_t$ is the overall age-specific mortality pattern (trajectory) of the twelve-population dataset. The log ASDRs are totally at variance with the average curves (the black curve), and they differ considerably. We also observe that the rates of the same age group in different populations are remarkably similar.
\subsection{The Model Formulation}\label{sec2.1}

The method of linear mixed-effects models is now well established. Let $m_{c,g,x,t}$ denote the ASDR at age $x$ and time $t$ of gender $g$ in country $c$, for $c=$ $1, 2,\cdots,M$; $g=1, 2$; $x = 0, 1, \cdots , \omega$; and $t=0,1, \cdots, T$. Let $y_{cgxt}= \log (m_{c,g,x,t})$. Let $k_{ct}$ represent the average of $y_{cgxt}$ for two distinct age range groups: the average for age groups 0 to 40 and the average for age groups 41 to $\omega$, encompassing both females and males in country $c$ at time $t$. Additionally, let $k_t$ denote the average of $y_{cgxt}$ at time $t$ across all age groups in all countries and genders:
\begin{align}
k_{ct} = &
\begin{cases} 
k_{ct}^{[0,40]} & \text{if}\ x  \in [0,40], \\ 
      k_{ct}^{[41,\omega]} & {\rm otherwise},
\end{cases};
\qquad
k_{t} = \dfrac{\sum\limits_{c=1}^{M}\sum\limits_{g=1}^{2}\sum\limits_{x=0}^{\omega} y_{cgxt}}{2 M(\omega+1)},
\label{ali6}
\end{align}
for $c=1, 2, \cdots, M$, and $t=0, 1,  \cdots, T$, where
\begin{align*}
k_{ct}^{[0,40]}=
\dfrac{\sum\limits_{g=1}^{2}\sum\limits_{x=0}^{40} y_{cgxt}}{2 (41)}  , \qquad
k_{ct}^{[41,\omega]}=\dfrac{\sum\limits_{g=1}^{2}\sum\limits_{x=41}^{\omega} y_{cgxt}}{2 (\omega-40)},
\end{align*}
 for $c=1, 2, \cdots, M$, and $t=0, 1,  \cdots, T$. $k_{ct}$ and $k_{t}$ will be included as predictors in our model. We will employ random walks with drift to forecast the future values of $k_{ct}$ for ages $0$ to $40$ for each country, as well as the future values of $k_{ct}$ for ages $41$ to $\omega$ for each country, and the future values of $k_t$. This approach allows us to consider the evolving values and unique temporal patterns exhibited by each country, effectively capturing the specific dynamics and variations in time trends across different countries. We refer to these covariates as the \say{mortality covariates} due to their significance in capturing the temporal patterns of mortality rates. By incorporating the country mortality covariate, namely $k_{ct}$, into our analysis, we can accurately account for the nuanced temporal dynamics observed within each country, thereby enhancing the accuracy and reliability of our mortality rate pattern analyses. The country mortality covariate play a crucial role in capturing the temporal patterns and enabling a comprehensive understanding of the mortality trends across the countries under study.
 
\citet{Alho1998Stochastic} studied neighboring ages for mortality rates and observed correlations of about $0.95$ between them. \citet{lee1994stochastic} say that a perfect correlation can be assumed. Therefore, we can say that $k_{ct}$ and $k_{t}$ are two significant covariates to regress log ASDRs. We propose the development of a linear mixed effects model to capture the relationships between the log ASDRs and the mortality covariates. Within this framework, the mortality covariates are considered as the drivers or exogenous series, while the log ASDRs are treated as dependent variables influenced by these drivers.

The subscripts $x$ and $t$ on $m$ indicate that each observation $t$ is nested within age group $x$. So, for each individual population, we have mortality data grouped by age range. The linear mixed-effects projection mortality model expresses the $(T+1)$-dimensional response vector $y_{cgx}$ for age group $x$ of gender $g$ in country $c$ as

\begin{align}
\begin{split}
 &\quad \qquad\qquad y_{cgx}=X_{cgx}\beta+Z_{cgx}\eta_{cgx}+\epsilon_{cgx},\\
& 
\quad \quad c=1, 2, \cdots M,\quad g=1, 2,\quad x=0, 1,\cdots,\omega,\\ &\quad  \quad\qquad
 \eta_{cgx}\overset{\mathrm{iid}}{\sim}{\cal N}(0\ ,\ \Psi),\
\epsilon_{cgx}\overset{\mathrm{iid}}{\sim}{\cal N}(0\ , \ \sigma ^2 I),
\end{split}
\label{ali7}
\end{align}
where
\begin{align}\nonumber
    y_{cgx}& = 
    \setlength{\tabcolsep}{4pt}
\renewcommand{\arraystretch}{1}
\begin{bmatrix}
		y_{cgx0} \\
		y_{cgx1}  \\
	\vdots    \\
		y_{cgxT}  
\end{bmatrix},
\\
 X_{cgx}&=
 \setlength{\tabcolsep}{3pt}
\renewcommand{\arraystretch}{1}
	\begin{bmatrix}
		1 &1 & 1 & k_{0} & k_{c0}& \cdots &  k_{0}^{n_{1}} & k_{c0}^{n_{1}}&0-x\\1 &1 & 1 & k_{1} & k_{c1}& \cdots &  k_{1}^{n_{1}} & k_{c1}^{n_{1}}&1-x\\
		\vdots  &\vdots  &\vdots  &\vdots  &\vdots   &\ddots  &\vdots &\vdots &\vdots    \\
		1 &1 & 1 & k_{T} & k_{cT}& \cdots &  k_{T}^{n_{1}} & k_{cT}^{n_{1}}&T-x
	\end{bmatrix},
&
 \beta &= 
   \setlength{\tabcolsep}{3pt}
\renewcommand{\arraystretch}{1}
\begin{bmatrix}
	\beta_{0}  \\
 \beta_{0x}  \\
 \beta_{0gx}  \\
    \beta_{1}  \\
 \beta_{1gx}  \\
     \vdots   \\
  \beta_{{n_{1}}} \\	
    \beta_{{n_{1}}gx} \\
 \gamma_{\rm{cohort}}
\end{bmatrix},
\nonumber \\ 
 Z_{cgx}&=
   \setlength{\tabcolsep}{3pt}
\renewcommand{\arraystretch}{1}
	\begin{bmatrix}
		1 & k_{0} & k_{0} ^{2}& \cdots &k_{0} ^{n_{2}}&0-x\\
		1 & k_{1} & k_{1} ^{2} &\cdots &k_{1} ^{n_{2}}&1-x\\
		\vdots  & \vdots& \vdots  & \ddots & \vdots &\vdots \\
		1 & k_{T} & k_{T}^{2}& \cdots &k_{T} ^{n_{2}}&T-x
	\end{bmatrix},
 &
\eta_{cgx} &= 
  \setlength{\tabcolsep}{3pt}
\renewcommand{\arraystretch}{1}
\begin{bmatrix}
	\eta_{0cgx}  \\
	\eta_{1cgx} \\
 \vdots   \\
	\eta_{n_{2}cgx}\\
 \eta_{{\rm{cohort}}_{cgx}}
\end{bmatrix},
\nonumber\\
\epsilon_{cgx}& = 
  \setlength{\tabcolsep}{3pt}
\renewcommand{\arraystretch}{1}
\begin{bmatrix}
	\epsilon_{cgx0} \\
	 \epsilon_{cgx1} \\
	\vdots   \\
	\epsilon_{cgxT}
\end{bmatrix},
\label{ali8}
\end{align}
where ${n_{1}}$ and ${n_{2}}$ are non-negative integers; $\Psi$ is a positive-definite, symmetric $(n_{2}+2)\times(n_{2}+2)$ matrix; $\beta$ is the fixed effect vector, and $\eta_{cgx}$ is the random effects vector; $X_{cgx}$ and $Z_{cgx}$ are observed fixed-effects and random-effects regressor matrices; and $\epsilon_{cgx}$ is the within-age-group error vector. We assume that the random effects $\eta_{cgx}$ and the within-age-group errors $\epsilon_{cgx}$ are independent for different age-gender-country groups. Additionally, we assume that they are independent of each other for the same age-gender-country group \citep[see][chap.~2]{pinheiro2006mixed}.

In the context of linear mixed effects modelling, random effects are systematically defined with an average value of 0. This characteristic signifies that, on average, these random effects do not significantly contribute to the anticipated outcome of the response variable; they are, essentially, clustered around zero. Any structured or non-zero variations observed in the response variable are attributed to the presence of fixed-effects terms within the model. By specifying a mean of 0 for random effects, it implies that, across the entire range of possible random effects within the true population, their combined influence on the response variable averages out to zero. Nevertheless, it is crucial to acknowledge that individual random effects can still take non-zero values. These non-zero values account for the inherent random fluctuations that are not accounted for by the fixed-effects part of the model. This modelling approach enables a distinct separation between systematic influences, effectively captured by the fixed effects, and the unexplained, random variability accounted for by random effects. Therefore, random effects are purposely defined with a mean of 0, and any deviation from this mean within random effects must be included as part of the fixed-effects terms. As a result, if there is a non-zero mean for any term within the random effects, it must be incorporated as a component of the fixed-effects terms. This often leads to a scenario where the columns in the regressor matrix pertaining to random effects form a subset of the columns in the regressor matrix linked to the fixed effects \citep[see][chap.~2]{pinheiro2006mixed}.

$\eta_{{\rm{cohort}}_{cgx}}$ 
  represents the cohort effect across age, gender, and country. This effect exhibits a modulating structure based on age, gender, and country. The cohort effect represents the impact or influence of the birth cohort to which individuals belong on mortality rates. It suggests that individuals born in different years (cohorts) may experience different mortality rates due to various historical, societal, or environmental factors specific to their birth years. We are inspired by \citet{renshaw2006cohort}, who introduced a cohort factor to the Lee-Carter model, along with an age-modulating coefficient. We define the LME model in a way that incorporates this cohort effect, structuring it to account for age, gender, and country influences.

 The model \eqref{ali7} is called linear mixed-effects model because it is a linear combination of fixed effects and random effects:

\begin{equation*}
 \text{E}(y|\eta)=X\beta+Z\eta.
\end{equation*}
\subsection{Understanding Identifiability in the LME Model}\label{sec2.2}
The model \eqref{ali7} is identifiable because we have assumed that the covariance matrix of $\epsilon$ is a multiple of a known positive definite matrix (the identity matrix) 
\begin{align*}
\epsilon\overset{\mathrm{iid}}{\sim}{\cal N}(0\ ,\ \sigma ^2 I),
\end{align*}
see \citet{wang2013identifiability}.

Inferences regarding mixed-effects models refer to the estimation of fixed effects, the estimation of variance and covariance components, and the prediction of random effects, which are discussed in the next section.

\section{The ASDRs Model for Multi Populations}\label{sec3}
\begin{figure}[H]
    \centering
    \includegraphics[width=\textwidth]{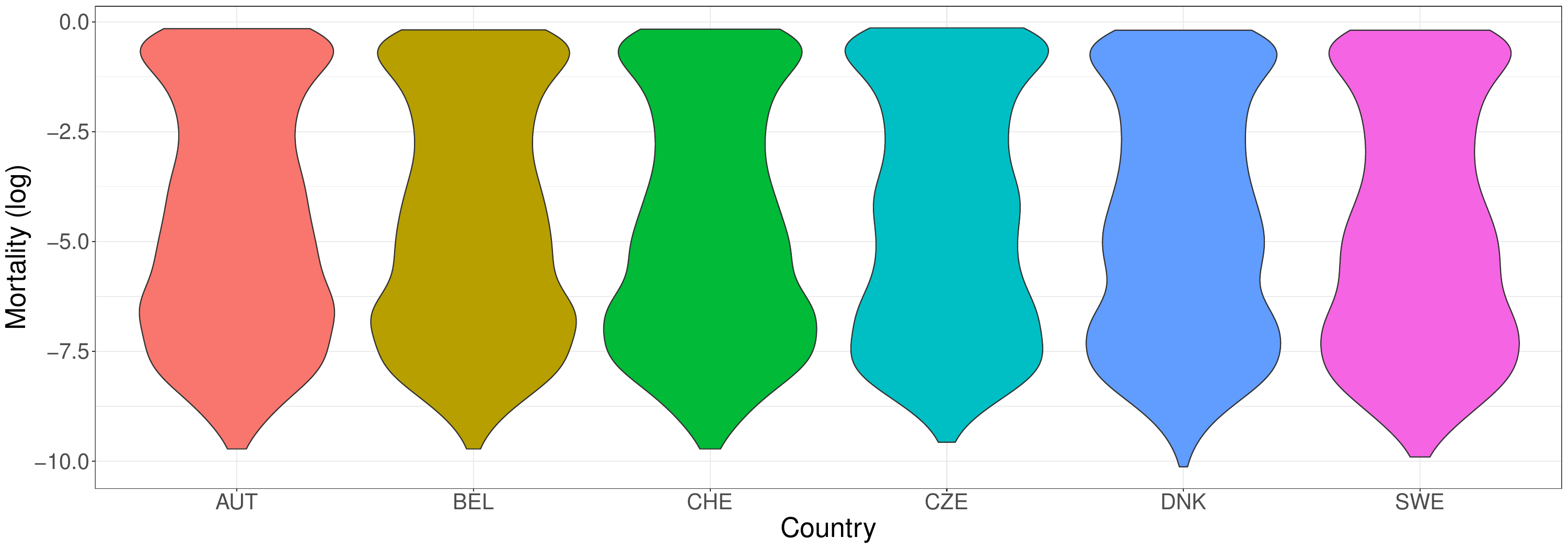}
    \caption{ Violin plots illustrating the distribution  and probability density of log mortality rates across the six European countries.}
    \label{fig4}
\end{figure}

In this section, we give an approach to modelling and forecasting ASDRs in an international context instead of looking at one country at a time. In the next section, we will consider ASDRs of a single population as a particular case of this section. 
We will consider the six countries chosen by \citet{enchev2017multi}. The population sizes of these countries are similar, their geographical locations are close, and there has been a remarkable drop in human mortality rates in the six countries since 1961. In Figure \ref{fig4}, vertical violin plots display the distribution of log mortality rates across the six European countries. Wider sections indicate a higher likelihood of observing values within that range, while narrower sections suggest a lower likelihood. The plots show remarkable similarity, indicating a shared distribution pattern for log mortality rates among the countries. The consistent lengths of the violin plots reflect the consistent frequency of observations for specific log mortality values across all countries. Notably, the Denmark plot exhibits a taller height, indicating greater variability in log mortality rates compared to the other countries. This implies a wider range of small mortality rate values in Denmark.

We will use ASDRs for conventional 5-year age groups up to $110^+$ by single calendar years $1961-2019$. We will group individuals' mortality rates by age, gender, and country.

\begin{figure}[H]
    \centering
    \includegraphics[width=\textwidth]{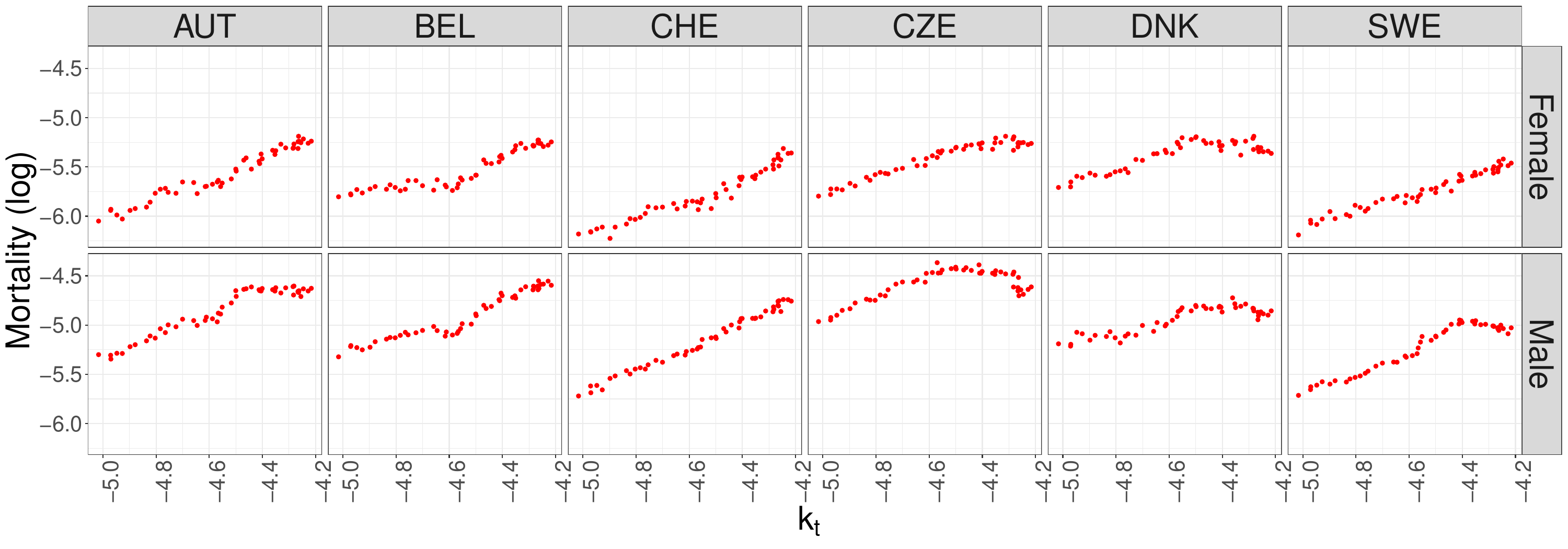}
    \caption{ $y_{cgxt}$ versus $k_{t}$ for age groups 50-54 of the twelve-population dataset.}
    \label{fig5}
\end{figure}

It is essential to look at the plot of $y_{cgxt}$ versus $k_t$ to find the correct values of $n_1$ and $n_2$ in \eqref{ali8} for the twelve-population dataset. We see that $y_{cgxt}$ versus $k_t$ combines a quadratic trend with a stochastic variation, see Figure \ref{fig5}.

In the matrices $X$ and $Z$ in \eqref{ali8}, set $n_1=n_2=2$. The linear mixed-effects model can then be expressed as
\begin{align}
\begin{split}
y_{cgxt}=&\beta_{0}+\beta_{0x}+\beta_{0gx}+\beta_{1}k_{t}+\beta_{1gx}k_{ct}+\beta_{2}k_{t}^2+\beta_{2gx}k_{ct}^2+\gamma_{\rm{cohort}}(t-x)+\\ &\eta_{0cgx}+\eta_{1cgx}k_{t}+\eta_{2cgx} k_{t}^2+\eta_{{\rm {cohort}}_ {cgx}}(t-x)+\epsilon_{cgxt},\\ &\quad
\eta_{cgx}\overset{\mathrm{iid}}{\sim}N(0\ , \ \sigma ^2_{\eta}), \qquad\qquad\qquad
\epsilon_{cgxt}\overset{\mathrm{iid}}{\sim}N(0\ ,\ \sigma ^2),
\end{split} 
\label{ali9}
\end{align}
where $c \in \{{\rm AUT}, {\rm BEL}, {\rm CHE}, {\rm CZE},{\rm DNK},{\rm SWE}\}$, $g={\rm Female},{\rm Male}$, $x=0, 1, 5$,$\cdots$, $110^+$, and  $t=1961,1962,\cdots, 2010$. Because of the quadratic trend that we can see in Figure \ref{fig5} it is the model \eqref{ali9}  that is conceptually most appropriate for the dataset.
The mixed-effects model combines a model for the fixed effects and a model for the random effects. 
\subsection{A Candidate Model Set}\label{sec3.1}

We want to build a linear mixed-effects model that can effectively use a set of explanatory variables to predict the values of the ASDRs. We consider the general class of models:

\begin{eqnarray}\label{Eq10} 
y_{cgxt} &=& 
x+g:x+{\rm{I}}(k_{t})+g:x:{\rm{I}}(k_{ct})+{\rm{I}}(k_{t}^2)+g:x:{\rm{I}}(k_{ct}^2)+\rm{cohort}+\nonumber\\&&({\rm{I}}(k_{t})+{\rm{I}}(k_{t}^2)+{\rm{cohort}}|c:g:x)=
\nonumber
\\ 
&& \beta_{0}+\beta_{0x}+\beta_{0gx}+\beta_{1}k_{t}+\beta_{1gx}k_{ct}+\beta_{2}k_{t}^2+\beta_{2gx}k_{ct}^2+\gamma_{\rm{cohort}}(t-x)+\nonumber\\ &&\eta_{0cgx}+\eta_{1cgx}k_{t}+\eta_{2cgx} k_{t}^2+\eta_{{\rm{cohort}}_{cgx}}(t-x)+\epsilon_{cgxt},
\end{eqnarray} 
where
\begin{eqnarray*}
\beta_{ 0x}&=& \beta_{0{1}}x_{\rm 1}+\beta_{0{5}}x_{\rm 5}+\cdots+ \beta_{0{110}}x_{\rm 110},\nonumber\\
\beta_{ 0gx}&=&(\beta_{0M{0}}x_{\rm 0}+  \beta_{0M{1}}x_{\rm 1}+\beta_{0M{5}}x_{\rm 5}+\cdots+ \beta_{0M{110}}x_{\rm 110})g_{M},\nonumber\\
\beta_{ jgx}&=& (\beta_{jF{0}}x_{\rm 0}+  \beta_{jF{1}}x_{\rm 1}+\beta_{jF{5}}x_{\rm 5}+\cdots+ \beta_{jF{110}}x_{\rm 110})g_{F}+\nonumber\\&&(\beta_{jM{0}}x_{\rm 0}+  \beta_{jM{1}}x_{\rm 1}+\beta_{jM{5}}x_{\rm 5}+\cdots+ \beta_{jM{110}}x_{\rm 110})g_{M}
\end{eqnarray*}
for $j=1,2$, where
\begin{align*}
x_{\rm n} &=\delta(x,{\rm n})=  \begin{cases} 
      1 & \text{if}\ x\ \text{is}\ {\rm n} ,\\ 
      0 & {\rm otherwise}
\end{cases},
&\\ \nonumber
g_{\rm F} &= \delta(g,{\rm F})= \begin{cases} 
      1 & \text{if}\ g\ \text{is Female},\\
      0 & {\rm otherwise} 
\end{cases},
&\\ \nonumber
g_{\rm M} &= \delta(g,{\rm M})= \begin{cases} 
      1 & \text{if}\ g\ \text{is Male},\\
      0 & {\rm otherwise} 
\end{cases},
\end{align*}
for ${\rm n} \in \{0, 1, 5, \cdots,110\}$. 

In this paper, females serve as the default gender, and age group $0$ is established as the reference point for comparison. Consequently, the parameter estimates derived for the interaction of gender and age, and age groups in the fixed effects of the intercept of the model are intended to quantify the disparities between males and females and evaluate the differences between each age group $1$, $5$,$\cdots$, and $110$ and age group $0$.

We have four distinct random effects for each age group, gender, and country. The first random effect corresponds to the intercept, while the second and third random effects account for the $k_{t}$ and $k_{t}^2$ coefficients, respectively. Additionally, the fourth random effect represents the cohort-specific random effects. To establish a unique age identification (ID) for each gender, we combine the gender and age variables  as ${\rm gender}:{\rm age}$. Similarly, to obtain a unique age ID for each gender in each country, we incorporate the country, gender, and age variables as ${\rm country}:{\rm gender}:{\rm age}$. We could also define another factor in the dataset, which enumerates the age groups from 1 to 288 instead from 1 to 24.

\subsection{Addressing Heteroskedasticity and Non-Linear QQ Plots in Mortality Models}\label{sec3.2}

\begin{figure}[H]
    \centering
    \includegraphics[width=\textwidth]{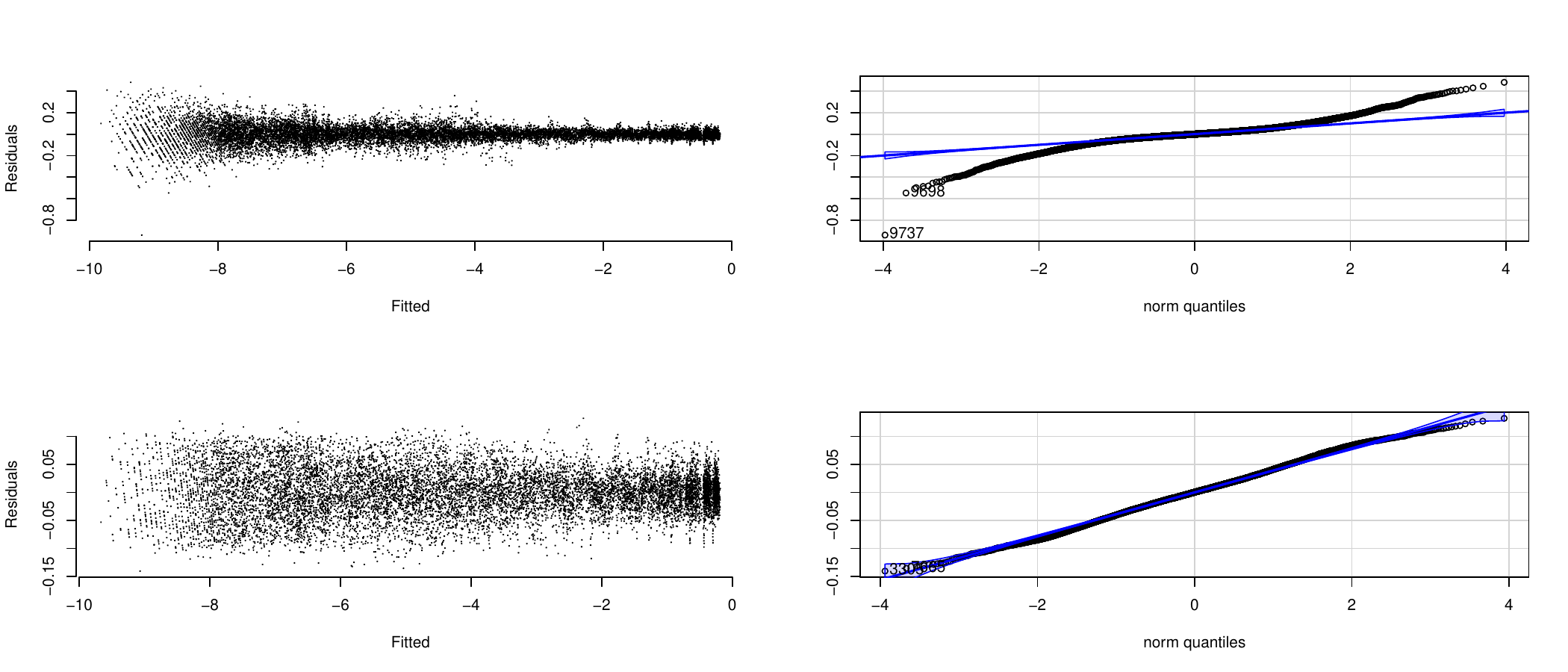}
    \caption{The four plots presented above illustrate various aspects of the LME model's residuals. The first two plots in the top row depict the residuals versus fitted values and the QQ plot of the residuals of the LME model (before correction), respectively. These plots reveal deviations from linearity in the residual pattern, and the QQ plot raises significant concerns about the normality of the residuals in the LME model (prior to correction) across the twelve populations. In contrast, the two plots in the bottom row represent the residuals versus fitted values and the QQ plot of the residuals of the LME model (after correction), respectively. In these plots, the quantile plot does not raise any significant concerns with the normality of the residuals. All points fall approximately along the reference line, allowing us to assume normality. Furthermore, as depicted in the second plot of residuals versus fitted values (after correction), the residual plot shows relatively constant variance across the fitted range.}
    \label{fig6}
\end{figure}

After fitting the candidate model, we conducted a thorough examination of the residuals to assess the presence of heteroskedasticity, which refers to the unequal variance of residuals across a range of observed values \citep[see][chap.~3]{james2013introduction}. Heteroskedasticity can manifest as a non-uniform scatter of residuals, which can impact the reliability of the model. Our analysis revealed that the residuals exhibited non-normality and heteroskedasticity, indicating that the variance of residuals was not consistent across different predicted values. To address this issue, we employed a data-cleaning procedure by excluding data points with absolute residuals greater than 0.1, signifying instances where the predicted and actual values diverged significantly. Subsequently, we re-estimated the LME model using this refined dataset, which typically retained about $85\%$ of the original data. This procedure resulted in improved characteristics of the residuals. We can observe the plot of the fitted values versus residuals and the QQ plot of residuals before and after these corrections in Figure \ref{fig6}. All findings presented in this paper are derived from the LME model post-critique, following the exclusion of data points and model refitting.

\begin{figure}[H]
    \centering
    \includegraphics[width=\textwidth]{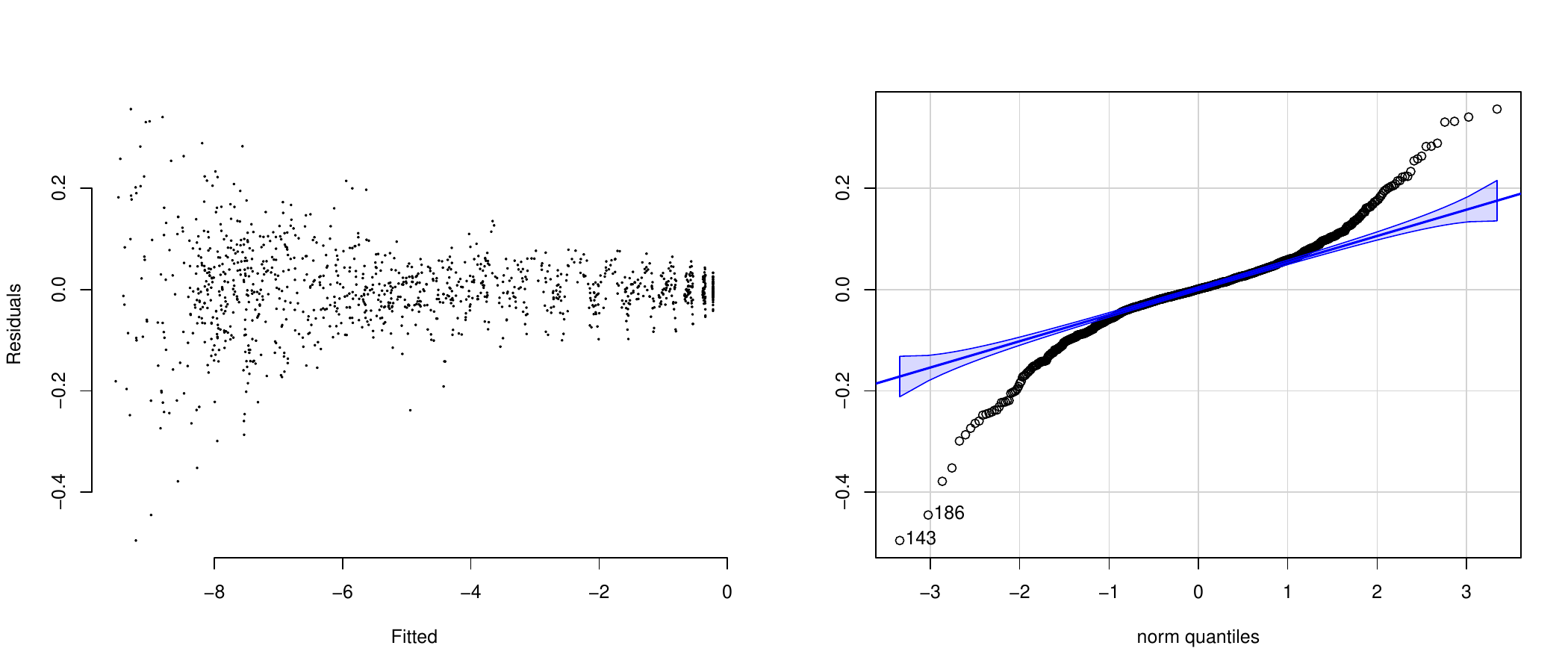}
    \caption{These plots reveal deviations from linearity in the residual pattern, and the QQ plot raises significant concerns about the normality of the residuals in the Lee-Carter model for Austrian female mortality rates.}
    \label{fig7}
\end{figure}

It is worth noting that these two issues—heteroskedasticity and nonlinearity of the QQ plot \citep{wilk1968probability, thode2002testing} of the residuals—are observed not only in the candidate model but also in other mortality models such as the LC model, LL model, and the Plat model. We conducted checks for these issues in all three mortality models. We can see the manifestation of these problems for the LC model in Figure \ref{fig7}.

Heteroskedasticity in regression models can have significant ramifications for both the model's validity and the interpretation of its results. It can lead to biased parameter estimates, particularly for fixed effects, potentially misrepresenting relationships between predictors and the response variable. The violation of the constant variance assumption results in incorrect standard errors, which, in turn, can distort p-values and confidence intervals, affecting hypothesis testing and variable significance assessments. Moreover, heteroskedasticity can inflate Type I error rates, leading to spurious findings, while also diminishing the model's power to detect true effects. Inefficient parameter estimation and model selection complications can further challenge the modelling process. When combined with nonlinearity in QQ plots of residuals, the consequences extend to potentially misleading residual analyses and unreliable predictions with wider confidence intervals, rendering the model's results less dependable for forecasting and inference. Addressing both heteroskedasticity and nonlinearity in residual patterns is crucial for ensuring the reliability and accuracy of regression model outcomes. 

We can also explore the application of a robust linear mixed-effects model using the robustlmm package \citep{koller2016robustlmm} as a robust approach to addressing heteroskedasticity in our dataset. Unlike traditional linear modeling methods, robustlmm is designed to handle situations where the variability of the data points (residuals) is not consistent, which is a common issue in many real-world datasets. By incorporating robustlmm, we can effectively account for and mitigate the impact of this non-constant variance in the residuals. Importantly, this approach eliminates the need to eliminate or exclude specific data points from our dataset.
\subsection{Stepwise Model Selection}\label{sec3.3}

We will use the refined dataset from $1961$ to $2010$ as the training set to estimate the parameters of the candidate model set. The REML estimation technique is used to fit the model \eqref{Eq10} to the dataset. To do so, we will use the lmer function in the lme4 package in R Programming language \citep{JSSv067i01}. How can we determine which terms in the model \eqref{Eq10} are necessary to fit our data better? Two goodness-of-fit measures that we can use are Akaike Information Criterion (AIC) \citep{sakamoto1986akaike}  and Schwarz's Bayesian Information Criterion (BIC) \citep{schwarz1978estimating}. The following formulas give AIC and BIC:
\begin{eqnarray}
   {\rm AIC }  &=& -2 \times {\rm loglikelihood }+ 2 \times d, \label{Eq11}\\ 
    {\rm BIC } &=& -2 \times {\rm loglikelihood }+  \log(N)\times d,\label{Eq12}
\end{eqnarray}
where $d$ is the total number of parameters, and $N$ is the sample size of the training set. To keep the model \eqref{Eq10} maximal and remove further terms based on AIC/BIC, one algorithm we can use is known as stepwise selection. Comparing a restricted set of models, it picks the best model \citep{hastie2009elements}. We implement a backward stepwise selection strategy that considers backward moves at each step and selects the best model \citep[see][]{kuznetsova2017lmertest}. According to the backward stepwise selection, the best-fit model for the log ASDRs of the twelve populations is the model
\begin{eqnarray} \label{Eq13} 
y_{cgxt} &=& 
x+g:x+g:x:{\rm{I}}(k_{ct})+{\rm{I}}(k_{t}^2)+g:x:{\rm{I}}(k_{ct}^2)+\rm{cohort}+\nonumber
\\ 
&&({\rm{I}}(k_{t}^2)+{\rm{cohort}}|c:g:x)=\nonumber
\\ 
&&
 \beta_{0}+\beta_{0x}+\beta_{0gx}+\beta_{1gx}k_{ct}+\beta_{2}k_{t}^2+\beta_{2gx}k_{ct}^2+\gamma_{\rm{cohort}}(t-x)+\nonumber\\ &&\eta_{0cgx}+\eta_{2cgx} k_{t}^2+\eta_{{\rm{cohort}}_{cgx}}(t-x)+\epsilon_{cgxt},
\end{eqnarray}  

\begin{figure}[H]
    \centering
    \includegraphics[width=\textwidth]{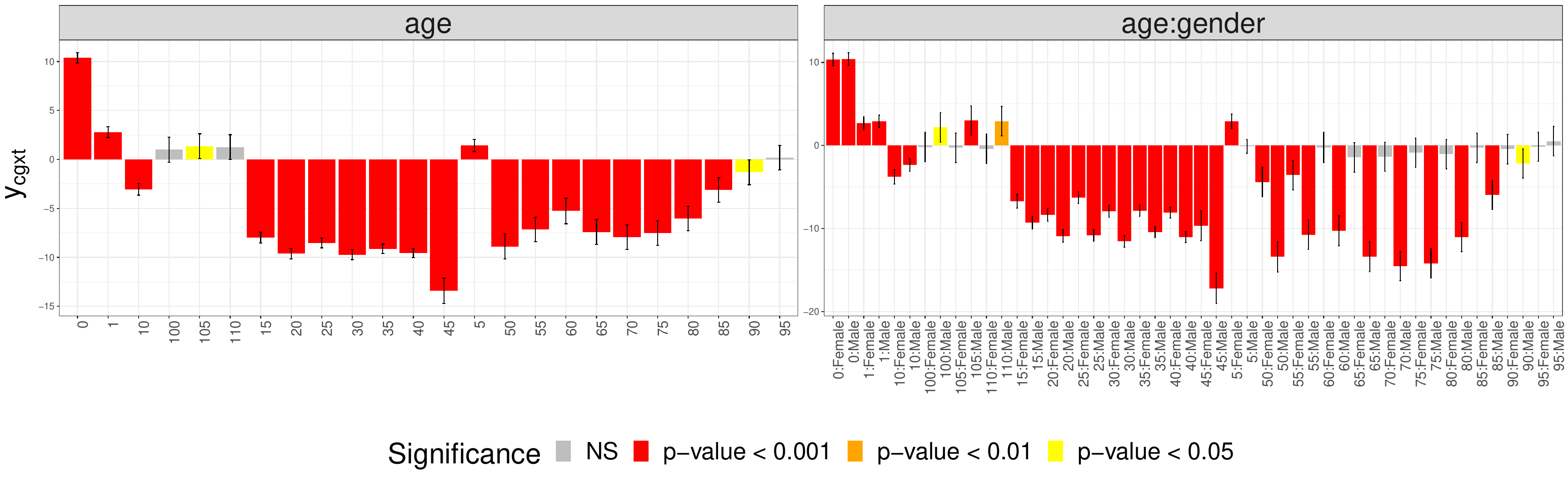}
    \caption{ Barplots displaying the differences of least square means for age and gender-age effects, along with their respective $95\%$ confidence intervals, for the ASDRs data.}
    \label{fig8}
\end{figure}

Our aim is to maintain consistency in the modelling approach across all age groups and genders in the six countries. By including the quadratic term $k_{ct}$ for all age groups, we are capturing a potential nonlinear relationship between $k_{ct}$ and mortality rates that might exist across different age groups of females and males. We intend to avoid introducing varying model degrees for different age groups, as this would introduce complexity and potential biases in the analysis. This decision is driven by our desire to establish a unified framework that treats all age groups equally and ensures a consistent modelling strategy, see Figure \ref{fig8}.

The parameters in the model (\ref{Eq13}) have specific interpretations: $\beta_{0x}$ and $\beta_{0gx}$ represent the fixed effects for age and age-gender on the intercept, respectively, while $\eta_{0cgx}$ represents the age-gender-country random effect on the intercept; $\beta_{j{gx}}$ for $j=1, 2$ are the age-gender fixed effects of $k_{ct}$ and $k_{ct}^2$;
$\beta_{2}$ represents the fixed effect of $k_{t}^2$, while $\eta_{2cgx}$ represents the age-gender-country random effect of $k_{t}^2$; Additionally, $\gamma_{\rm{cohort}}$ represents the fixed effect of cohort, and $\eta_{{\rm{cohort}}_{cgx}}$  represents the age-gender-country cohort random effect. The model selection reveals that age and age-gender effects play crucial roles as fixed effects, while age-gender-country is identified as a critical random effect on the intercept of the model. The model captures a shared age effect in the intercept across the twelve populations, enabling the projection of ASDRs. Similarly, a common age-gender effect in the intercept is observed, contributing to ASDRs projections. The results obtained from the model selection analysis indicate that the inclusion of fixed effect term $k_t$ does not yield a statistically significant improvement in the capability of the model to explain the observed variability in mortality rates. Furthermore, when taking into account the random variation in the effect of $k_t$ across age-gender-country, there is no significant enhancement observed in the capability of the model to provide a more comprehensive explanation for the observed variations in mortality rates.

We have the flexibility to incorporate a covariate, denoted as $k_{cgt}$, which represents the overall mortality trend movement across age groups for each gender in each country at time $t$. Additionally, we encourage the exploration of other biologically relevant covariates that could enhance the model.

We can also consider incorporating a covariate region into the LME model. For instance, we may group certain countries, such as the US and Canada in North America, and the UK and France in Europe, into distinct regions. In this expanded model, the LME now encompasses four levels of nesting, resulting in random effects at four distinct levels. Consequently, there are four levels of random variation in the model, structured as ${\rm region}:{\rm country}:{\rm gender}:{\rm age}$. This approach enables us to capture the regional effect on mortality rates.

One of the notable strengths of the LME model is its capability for stepwise model selection, allowing us to identify which terms are essential for improving data fit. However, it is important to keep in mind that the random effects are designed with a mean of 0, which means that any non-zero mean for a term in the random effects must be integrated into the fixed-effects terms \citep[see][chap.~2]{pinheiro2006mixed}. Consequently, we may need to consider incorporating new covariates as fixed effects if warranted.
\subsection{Estimation of Model Parameters}\label{sec3.4}

The estimated fixed effects of $\beta_{0}$ , $\beta_{2}$, $\gamma_{\rm{cohort}}$, and their standard deviations are as follows:
\begin{align}\label{Eq14}
\hat{\beta}_{0} &= 49.874, & \mathrm{sd}(\hat{\beta}_{0}) &= 1.410, \nonumber \\
\hat{\beta}_{2} &= 0.0061, & \mathrm{sd}(\hat{\beta}_{2}) &= 0.003, \\
\hat{\gamma}_{\mathrm{cohort}} &= -0.00103, & \mathrm{sd}(\hat{\gamma}_{\mathrm{cohort}}) &= 0.00023. \nonumber
\end{align}

The other estimated fixed effects of the model \eqref{Eq13} are shown in Table \ref{tab1}.

\begin{table}[H]
\Huge
    \centering
\resizebox{15 cm}{!}{
\setlength{\tabcolsep}{8pt}
\renewcommand{\arraystretch}{1}
    \centering
\begin{tabular}{|c|c|c|c|c|c|c|c|c|c|c|c|c|}
%\toprule
 \cmidrule{1-13}
%\midrule
\multicolumn{1}{|c}{$x$}&\multicolumn{1}{|c}{$\hat{\beta}_{0x}$} & \multicolumn{1}{c}{{\rm sd}$(\hat{\beta}_{0x})$} &\multicolumn{1}{|c}{$\hat{\beta}_{0Mx}$} & \multicolumn{1}{c}{{\rm sd}$(\hat{\beta}_{0Mx})$} &\multicolumn{1}{|c}{$\hat{\beta}_{1Fx}$} & \multicolumn{1}{c}{{\rm sd}$(\hat{\beta}_{1Fx})$} &\multicolumn{1}{|c}{$\hat{\beta}_{1Mx}$} & \multicolumn{1}{c|}{{\rm sd}$(\hat{\beta}_{1Mx})$}&\multicolumn{1}{|c}{$\hat{\beta}_{2Fx}$} & \multicolumn{1}{c|}{{\rm sd}$(\hat{\beta}_{2Fx})$}&\multicolumn{1}{|c}{$\hat{\beta}_{2Mx}$} & \multicolumn{1}{c|}{{\rm sd}$(\hat{\beta}_{2Mx})$}\\
\cmidrule{2-13}
0 &  &  & -0.668 & 2.016 & 13.542 & 0.395 & 13.242 & 0.411 & 0.857 & 0.028 & 0.833 & 0.029 \\ 
  1 & -22.552 & 2.022 & 0.920 & 2.016 & 7.705 & 0.415 & 8.040 & 0.389 & 0.425 & 0.030 & 0.458 & 0.027 \\ 
  5 & -18.498 & 2.147 & -12.793 & 2.267 & 9.263 & 0.458 & 5.546 & 0.440 & 0.549 & 0.032 & 0.287 & 0.031 \\ 
  10 & -42.417 & 2.169 & 3.220 & 2.202 & 2.956 & 0.465 & 3.492 & 0.408 & 0.135 & 0.033 & 0.155 & 0.029 \\ 
  15 & -54.549 & 2.048 & -12.347 & 2.039 & -0.364 & 0.417 & -4.086 & 0.388 & -0.078 & 0.029 & -0.339 & 0.027 \\ 
  20 & -61.481 & 2.029 & -12.875 & 2.013 & -2.414 & 0.423 & -6.331 & 0.389 & -0.226 & 0.030 & -0.500 & 0.028 \\ 
  25 & -53.999 & 1.929 & -20.069 & 1.864 & -0.398 & 0.380 & -6.272 & 0.373 & -0.088 & 0.027 & -0.498 & 0.027 \\ 
 30 & -60.745 & 1.913 & -16.276 & 1.852 & -2.257 & 0.373 & -7.082 & 0.371 & -0.210 & 0.027 & -0.551 & 0.026 \\ 
  35 & -61.290 & 1.899 & -11.888 & 1.755 & -2.458 & 0.367 & -5.999 & 0.337 & -0.219 & 0.026 & -0.469 & 0.024 \\ 
  40 &\cellcolor{blue!25} -62.818 & 1.843 &\cellcolor{blue!25}  -13.092 & 1.736 & -2.836 & 0.342 &\cellcolor{blue!25} -6.697 & 0.353 &-0.233 & 0.024 &\cellcolor{blue!25}  -0.505 & 0.025 \\ 
  45 & -53.617 & 1.630 & -5.245 & 1.200 & -0.746 & 0.664 & -5.821 & 0.673 & -0.329 & 0.130 & -1.406 & 0.132 \\ 
  50 &\cellcolor{blue!25} -48.080 & 1.623 & -6.697 & 1.190 &\cellcolor{blue!25}3.112 & 0.653 & -3.121 & 0.673 &\cellcolor{blue!25} 0.396 & 0.128 &-0.903 & 0.132 \\ 
  55 & -46.059 & 1.623 & -6.570 & 1.175 & 4.039 & 0.655 & -1.701 & 0.658 & 0.520 & 0.128 & -0.608 & 0.129 \\ 
  60 & -42.398 & 1.633 & -11.011 & 1.194 & 6.420 & 0.673 & -2.348 & 0.664 & 0.959 & 0.132 & -0.674 & 0.130 \\ 
  65 & -43.057 & 1.625 & -13.808 & 1.181 & 5.397 & 0.660 & -5.278 & 0.661 & 0.739 & 0.129 & -1.219 & 0.129 \\ 
  70 & -43.084 & 1.617 & -15.237 & 1.165 & 4.978 & 0.647 & -6.669 & 0.655 & 0.663 & 0.127 & -1.470 & 0.128 \\ 
  75 & -43.046 & 1.619 & -14.700 & 1.160 & 4.677 & 0.649 & -6.645 & 0.648 & 0.627 & 0.127 & -1.477 & 0.127 \\ 
  80 & -44.040 & 1.619 & -10.831 & 1.160 & 3.665 & 0.649 & -4.718 & 0.647 & 0.471 & 0.127 & -1.095 & 0.127 \\ 
  85 & -43.758 & 1.617 & -6.387 & 1.158 & 3.579 & 0.648 & -1.334 & 0.648 & 0.477 & 0.127 & -0.430 & 0.127 \\ 
  90 & -44.460 & 1.618 & -2.444 & 1.159 & 2.804 & 0.649 & 0.969 & 0.648 & 0.353 & 0.127 & 0.034 & 0.127 \\ 
  95 & -44.843 & 1.617 & 0.080 & 1.159 & 2.365 & 0.648 & 2.502 & 0.649 & 0.300 & 0.127 & 0.353 & 0.127 \\ 
  100 & -45.652 & 1.617 & 1.961 & 1.160 & 1.692 & 0.648 & 3.268 & 0.650 & 0.207 & 0.127 & 0.527 & 0.127 \\ 
 105 & -46.505 & 1.618 & 3.035 & 1.159 & 1.045 & 0.649 & 3.436 & 0.649 & 0.118 & 0.127 & 0.587 & 0.127 \\ 
 $ 110^{+}$ & -47.156 & 1.618 & 3.210 & 1.160 & 0.570 & 0.649 & 3.079 & 0.649 & 0.051 & 0.127 & 0.537 & 0.127 \\ 
     \hline
 \end{tabular}
    }
    \caption{Estimated fixed effects and their standard deviations.}
    \label{tab1}
\end{table}

The estimated variances and correlations of the random effects are:

\begin{align}
    \begin{split}
\widehat{\rm Var}({\eta}_{0cgx})&=4.408*10^{-3},  \widehat {\rm Var}({\eta}_{2cgx})=1.264*10^{-3},\widehat {\rm Var}({\eta}_{{\rm{cohort}}_{cgx}})=1.476*10^{-7}\\
    &\widehat{\rm corr}(\eta_{0cgx},\eta_{2cgx}) = -0.15 ,\widehat{\rm corr}(\eta_{0cgx},\eta_{{\rm{cohort}}_{cgx}}) = -0.03 ,\\
    &\widehat{\rm corr}(\eta_{2cgx},\eta_{{\rm{cohort}}_{cgx}}) =-0.98 .
\end{split}
\label{Eq15}
\end{align}

$\widehat{\rm Var}({\eta}_{0cgx})=4.408*10^{-3}$ indicates the extent to which the intercept values differ among different combinations of age, gender, and country within the LME model. $\widehat{\rm Var}({\eta}_{2cgx})=1.264*10^{-3}$ signifies the degree to which the effect of $k_t^2$ varies across different combinations of age, gender, and country in the LME model.
$\widehat{\rm Var}({\eta}_{{\rm{cohort}}_{ cgx}})=1.476*10^{-7}$ reflects the variability in the cohort effect across age-gender-country. 

The estimated variance-covariance matrix of the random effects is therefore
\begin{align}
\begin{split}
    \hat{\Psi}& = 
\begin{bmatrix}
      \quad 4.408*10^{-3} & -3.424*10^{-4}& -8.856*10^{-7}\\
       -3.424*10^{-4} &\quad 1.264*10^{-3} &-1.344*10^{-5}\\
        -8.856*10^{-7}& -1.344*10^{-5} & \quad 1.476*10^{-7}\\
\end{bmatrix}.
\end{split} 
\label{ali16} 
\end{align}

Finally, the estimated variance of the residual errors is
\begin{equation}\label{Eq17}
    \hat{\sigma}^2 =   1.786*10^{-3}.
\end{equation}

This is the within-age-group variance. The variance of $y_{cgxt}$ is therefore estimated as
\begin{eqnarray}\nonumber
\widehat{\rm Var}(y_{cgxt})&=&  \widehat{\rm Var}({\eta}_{0cgx})+ \widehat{\rm Var}({\eta}_{2cgx})+\widehat {\rm Var}({\eta}_{{\rm{cohort}}_{ cgx}})+\hat{\sigma}^2\\ \nonumber 
   &=&4.408*10^{-3}+1.264*10^{-3}+1.476*10^{-7} +1.786*10^{-3}\label{Eq18}\\ 
   &=&7.459*10^{-3}.
\end{eqnarray}

Hence, only about $0.24$ of the total variance of the log ASDRs, $y_{cgxt}$, is attributed to residual variation, resulting in an ICC of approximately $0.76$. This observation aligns with our visual inspection from Figures \ref{fig1} to \ref{fig3}. Therefore, it is evident that while there is some variation within age groups across different years, the variation across age-gender-country combinations is comparatively greater.

Now, after studying the variability of the random effects, we would like to discuss individual random effects too. The predicted random effects $\hat{\eta}_{0cgx}$, $\hat{\eta}_{2cgx}$, $\hat{\eta}_{{\rm{cohort}}_{ cgx}}$ are shown in Tables \ref{tab5}, \ref{tab6}, and \ref{tab7} in \ref{A}. We can use this information to make a separate regression equation for each age group of each gender in each country. For example, in the year $2019$, for females at age $50$ and males at age $40$ in Austria (AUT), respectively, we have
\begin{eqnarray*}\nonumber
    y_{{\rm AUT}\ {\rm Female}\ 50\ 2019}&=& 49.874-48.080+3.112k_{{\rm{AUT}}\ 2019}^{[41,110]}+0.0061k_{2019}^2+\\ \nonumber
    &&0.396{k_{{\rm{AUT}}\ 2019}^{[41,110]}}^2-0.00103(2019-50)+\\ \nonumber
    && 0.0429-0.02821k_{2019}^2+0.000263(2019-50),\\ \nonumber
    y_{{\rm AUT}\ {\rm Male}\ 40\ 2019}&=& 49.874-62.818-13.092 -6.697  {k_{{\rm{AUT}}\ 2019}^{[1,40]}}+\\ \nonumber
    &&0.0061k_{2019}^2 -0.505  {k_{{\rm{AUT}}\ 2019}^{[1,40]}}^{2}-0.00103(2019-40)+\\ \nonumber
    &&-0.07183-0.05167 k_{2019}^2+ 0.000639(2019-40).
\end{eqnarray*}

\begin{figure}[H]
    \centering
    \includegraphics[width=\textwidth]{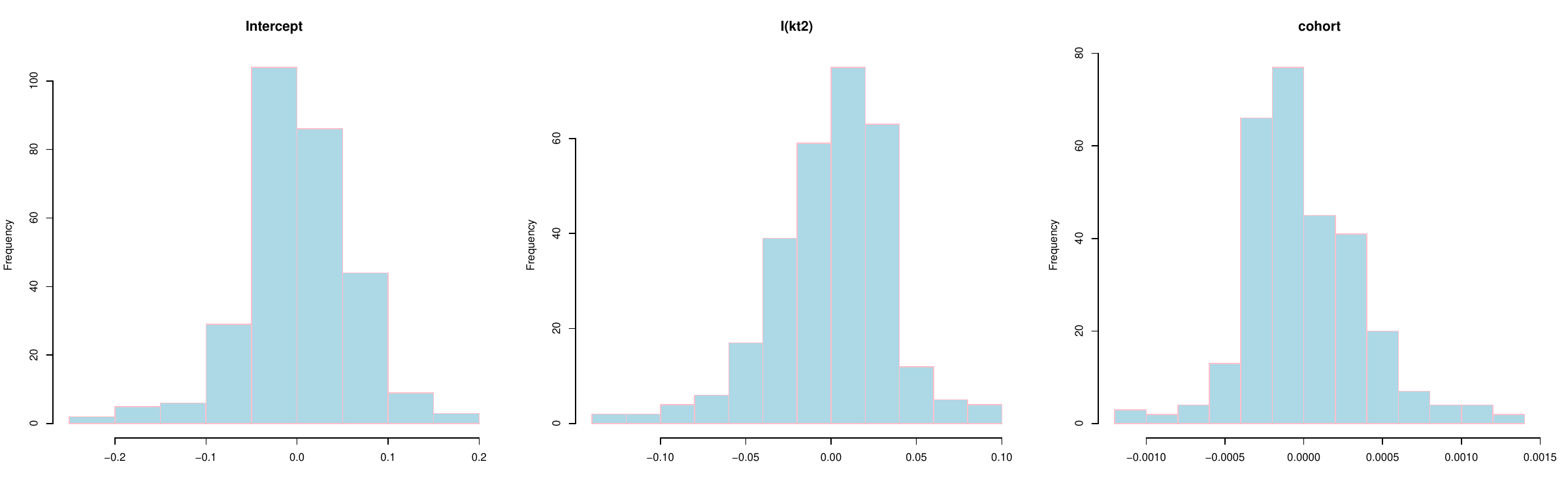}
    \caption{Histograms of random effects.}
    \label{fig9}
\end{figure}

The histograms displayed in Figure \ref{fig9} illustrate normal distributions with mean values that closely approximate zero for the intercept, $k_{t}^2$, and cohort random effects. Specifically, the means for these effects are approximately $3.76*10^{-12}$, $-1.96*10^{-11}$, and $2.10*10^{-13}$, respectively.

Table \ref{tab2} presents a comprehensive overview of the ANOVA analysis conducted to evaluate the significance of the fixed terms within the LME model \citep[see][]{kuznetsova2017lmertest}. The p-values associated with all six fixed terms indicate their statistical significance.
\begin{table}[H]
\resizebox{\textwidth}{!}{
    \centering
\begin{tabular}{lrrrrrr}
  \hline
 Fixed-Effect Term& Sum Sq & Mean Sq & NumDF & DenDF & F value & Pr($>$F) \\ 
  \hline
$x$ & 4.34 & 0.19 & 23.00 & 10710.71 & 105.74 & $<$ 2.2e-16 $***$ \\ 
  $g:x$ & 2.11 & 0.09 & 24.00 & 10698.14 & 49.28 & $<$ 2.2e-16 $***$ \\ 
  $g:x:{\rm{I}}(k_{ct})$ & 11.91 & 0.25 & 48.00 & 9999.26 & 138.91 & $<$ 2.2e-16 $***$ \\ 
  ${\rm{I}}(k_{t}^2)$ & 0.01 & 0.01 & 1.00 & 593.35 & 3.95 & 0.0473\enspace $*$\quad\enspace \\ 
 $ g:x:{\rm{I}}(k_{ct}^2)$ & 10.83 & 0.23 & 48.00 & 8825.68 & 126.34 & $<$ 2.2e-16 $***$ \\ 
   $\rm{cohrot}$ & 0.04 & 0.04 & 1.00 & 11084.33 & 19.60 & $<$ 2.2e-16 $***$ \\ 
   \hline
\end{tabular}
}
\caption{Summary of Type III Analysis of Variance (ANOVA) Results:
This table provides a comprehensive overview of the ANOVA analysis conducted to assess the significance of the fixed terms in the LME model. The p-values associated with each fixed term indicate the strength of evidence against the null hypothesis of no significant effect. P-values less than 2.2e-16 are denoted as '***,' signifying extremely high statistical significance. The ANOVA analysis confirms the significance of the quadratic terms, aligning with our visual inspection in Figure \ref{fig5}.}
\label{tab2}
\end{table}

\begin{figure}[H]
    \centering
    \includegraphics[width=\textwidth]{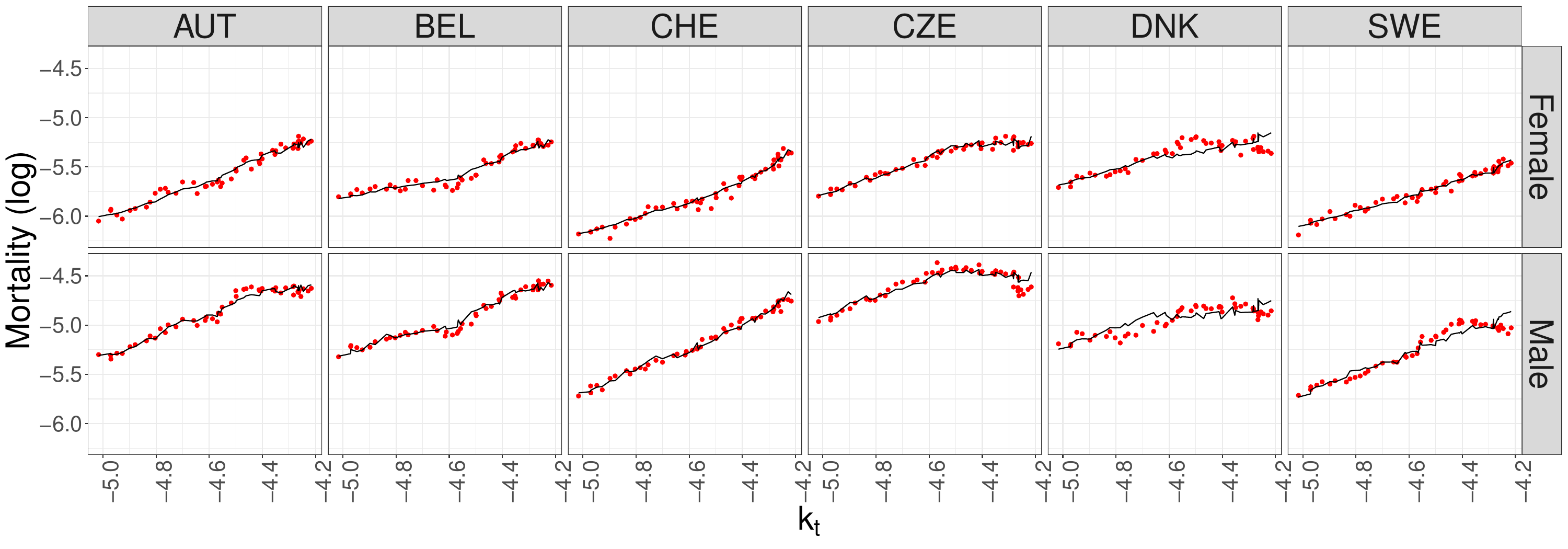}
    \caption{LME fits of $y_{cgxt}$ versus $k_{t}$ for age groups 50-54 of the twelve-population dataset.}
    \label{fig10}
\end{figure}

\begin{figure}[H]
    \centering
    \includegraphics[width=\textwidth]{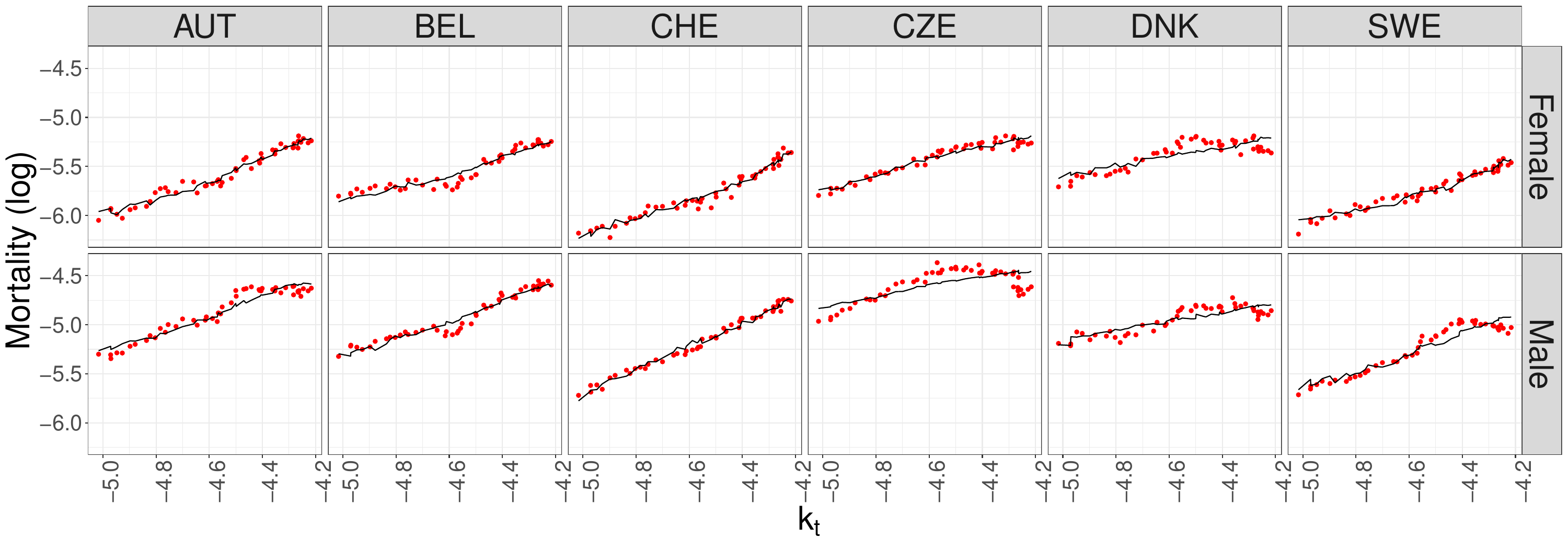}
    \caption{LC fits of $y_{cgxt}$ versus $k_{t}$ for age groups 50-54 of the twelve-population dataset.}
    \label{fig11}
\end{figure}

Figures \ref{fig10} and \ref{fig11} depict the fits of the LME and LC models for mortality rates versus $k_t$ in the age group 50-54 for each population. The LME model demonstrates good performance in capturing the patterns. Specifically, when comparing the plots for Czech males in both models, it is evident that the quadratic term in the LME model (\ref{Eq13}) is crucial for accurately fitting the model to the dataset. This highlights the importance of including the quadratic term to capture the non-linear relationship between mortality rates and $k_t$ in the Czech male population. 

\begin{figure}[H]
    \centering
    \includegraphics[width=\textwidth]{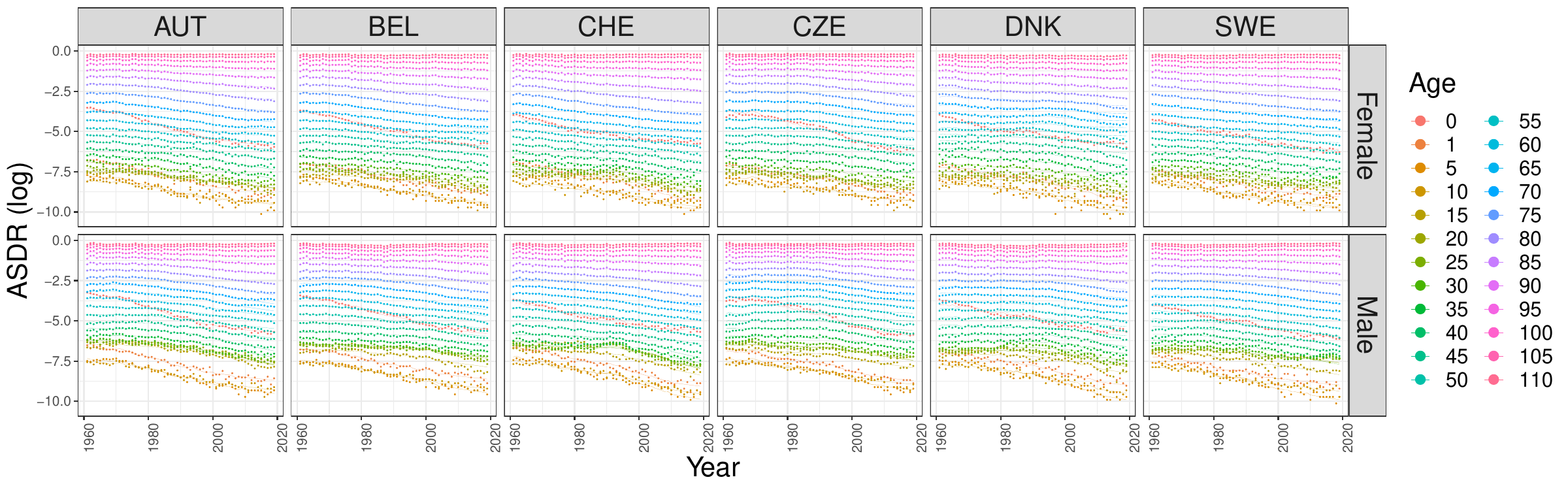}
    \caption{LME fits and predictions of log ASDRs for the twelve-population dataset ($1961 - 2019$).}
    \label{fig12}
\end{figure}

\begin{figure}[H]
    \centering
    \includegraphics[width=\textwidth]{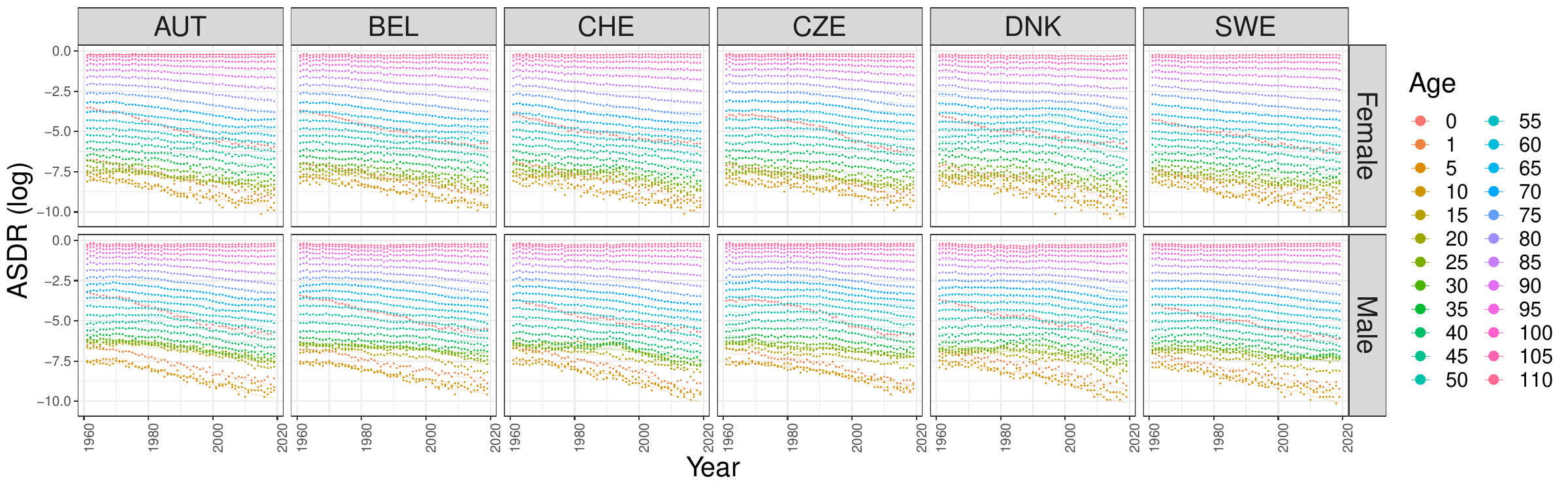}
    \caption{LC fits and predictions of log ASDRs for the twelve-population dataset ($1961 - 2019$).}
    \label{fig13}
\end{figure}

Figures \ref{fig12} and \ref{fig13} display the LME and LC fits for log ASDRs of the twelve-population dataset, respectively, for both the training and test sets covering the years $1961-2019$. The models effectively capture the mortality rate patterns associated with age, gender, and country, indicating their overall adequacy. Remarkably, the LME model outperforms the LC model in accurately modelling infant mortality rates. The LME model demonstrates a better ability to capture the rates and patterns of infant mortality, resulting in more precise predictions of life expectancies, as we will see in section \ref{sec3.6}. 

\begin{figure}[H]
    \centering
    \includegraphics[width=\textwidth]{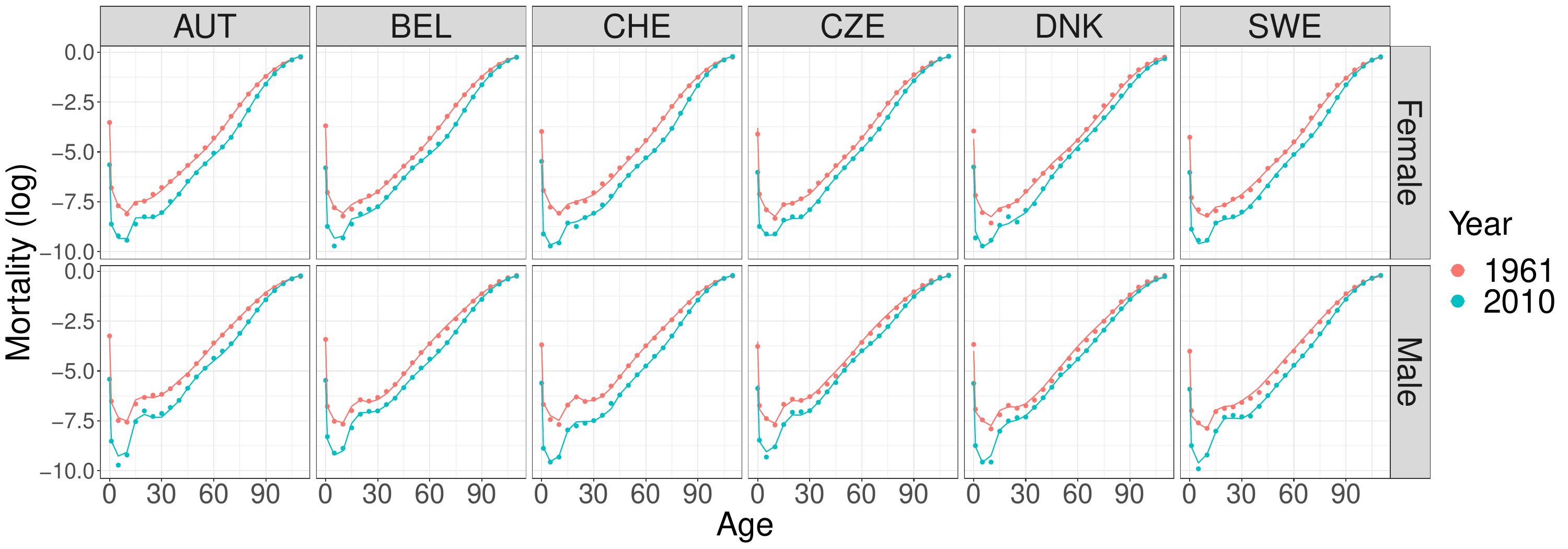}
    \caption{LME fits of log death rates in 1961 and 2010.}
    \label{fig14}
\end{figure}

\begin{figure}[H]
    \centering
    \includegraphics[width=\textwidth]{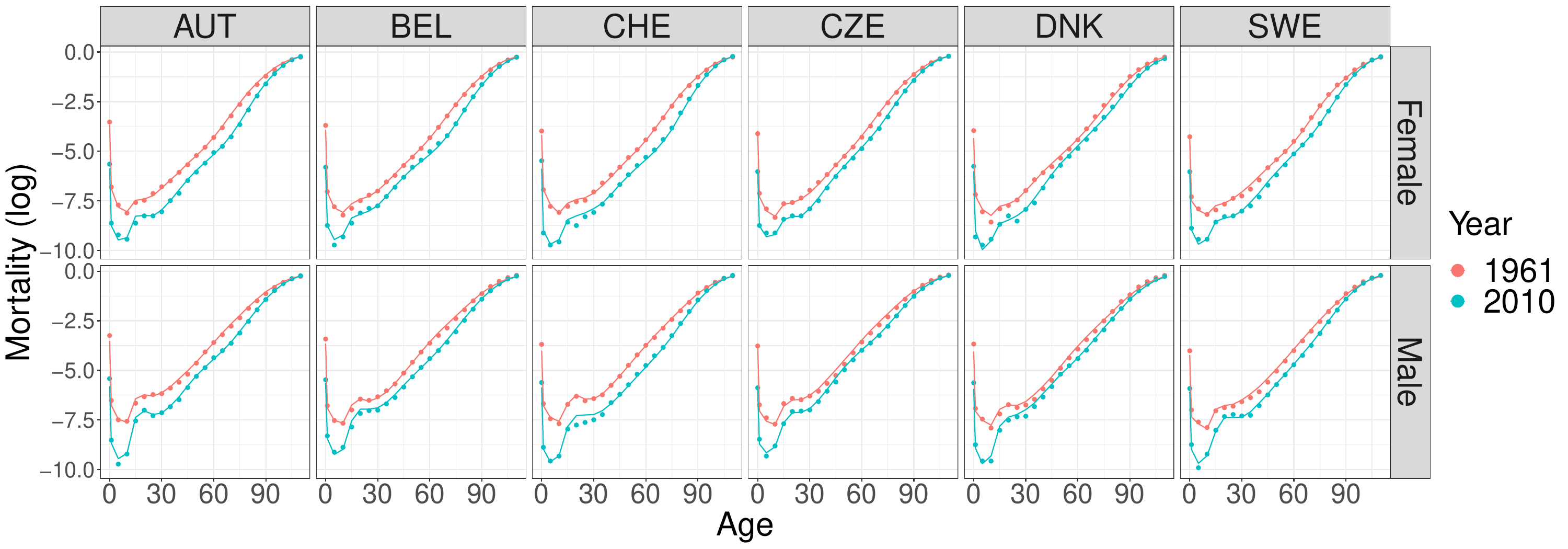}
    \caption{LC fits of log death rates in 1961 and 2010.}
    \label{fig15}
\end{figure}

Figures \ref{fig14} and \ref{fig15} depict the observed rates and fitted values obtained from the LME and LC models, respectively. These rates represent the log probabilities of dying at each age for individuals who have survived until that age. Notably, there has been a mortality improvement over time, as evidenced by the curves of the mortality rates in 2010 being lower than those in 1961.
Analyzing the mortality curves, we observe a distinct pattern. During childhood, mortality rates are relatively low, but from around age 30, they begin to increase exponentially. Interestingly, the mortality rates appear to stabilize at late ages, while they significantly decline at early ages.
Both the LME and LC models demonstrate their efficacy in identifying age patterns and capturing the evolution of mortality. The fitted values closely align with the observed log probabilities, indicating a successful modelling process. However, when considering males from Switzerland, it appears that the LME model performs better in capturing their mortality trends.

\subsection{The ASDRs Model Forecasts}\label{sec3.5}

\begin{figure}[H]
    \centering
    \includegraphics[width=\textwidth]{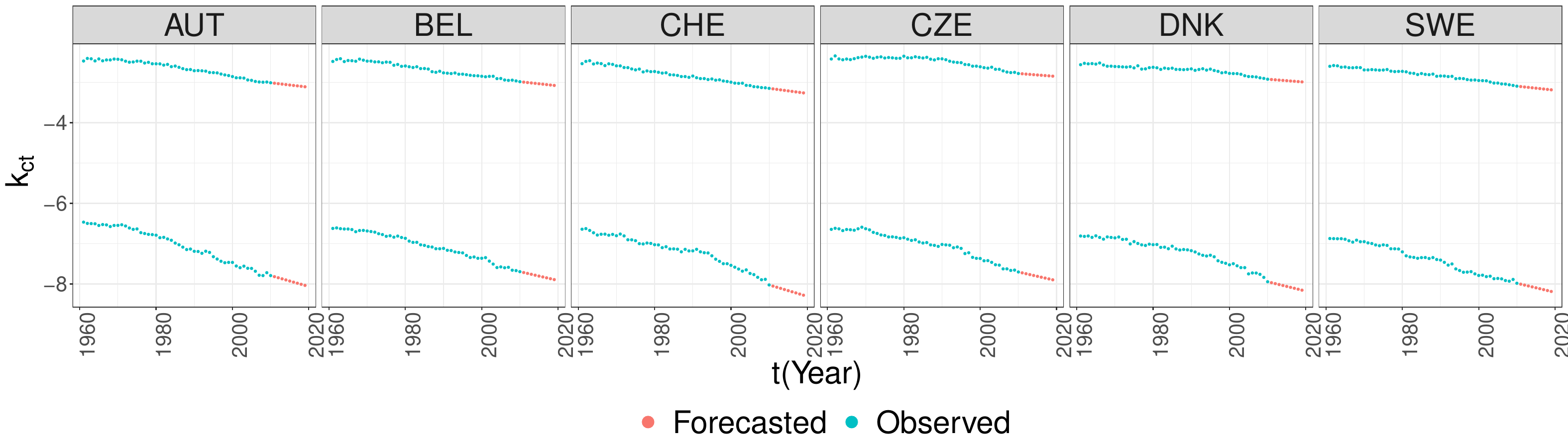}
    \caption{Observed and predicted values of $k_{ct}$ ($1961-2019$).}
    \label{fig16}
\end{figure}

\begin{figure}[H]
    \centering
    \includegraphics[width=\textwidth]{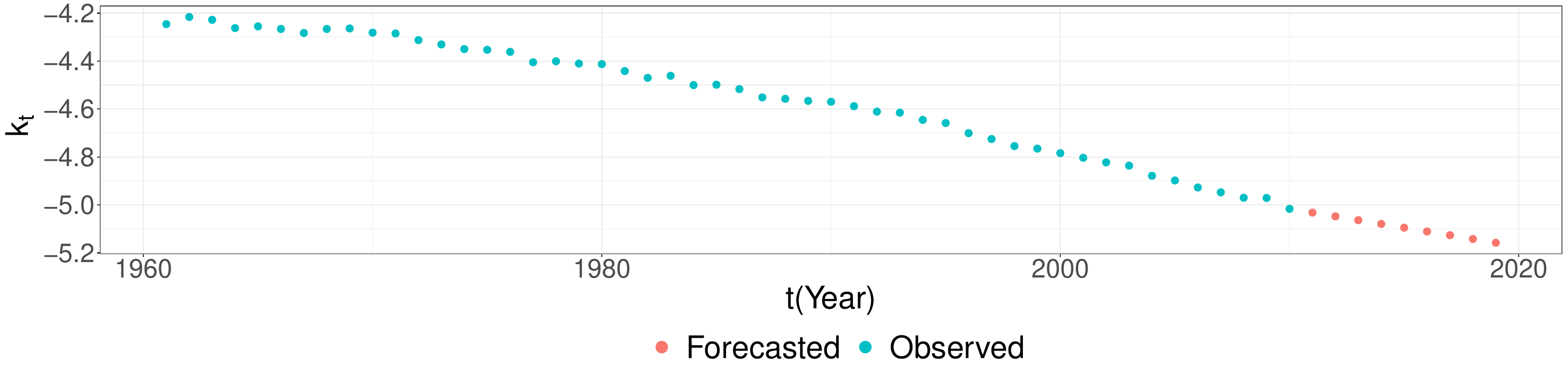}
    \caption{Observed and predicted values of $k_{t}$ ($1961-2019$).}
    \label{fig17}
\end{figure}

We will use the data of $2011-2019$ to assess the predictive quality of the model \eqref{Eq13} that the stepwise algorithm has selected. We have found that the time-dependent covariates $k_{ct}$ and $k_{t}$ can be modeled and forecast using ARIMA models \citep[see][]{box2015time}. We assume random walks with drift (i.e., ARIMA(0,1,0)) for $k_{ct}$
and $k_{t}$:
\begin{align*}
&k_{c{t}}^{[0,40]} =   k_{c{t-1}}^{[0,40]}+\delta_{c}^{[0,40]}+\epsilon_{ct}^{[0,40]} \qquad
\epsilon_{ct}^{[0,40]}\overset{\mathrm{iid}}{\sim}{ N}(0\ ,\ \sigma_{c[0,40]} ^2),\\ 
&  k_{c{t}}^{[41,\omega]}=    k_{c{t-1}}^{[41,\omega]}+\delta_{c}^{[41,\omega]}+\epsilon_{ct}^{[41,\omega]} \qquad
\epsilon_{ct}^{[41,\omega]}\overset{\mathrm{iid}}{\sim}{ N}(0\ ,\ \sigma_{c[41,\omega]} ^2),
\end{align*}
\begin{align*}
\begin{split}
 k_t&=k_{t-1}+\delta+\epsilon_{t},\qquad
\epsilon_{t}\overset{\mathrm{iid}}{\sim}{ N}(0\ ,\ \sigma ^2),
\end{split} 
\end{align*}
where $\delta_{c}^{[0,40]},\delta_{c}^{[41,\omega]}$, and $\delta$ are the drift parameters, and $\epsilon_{ct}^{[0,40]},\epsilon_{ct}^{[41,\omega]}$, and $\epsilon_{t}$ are error terms. The observed and forecasted values of $k_{c{t}}^{[0,41]}$, $k_{c{t}}^{[41,\omega]}$, and $k_{t}$ are presented in Figures \ref{fig16} and \ref{fig17}, respectively. In these figures, the observed values are represented by blue points, while the forecasted values are denoted by red points. Furthermore, the observed and forecasted values for these variables can also be found in Tables \ref{tab8} and \ref{tab9}  in \ref{B}.

Log ASDRs forecasts are generated using estimated parameters: $\beta_{0}$, $\beta_{0x}$, $\beta_{0gx}$, $\beta_{1gx}$, $\beta_{2}$, $\beta_{2gx}$, $\gamma_{{\rm{cohort}}}$, $\eta_{0cgx}$, $\eta_{2cgx}$, and $\eta_{{\rm{cohort}}_{cgx}}$. These parameters, along with the forecasted values for $k_{ct}$ and $k_{t}$, are utilized in the forecasting process. It is assumed that these parameters remain constant and do not change over time. Actuarial applications have shown the importance of estimating uncertainty in mortality forecasting, see \citet{olivieri2001uncertainty, li2009uncertainty}. 

To predict mortality rates using the LME model for the year 2019 and quantify their uncertainty, we employ a simulation approach utilizing the predictInterval function in the R programming language's merTools package \citep{knowles2023package}. We calculate prediction intervals for 2019, which incorporate the uncertainty in fixed coefficients, random effects, and residual variance, with $95\%$ coverage based on 1,000 simulations, and we use the median as the point estimate.

An alternative approach is to perform a bootstrap \citep[see][]{Davison1997Bootstrap, efron1994introduction} simulation utilizing the bootMer function from the R programming language's lme4 package \citep[see][]{JSSv067i01}. This method allows for the estimation of prediction intervals by resampling and is considered a gold-standard approach for capturing the uncertainty inherent in LMMs. However, it is worth noting that obtaining a sufficient number of replications from bootMer can be computationally time-consuming, especially when the initial model fit requires a substantial amount of time, such as hours instead of seconds. 

\begin{figure}[H]
    \centering
    \includegraphics[width=\textwidth]{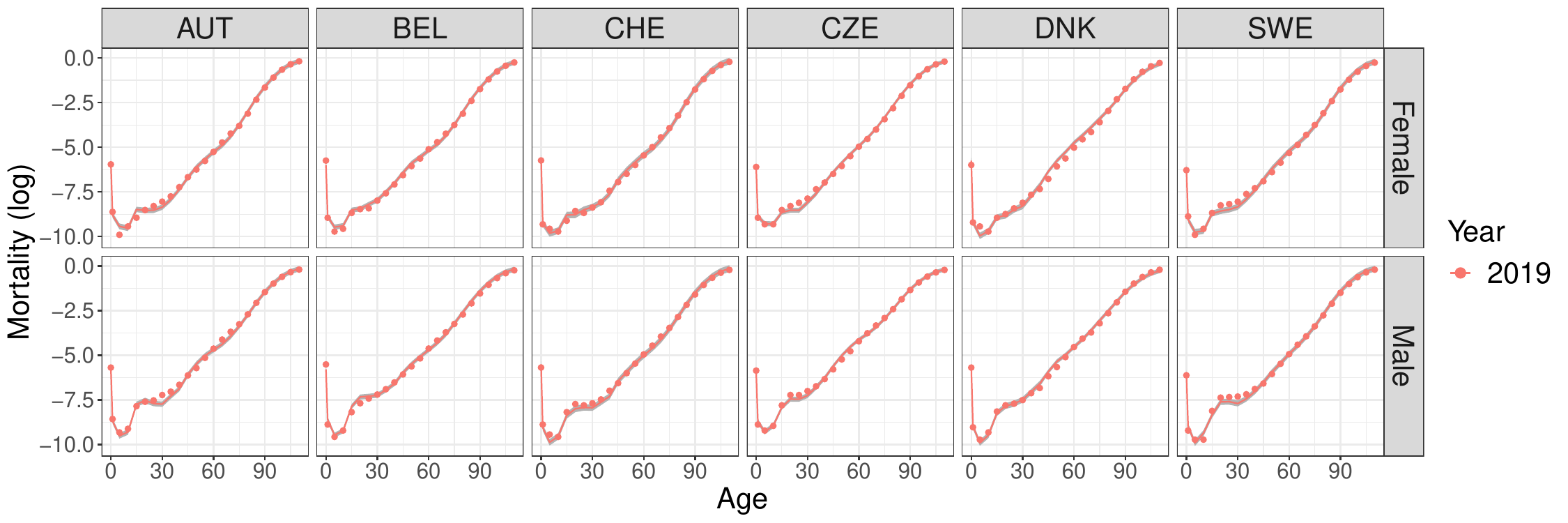}
    \caption{Predicted log mortality rates generated by the LME model for the year 2019. The grey bands represent the $95\%$ uncertainty intervals surrounding our estimates, providing a measure of the model's confidence in the predictions.}
    \label{fig18}
\end{figure}

\begin{figure}[H]
    \centering
    \includegraphics[width=\textwidth]{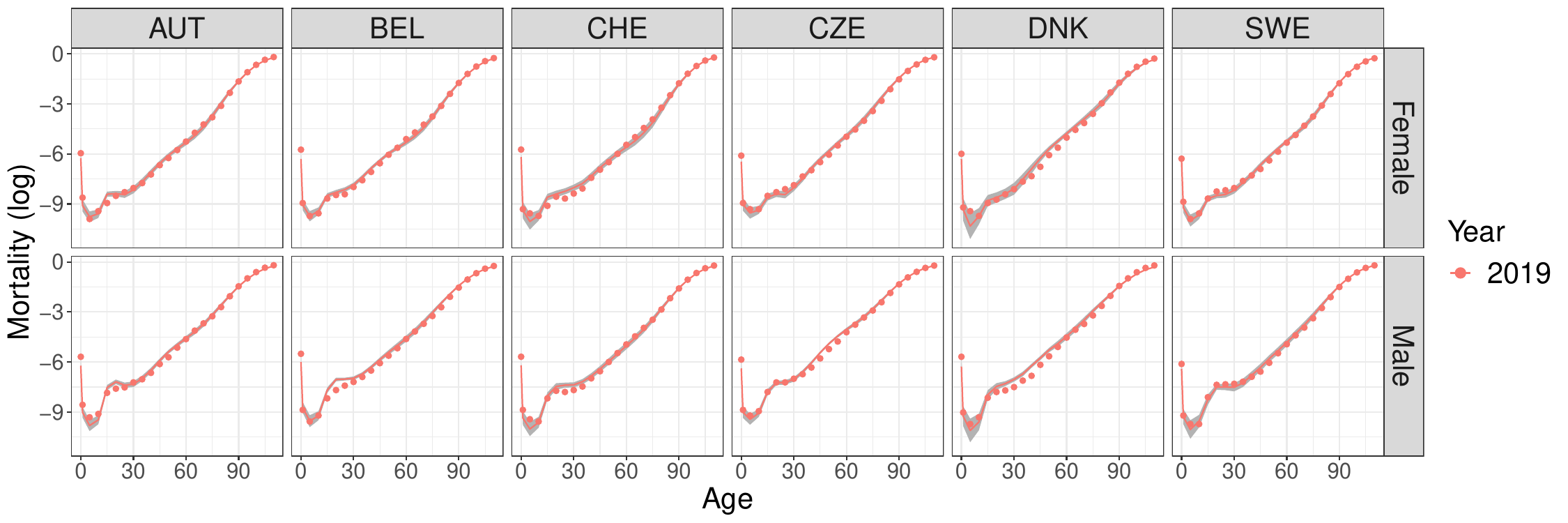}
    \caption{Predicted log mortality rates generated by the LC models for the year 2019. The grey bands represent the $95\%$ uncertainty intervals surrounding the estimates, offering a measure of the LC models' confidence in the predictions.}
    \label{fig19}
\end{figure}

In Figure \ref{fig18} we can see both the predicted mortality rates and their accompanying $95\%$ confidence intervals. Additionally, in Figure \ref{fig19}, the observed and predicted log mortality rates of the LC models for 2019 are shown, along with a $95\%$ confidence interval obtained from the random walk in the LC model. Notably, the confidence intervals of the LME model, particularly for both females and males in CHE, better encompass and capture the log mortality rates compared to the confidence intervals of the LC model.

 As a measure of predictive accuracy, mean square error (MSE) is discussed. The MSE of the twelve populations for the test set ($2011-2019$) are shown in Table \ref{tab3}. We have fitted the LC model to each population separately, and we will use the demographic LC model as a benchmark model to evaluate the performance of the LME model. In the test datasets, the MSEs of the LME model are lower than those of the LC models for 11 out of the 12 populations.
\begin{table}[H]
\huge
    \centering
\resizebox{\textwidth}{!}{
    \centering
\begin{tabular}{|c|c|c|c|c|c|c|c|c|c|c|c|c|}
%\toprule
\multicolumn{13}{c}{} \\\cmidrule{1-13}
%\midrule
\multicolumn{1}{|c}{}&\multicolumn{2}{|c}{AUT} & \multicolumn{2}{|c}{BEL} & \multicolumn{2}{|c}{CHE} & \multicolumn{2}{|c}{CZE} & \multicolumn{2}{|c}{DNK} & \multicolumn{2}{|c|}{SWE} \\

\multicolumn{1}{|c}{Error}&\multicolumn{1}{|c}{Female} & \multicolumn{1}{c}{Male} &\multicolumn{1}{|c}{Female} & \multicolumn{1}{c}{Male} &\multicolumn{1}{|c}{Female} & \multicolumn{1}{c}{Male} &\multicolumn{1}{|c}{Female} & \multicolumn{1}{c}{Male} &\multicolumn{1}{|c}{Female} & \multicolumn{1}{c}{Male} &\multicolumn{1}{|c}{Female} & \multicolumn{1}{c|}{Male}  \\
 \cmidrule{2-13}
LC test set  & 0.0231 & 0.0362 & 0.0237 & 0.0397 & 0.0484 & 0.0457 & 0.0211 & 0.0278 & 0.0615 & 0.0825 & 0.0236 & 0.0230 \\
LME test set & 0.0235 & 0.0191 & 0.0130 & 0.0170 & 0.0182 & 0.0185 & 0.0187 & 0.0154 & 0.0369 & 0.0219 & 0.0232 & 0.0191 \\
\cmidrule{2-13}
LC test set/LME test set& \cellcolor{red!25} 0.9804 &\cellcolor{green!25} 1.8948 &\cellcolor{green!25} 1.8307 &\cellcolor{green!25} 2.3274 &\cellcolor{green!25} 2.6587 &\cellcolor{green!25} 2.4685 & \cellcolor{green!25}1.1289 &\cellcolor{green!25} 1.8064 &\cellcolor{green!25} 1.6667 & \cellcolor{green!25}3.7648 &\cellcolor{green!25} 1.0195 &\cellcolor{green!25} 1.2007\\
\hline
\end{tabular}
}
\caption{MSE comparison between the LME and LC Models on predicted log ASDRs.}
\label{tab3}
\end{table}

\subsection{Exploring Life Expectancy: Trends, 
Forecasts, and Gender Analysis}\label{sec3.6}

\begin{figure}[H]
    \centering
    \includegraphics[width=\textwidth]{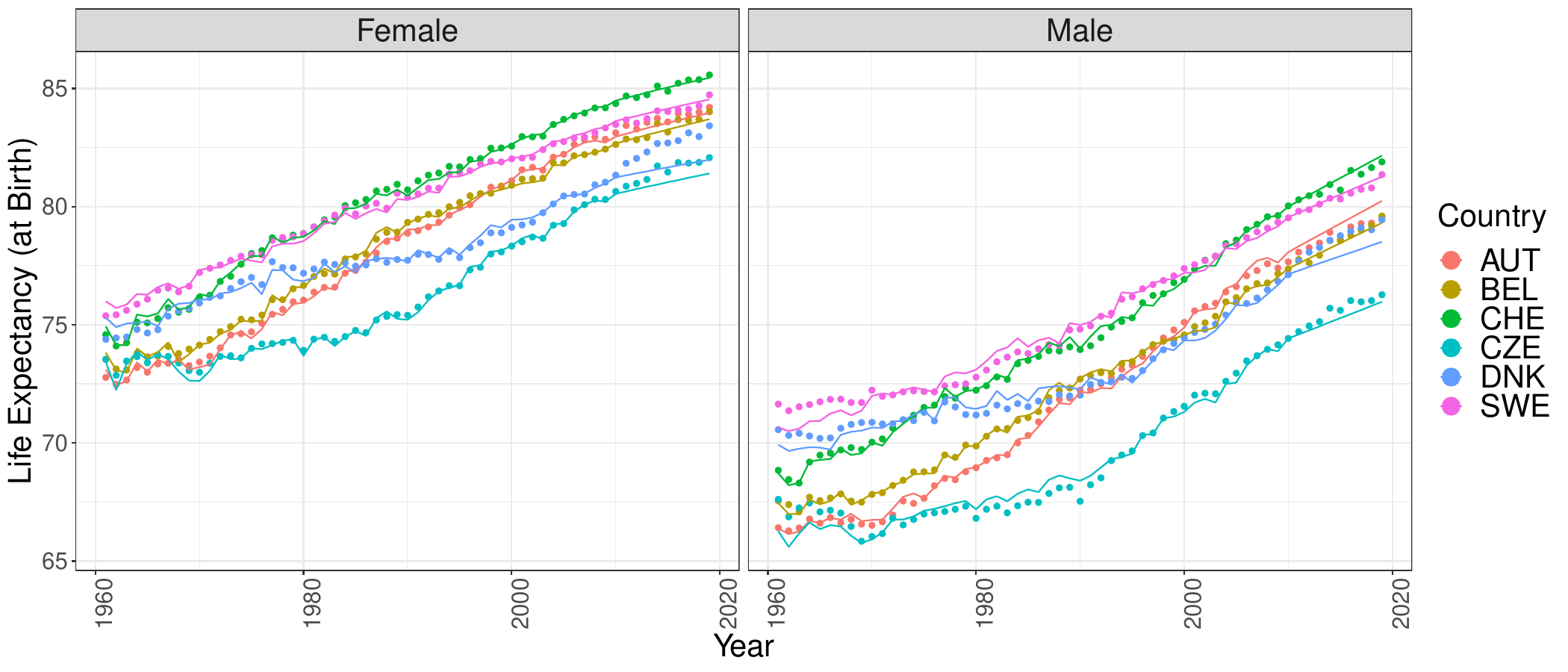}
    \caption{Life expectancy trends (the LME model): A gender perspective across the six European countries}
    \label{fig20}
\end{figure}

\begin{figure}[H]
    \centering
    \includegraphics[width=\textwidth]{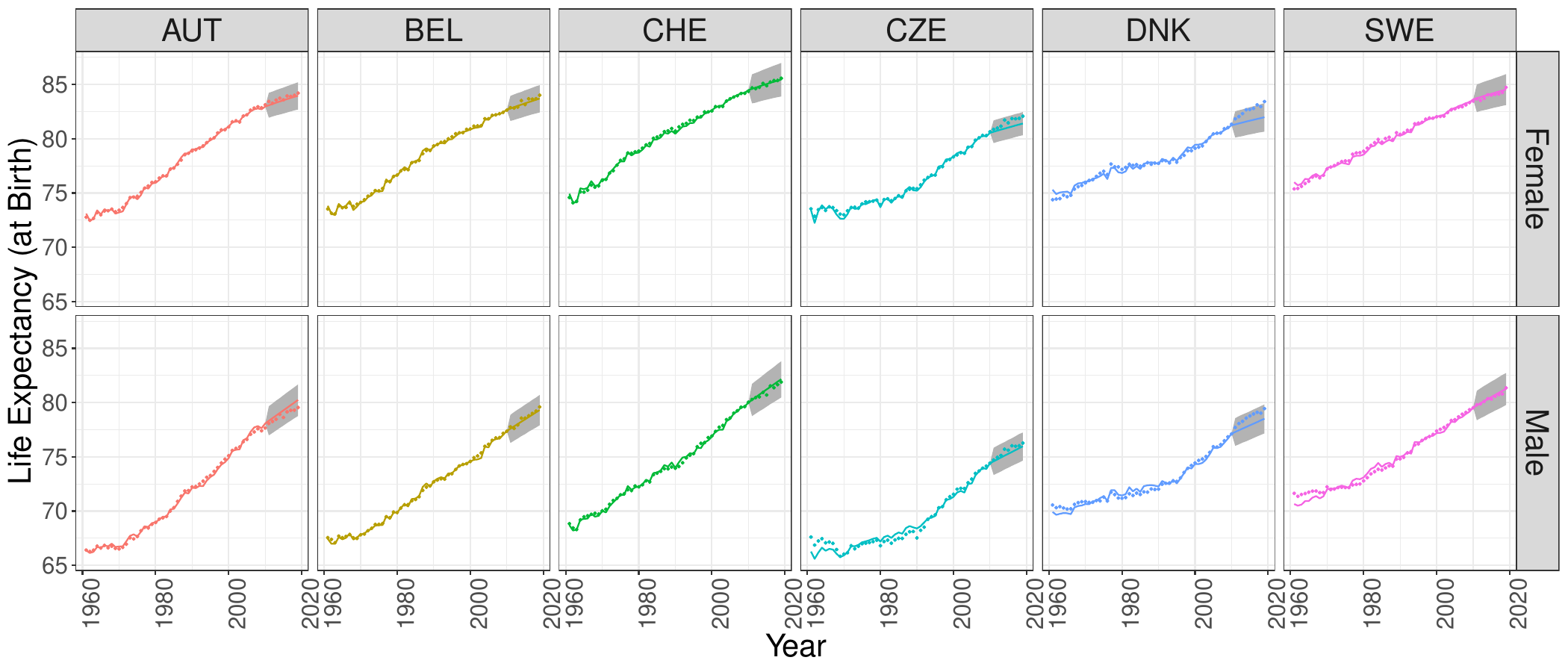}
    \caption{Life expectancy and forecasts (the LME model): Predicted with confidence intervals and uncertainty.}
    \label{fig21}
\end{figure}

Within Figures \ref{fig20} and \ref{fig21}, two sets of forecasts are displayed: Figure \ref{fig20} presents forecasts without the incorporation of $95\%$ confidence intervals, while Figure\ref{fig21} showcases forecasts along with $95\%$ confidence intervals. These intervals are derived from the $95\%$ uncertainty intervals for death rate forecasts. This visualization
provides valuable insights into projected life expectancy, highlighting potential
variations and forecast confidence. Separate visualizations for females and males enable a comparative analysis of their life expectancy trajectories, offering insights into evolving dynamics and potential gender disparities. Notably, the analysis reveals a remarkable growth in life expectancy for Switzerland (CHE), surpassing the other five European countries studied. Belgium exhibits a consistent average life expectancy from 1961 to 2019, while Czech females experience an increasing trend with a noticeable divergence from other countries between $1975$ and $1995$. However, from $1995$ to $2019$, Czech female life expectancy becomes comparable to the other countries. Czech males also witness an increasing life expectancy, but the gap with the other countries observed since $1975$ persists, indicating a continued disparity. Additionally, it is worth noting that the lowest life expectancy values among females across the six European countries consistently surpass the highest life expectancy values observed among males within the same six European countries for each respective year.

Figure \ref{fig21} presents life expectancy trends and forecasts based on the LME model for each gender in the six studied countries. The forecasts closely align with the target values, except for females and males in Denmark between 2011 and 2019. However, it is important to note that these forecasts, deviating by approximately 1 year from the target values, fall within the corresponding confidence intervals.

The plotted life expectancies for the six European countries obtained from the LME and LC models reveal important insights. The LME model demonstrates superior performance, with its predicted values closely aligning with the target values. Notably, the confidence intervals for life expectancies derived from the LME model effectively capture the target values, resolving the issue of the LME model's earlier discrepancy in forecasting years 2011 to 2019 for DNK, where the predicted values were approximately one year off.

\begin{figure}[H]
    \centering
    \includegraphics[width=\textwidth]{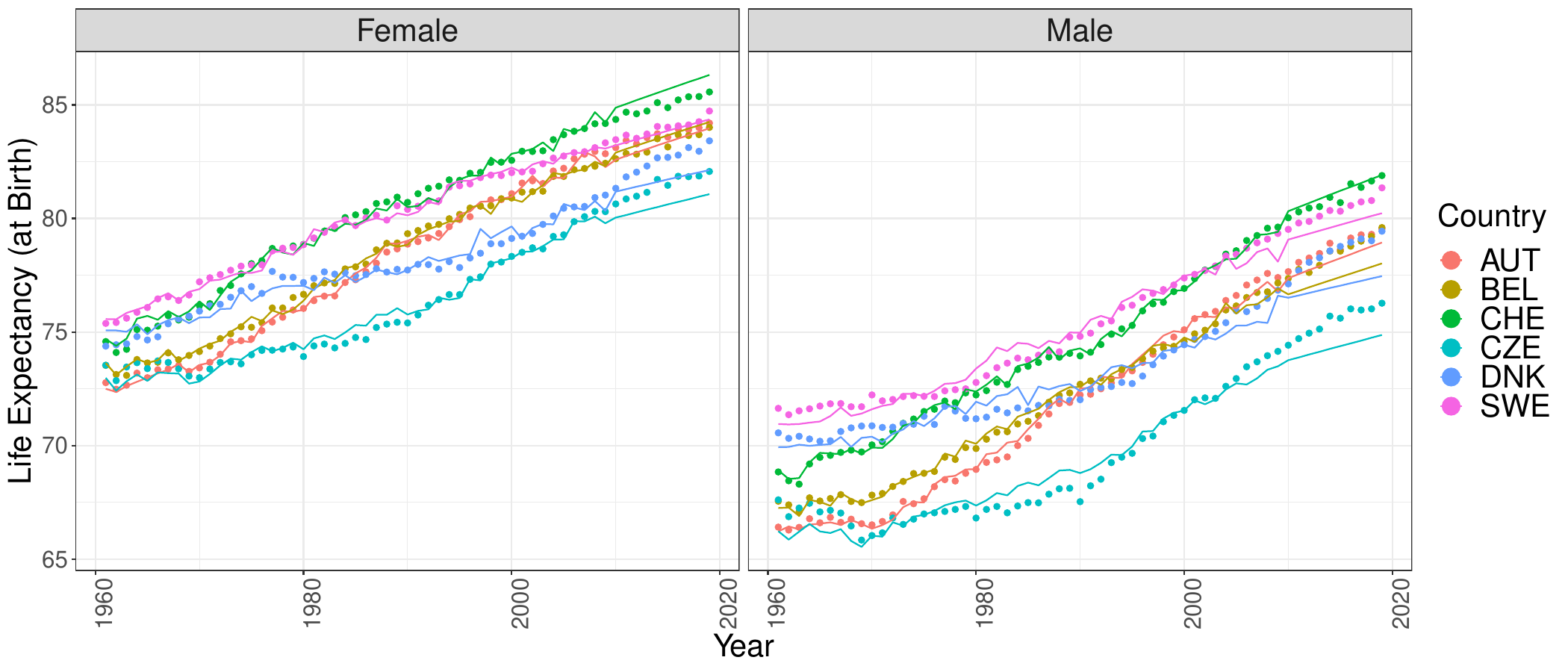}
    \caption{Life expectancy trends (the LC models): A gender perspective across the six European countries.}
    \label{fig22}
\end{figure}

\begin{figure}[H]
    \centering
    \includegraphics[width=\textwidth]{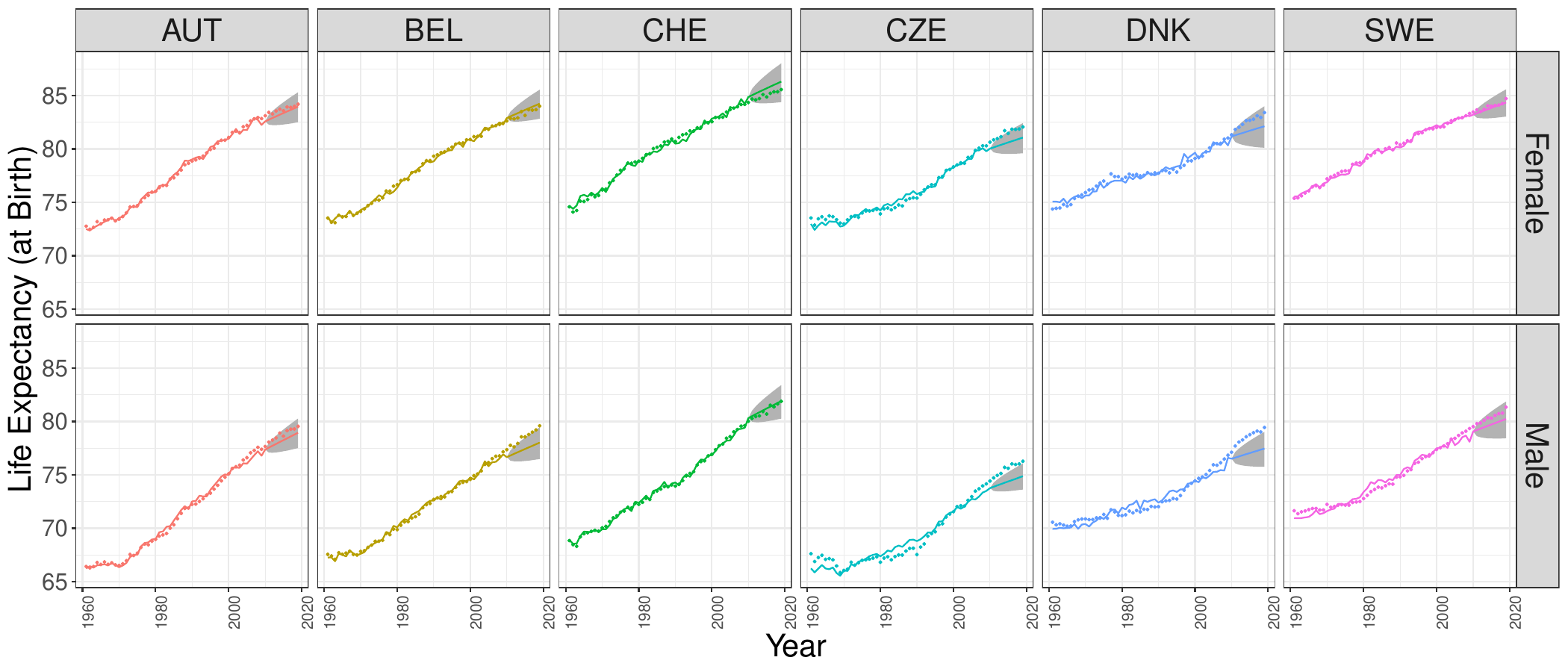}
    \caption{Life expectancy and forecasts (the LC models): Predicted with confidence intervals and uncertainty.}
    \label{fig23}
\end{figure}

Conversely, the life expectancy values obtained from the LC model for all six European countries are not as accurate when compared to the LME model. Particularly concerning is the fact that the confidence intervals for life expectancies derived from the LC model, specifically for males in BEL, CZE, and DNK, fail to encompass the target values. This discrepancy highlights the limitations of the LC model in accurately predicting life expectancies. Therefore, the comparison of the plots reveals that the LME model outperforms the LC model in estimating life expectancies, see Figures \ref{fig22} and \ref{fig23}. 

\subsection{Comparing the Prediction Accuracy of the LME and LL Models}\label{sec3.7}
We aim to utilize both the LME and LL mortality models to analyze mortality rates of the twelve populations, spanning age groups from 0 to 100 years. Our primary objective is to compare the forecast accuracy of these two models. Our analysis will involve two distinct datasets: a training dataset covering the period from 1961 to 2010, which will serve as the foundation for model development, and a test dataset encompassing the years 2011 to 2019, designed to assess the performance of the models. The dataset is organized into 5-year age intervals (5 × 1 intervals).

According to the hybrid stepwise selection, the best-fit model for the log ASDRs of the twelve populations is the model

\begin{eqnarray}\label{Eq19}
y_{cgxt} &=& 
x+g:x+{\rm{I}}(k_{t})+g:x:{\rm{I}}(k_{ct})+{\rm{I}}(k_{t}^2)+g:x:{\rm{I}}(k_{ct}^2)+\rm{cohort}+\nonumber\\&&({\rm{I}}(k_{t}^2)+{\rm{cohort}}|c:g:x)=
\nonumber
\\ 
&& \beta_{0}+\beta_{0x}+\beta_{0gx}+\beta_{1}k_{t}+\beta_{1gx}k_{ct}+\beta_{2}k_{t}^2+\beta_{2gx}k_{ct}^2+\gamma_{\rm{cohort}}(t-x)+\nonumber\\ &&\eta_{0cgx}+\eta_{2cgx} k_{t}^2+\eta_{{\rm{cohort}}_{cgx}}(t-x)+\epsilon_{cgxt}.
\end{eqnarray}

We utilize the MultiMoMo package (Copyright (c) 2020 Katrien Antonio; Sander Devriendt; Jens Robben) to fit the LL model to the dataset. In terms of forecasting accuracy, the LME model exhibit superior predictive performance, with their MSEs surpassing those of the LL models for 12 out of 12 populations. This notable improvement is especially pronounced for females in Denmark, where the forecast accuracy of the LME model is approximately 20 times better than that of the LL model. These results are summarized in Table \ref{tab4}.
\begin{table}[H]
\Huge
    \centering
\resizebox{\textwidth}{!}{
    \centering
\begin{tabular}{|c|c|c|c|c|c|c|c|c|c|c|c|c|}
%\toprule
\multicolumn{13}{c}{} \\\cmidrule{1-13}
%\midrule
\multicolumn{1}{|c}{}&\multicolumn{2}{|c}{AUT} & \multicolumn{2}{|c}{BEL} & \multicolumn{2}{|c}{CHE} & \multicolumn{2}{|c}{CZE} & \multicolumn{2}{|c}{DNK} & \multicolumn{2}{|c|}{SWE} \\

\multicolumn{1}{|c}{Error}&\multicolumn{1}{|c}{Female} & \multicolumn{1}{c}{Male} &\multicolumn{1}{|c}{Female} & \multicolumn{1}{c}{Male} &\multicolumn{1}{|c}{Female} & \multicolumn{1}{c}{Male} &\multicolumn{1}{|c}{Female} & \multicolumn{1}{c}{Male} &\multicolumn{1}{|c}{Female} & \multicolumn{1}{c}{Male} &\multicolumn{1}{|c}{Female} & \multicolumn{1}{c|}{Male}  \\
 \cmidrule{2-13}
LL test set  & 0.000518& 0.000606& 0.000524& 0.001048& 0.000466&
 0.000970& 0.001018& 0.002316 &0.000574& 0.000889& 0.000807& 0.001092\\
LME test set & 0.000205& 0.000451&0.000042& 0.000290 &0.000121&
  0.000404& 0.000111 &0.000521& 0.000028& 0.000221&
 0.000051& 0.000213\\
\cmidrule{2-13}
LL test set/LME test set&\cellcolor{green!25}2.525865 &\cellcolor{green!25} 1.344800&\cellcolor{green!25} 12.456332&\cellcolor{green!25}  3.618779&\cellcolor{green!25}  3.849733&\cellcolor{green!25}  2.402769&\cellcolor{green!25}  9.175995&\cellcolor{green!25}
  4.441440 &\cellcolor{green!25}20.493963 &\cellcolor{green!25} 4.027966 &\cellcolor{green!25}15.741459&\cellcolor{green!25}  5.125900\\
  \hline
\end{tabular}
}
\caption{MSE comparison between the LME and LL Models on predicted ASDRs.}
\label{tab4}
\end{table}

\subsection{The All-in-One LME Model: Integrating Age, Gender, Country, Cohort, and GDP per Capita Effects}\label{sec3.8}

\citet{boonen2017modeling} have introduced a novel approach to mortality modelling for multiple populations. Instead of relying solely on historical mortality data and latent factors, the authors propose a multipopulation stochastic mortality model that incorporates the influence of economic growth, specifically using real GDP per capita. Building upon the LL model, this extended framework aims to capture common mortality trends among populations with similar socioeconomic conditions. The incorporation of GDP into mortality models is of paramount importance as it recognizes the substantial impact of economic factors on mortality patterns, providing a more comprehensive understanding of population dynamics and enhancing the accuracy of mortality projections. 

Here, our objective is to demonstrate how we can augment LME model by introducing the influence of economic growth, specifically by integrating real GDP per capita as a variable within the LME framework. We will outline the modifications needed to incorporate this economic factor into the LME model. The modified LME model, which incorporates this economic factor, will be structured as follows:
\begin{align*}
\begin{split}
 & \quad\quad y_{cgx}=X_{cgx}\beta+Z_{cgx}\eta_{cgx}+Z_{c,gx}\eta_{c}+\epsilon_{cgx},\\
& 
 \quad c=1, 2, \cdots M,\quad g=1, 2,\quad x=0, 1,\cdots,\omega,\\ &\quad  \quad\quad
 \eta_{cgx}\overset{\mathrm{iid}}{\sim}{\cal N}(0\ ,\ \Psi),\
\epsilon_{cgx}\overset{\mathrm{iid}}{\sim}{\cal N}(0\ , \ \sigma ^2 I),
\end{split}
\end{align*}
where

\begin{align*}\nonumber
    y_{cgx}& = 
    \setlength{\tabcolsep}{4pt}
\renewcommand{\arraystretch}{1}
\begin{bmatrix}
		y_{cgx0} \\
		y_{cgx1}  \\
	\vdots    \\
		y_{cgxT}  
\end{bmatrix},
\\
 X_{cgx}&=
 \setlength{\tabcolsep}{3pt}
\renewcommand{\arraystretch}{1}
	\begin{bmatrix}
		1 &1 & 1 & k_{0} & k_{c0}& \cdots &  k_{0}^{n_{1}} & k_{c0}^{n_{1}}&0-x&{\rm{GDP}}\\1 &1 & 1 & k_{1} & k_{c1}& \cdots &  k_{1}^{n_{1}} & k_{c1}^{n_{1}}&1-x&{\rm{GDP}}\\
		\vdots  &\vdots  &\vdots  &\vdots  &\vdots   &\ddots  &\vdots &\vdots &\vdots    \\
		1 &1 & 1 & k_{T} & k_{cT}& \cdots &  k_{T}^{n_{1}} & k_{cT}^{n_{1}}&T-x&{\rm{GDP}}
	\end{bmatrix},
&
 \beta &= 
   \setlength{\tabcolsep}{3pt}
\renewcommand{\arraystretch}{1}
\begin{bmatrix}
	\beta_{0}  \\
 \beta_{0x}  \\
 \beta_{0gx}  \\
    \beta_{1}  \\
     \beta_{1gx}  \\
     \vdots   \\
  \beta_{{n_{1}}} \\	
   \beta_{{n_{1}}gx} \\
 \gamma_{\rm{cohort}}\\
 \gamma_{\rm{GDP}}
\end{bmatrix},
\nonumber \\ 
 Z_{cgx}&=
   \setlength{\tabcolsep}{3pt}
\renewcommand{\arraystretch}{1}
	\begin{bmatrix}
		1 & k_{0} & k_{0} ^{2}& \cdots &k_{0} ^{n_{2}}&0-x\\
		1 & k_{1} & k_{1} ^{2} &\cdots &k_{1} ^{n_{2}}&1-x\\
		\vdots  & \vdots& \vdots  & \ddots & \vdots &\vdots \\
		1 & k_{T} & k_{T}^{2}& \cdots &k_{T} ^{n_{2}}&T-x
	\end{bmatrix},
 &
\eta_{cgx} &= 
  \setlength{\tabcolsep}{3pt}
\renewcommand{\arraystretch}{1}
\begin{bmatrix}
	\eta_{0cgx}  \\
	\eta_{1cgx} \\
 \vdots   \\
	\eta_{n_{2}cgx}\\
 \eta_{{\rm{cohort}}_{cgx}}
\end{bmatrix},
\nonumber \\ 
 Z_{c,gx}&=
   \setlength{\tabcolsep}{3pt}
\renewcommand{\arraystretch}{1}
	\begin{bmatrix}
		\rm{GDP} \\
		\rm{GDP} \\
		\vdots   \\
		\rm{GDP} 
	\end{bmatrix},
 &
\eta_{c,gx} &= 
  \setlength{\tabcolsep}{3pt}
\renewcommand{\arraystretch}{1}
\begin{bmatrix}
	\eta_{{\rm{GDP}}_c}  
\end{bmatrix},
\nonumber\\
\epsilon_{cgx}& = 
  \setlength{\tabcolsep}{3pt}
\renewcommand{\arraystretch}{1}
\begin{bmatrix}
	\epsilon_{cgx0} \\
	 \epsilon_{cgx1} \\
	\vdots   \\
	\epsilon_{cgxT}
\end{bmatrix},
\end{align*}

To fit the model using the lmer function in the lme4 package in the R programming language \citep[see][]{JSSv067i01}, the LME model can be specified as follows:

\begin{eqnarray}\nonumber 
y_{cgxt} &=& 
x+g:x+{\rm{I}}(k_{t})+g:x:{\rm{I}}(k_{ct})+{\rm{I}}(k_{t}^2)+g:x:{\rm{I}}(k_{ct}^2)+\rm{cohort}+{\rm{GDP}}+\nonumber\\&&({\rm{I}}(k_{t})+{\rm{I}}(k_{t}^2)+{\rm{cohort}}|c:g:x)+({\rm{GDP}}|c)\nonumber
\end{eqnarray} 
Given that GDP per capita varies across countries but remains constant across age and gender, we incorporate the term $({\rm{GDP}}|c)$ into the LME model to capture random effects for GDP across different countries.
\section{The ASDRs Model for Single Population}\label{sec4}

Let $m_{x,t}$ denote the ASDR at age $x$ and time $t$, for $x = 0,1, \cdots , \omega$ and $t=0,1, \cdots, T$, see \citet{dickson2019actuarial}. Let $y_{xt}= \log (m_{x,t})$, the log of the observed age-specific death rates in a given year $t$. Let $k_{2_t}$ represent the average of $y_{xt}$ for two distinct age range groups: the average for age groups 0 to 40 and the average for age groups 41 to $\omega$ at time $t$. Additionally, let the average of $y_{xt}$ at time $t$ be $k_t$:

\begin{align}
k_{2_t}  = &
\begin{cases} 
k_{2_t} ^{[0,40]} & \text{if}\ x \in [0,40], \\ 
     k_{2_t} ^{[41,\omega]} & {\rm otherwise},
\end{cases};
\qquad
k_{t} = \dfrac{\sum\limits_{x=0}^{\omega} y_{xt}}{\omega +1},
\label{ali20}
\end{align}
for $t=0, 1,  \cdots, T$, where
\begin{align*}
k_{2_t}^{[0,40]}=
\dfrac{\sum\limits_{x=0}^{40} y_{xt}}{41}  , \qquad
k_{2_t}^{[41,\omega]}=\dfrac{\sum\limits_{x=41}^{\omega} y_{xt}}{\omega-40},
\end{align*}
for $t=0, 1,  \cdots, T$.

The linear mixed-effects model of each individual population expresses the $(T+1)$-dimensional response vector $y_x$ for age group $x$ as

\begin{align}
\begin{split}
y_{x}=X_{x}\beta+Z_{x}\eta_{x}+\epsilon_{x},\quad x=0,1,\cdots,\omega,\\ 
\eta_x\overset{\mathrm{iid}}{\sim}{\cal N}(0_{(n_2+1)}\ , \ \Psi),\quad
\epsilon_{x}\overset{\mathrm{iid}}{\sim}{\cal N}(0_{(T+1)}\ ,\ \sigma ^2 I),
\end{split}
\label{ali21}
\end{align}
where

\begin{align}\nonumber
    y_{x}& = 
    \setlength{\tabcolsep}{4pt}
\renewcommand{\arraystretch}{1}
\begin{bmatrix}
		y_{x0} \\
		y_{x1}  \\
	\vdots    \\
		y_{xT}  
\end{bmatrix},
\\
 X_{x}&=
 \setlength{\tabcolsep}{3pt}
\renewcommand{\arraystretch}{1}
	\begin{bmatrix}
		1  & k_{0} & k_{2_0}& \cdots &  k_{0}^{n_{1}} & k_{2_0}^{n_{1}}&0-x\\1 & k_{1} & k_{2_1}& \cdots &  k_{1}^{n_{1}} & k_{2_1}^{n_{1}}&1-x\\
		\vdots  &\vdots  &\vdots   &\ddots  &\vdots &\vdots &\vdots    \\
		1  & k_{T} & k_{2_T}& \cdots &  k_{T}^{n_{1}} & k_{2_T}^{n_{1}}&T-x
	\end{bmatrix},
&
 \beta &= 
   \setlength{\tabcolsep}{3pt}
\renewcommand{\arraystretch}{1}
\begin{bmatrix}
	\beta_{0}  \\
     \beta_{1}  \\
    \beta_{2_1}  \\
      \vdots   \\
  \beta_{{n_{1}}} \\	
   \beta_{2_{n_{1}}}  \\
  \gamma_{\rm{cohort}}
\end{bmatrix},
\nonumber \\ 
 Z_{x}&=
   \setlength{\tabcolsep}{3pt}
\renewcommand{\arraystretch}{1}
	\begin{bmatrix}
		1 & k_{0} & k_{0} ^{2}& \cdots &k_{0} ^{n_{2}}&0-x\\
		1 & k_{1} & k_{1} ^{2} &\cdots &k_{1} ^{n_{2}}&1-x\\
		\vdots  & \vdots& \vdots  & \ddots & \vdots &\vdots \\
		1 & k_{T} & k_{T}^{2}& \cdots &k_{T} ^{n_{2}}&T-x
	\end{bmatrix},
 &
\eta_{x} &= 
  \setlength{\tabcolsep}{3pt}
\renewcommand{\arraystretch}{1}
\begin{bmatrix}
	\eta_{0x}  \\
	\eta_{1x} \\
 \vdots   \\
	\eta_{n_{2}x}\\
 \eta_{{\rm{cohort}}_{x}}
\end{bmatrix},
\nonumber \\ 
 \epsilon_{cg}& = 
  \setlength{\tabcolsep}{3pt}
\renewcommand{\arraystretch}{1}
\begin{bmatrix}
	\epsilon_{cgx0} \\
	 \epsilon_{cgx1} \\
	\vdots   \\
	\epsilon_{cgxT}
\end{bmatrix},
\label{ali22}
\end{align}
where $n_1$ and $n_2$ are non-negative integers; $\Psi$ is a positive-definite, symmetric $(n_2+2)\times(n_2+2)$ matrix; $\beta$ is the fixed effects vector, and $\eta_{x}$ is the random effects vector; $X_x$ and $Z_x$ are known regressor matrices for the fixed-effects and random-effects; and $\epsilon_{x}$ is the within-age-group error vector. We assume that the random effects $\eta_{x}$ and the within-age-group errors $\epsilon_{x}$ are independent for different age groups. We also assume they are independent of one another for the same age group \citep[see][chap.~2]{pinheiro2006mixed}.

The model \eqref{ali21} exhibits identifiability due to our assumption that the covariance matrix of $\epsilon$ is a scalar multiple of a known positive definite matrix (specifically, the identity matrix). Mathematically, it can be expressed as:
\begin{align*}
\epsilon \overset{\mathrm{iid}}{\sim} \mathcal{N}(0_{(\omega+1)(T+1)}, \sigma^2 I),
\end{align*}
for further details, see \citet{wang2013identifiability}.

In the previous section, we conducted a joint estimation of the ASDRs for twelve populations using a single model. However, in this section, our analysis expands to include a larger dataset consisting of fifty-eight populations. To be included in our analysis, populations are required to have a dataset spanning at least 45 years before the year 2000. We obtained this data from the HMD, and it is organized into five-year age intervals ($5\times1$ intervals). Our goal is to examine the ASDRs of these fifty-eight populations individually and explore any patterns or trends that may emerge. We consider both females and males of AUS, AUT, BEL, BGR, CAN, CHE, CZE, DEUTE, DEUTW, DNK, ESP, FIN, FRATNP, GBR-NIR, GBR-NP, GBR-SCO, GBRTENW, HUN, IRL, ISL, ITA, JPN, NLD, NOR, NZL-NM, PRT, SVK, SWE, and USA. For our analysis, we have chosen a sample of age groups ranging from 45 to 90 years. The mortality rates prior to the year 2000 serve as our training dataset, which we use to estimate the parameters of the ASDRs models. From the year 2000 to 2019 (prior to the onset of the Covid-19 pandemic), we forecast the ASDRs for each individual population. The data are repeated measurements made on the same age groups. Our objective is to individually apply a specific variant of the LME model \eqref{ali21}, which does not account for cohort effects, to each unique population, as follows:
\begin{eqnarray}\label{Eq23}
y_{xt} &=&  {\rm{I}}(k_{t})+ ({\rm I}(k_{t})|x)=\nonumber\\  
 && \beta_{0}+\beta_{1}k_{t}+\eta_{0x} + \eta_{1x} k_{t}+ \epsilon_{xt}.
\end{eqnarray}

The REML estimation technique is used to fit the model \eqref{Eq23} to each individual dataset \citep[see][]{JSSv067i01} .

The model \eqref{Eq23} assumes different intercepts and slopes for different age groups. Hence, the intercept and the slope may change from one age group to another. It also includes two random effects, one for intercept and the other for slope. The parameters $\beta_{0}$ and $ \beta_{1}$ are, respectively, the intercept and slope fixed effects, and they are the same across ages. The parameters $\eta_{0x}$ and $ \eta_{1x}$ are, respectively, the intercept and slope random effects, and they change across age groups.  

The model \eqref{Eq23} can be expressed as 

\begin{eqnarray} \label{Eq24}
\log (m_{x,t})&=& a_x+b_x k_t+\epsilon_{xt},
\end{eqnarray}
where
\begin{equation*}
     a_{x}=\beta_0 + \eta_{0x}, \quad
    b_{x}= \beta_{1}+ \eta_{1x}.
\end{equation*}

The model \eqref{Eq24} closely resembles the LC model \eqref{Eq2} in its mathematical structure. However, it is important to note that the meanings we assign to the terms of the model \eqref{Eq24} are completely different from those used in the original LC model. While the equations may look the same, the model \eqref{Eq24} interprets each term in a distinct way, leading to unique implications. Essentially, the model \eqref{Eq24} and the LC model share the same form, but the underlying concepts and interpretations are entirely separate. In the model \eqref{Eq24}, $k_t$ is the known mortality covariate obtained from the equation \eqref{ali20}. $a_x$ and $b_x$ are obtained by adding the fixed effects of the intercept and slope to the random effects of the intercept and slope, respectively. We can write the model \eqref{Eq24} as 

\begin{eqnarray*}
y_{xt} &=&  \beta_{0}+\eta_{0x}+(\beta_{1} + \eta_{1x}) k_{t}+ \epsilon_{xt} \\ \nonumber &=& \text{fixed-effect intercept
+ by-age-group random variation in the intercept}+\\  && (\text{fixed-effect slope + by-age-group random variation in the slope})k_t+ \\ \nonumber
& & \epsilon_{xt}.
\end{eqnarray*}

\begin{figure}[H]
    \centering
    \includegraphics[width=\textwidth]{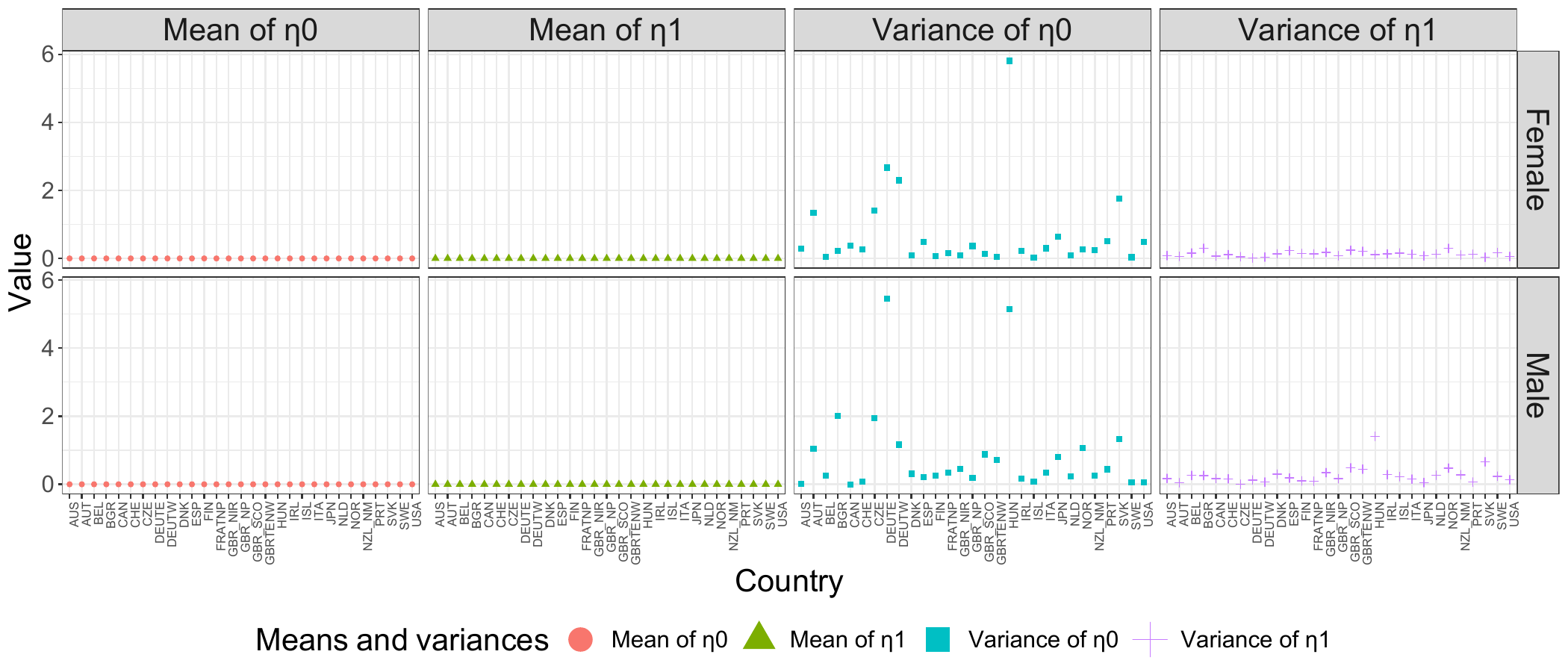}
    \caption{Mean and variance of random effects in the $58$ LME models for males and females across the $29$ countries.}
    \label{fig24}
\end{figure}

\begin{figure}[H]
    \centering
    \includegraphics[width=\textwidth]{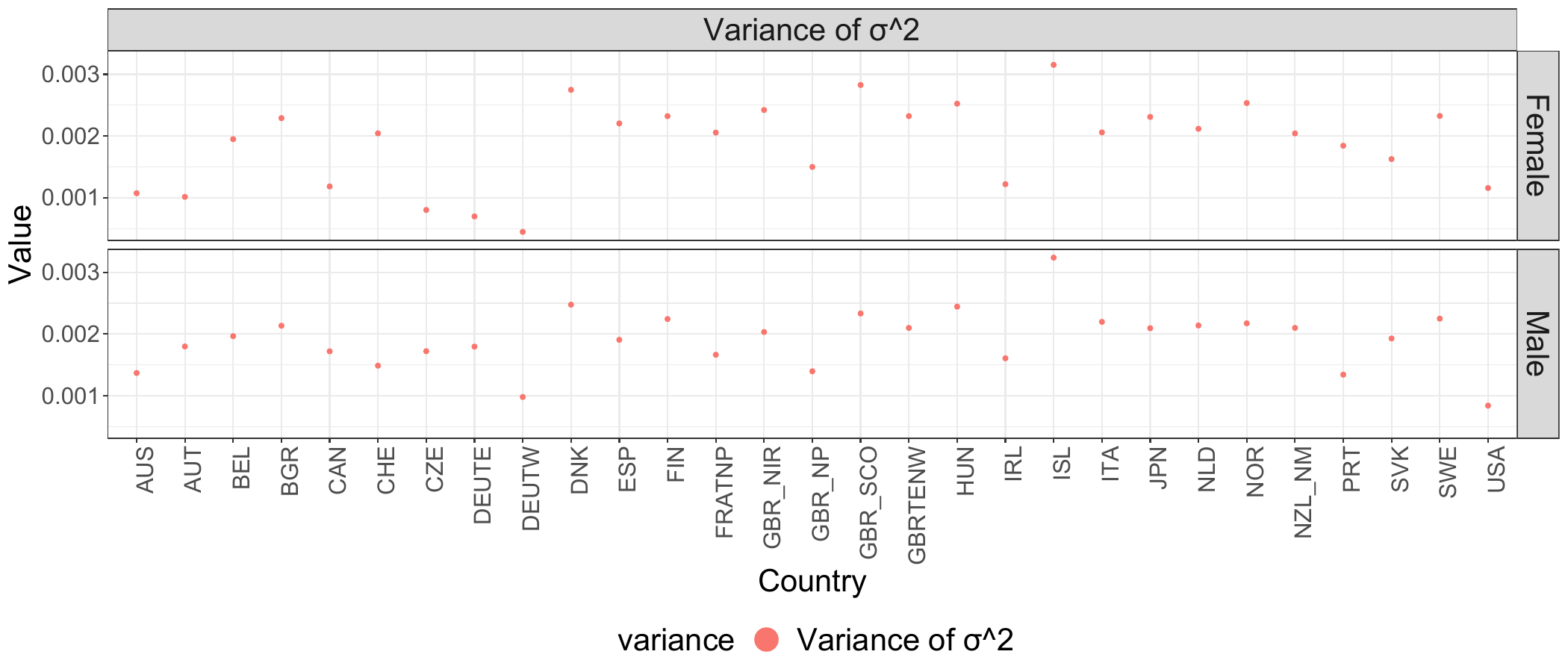}
    \caption{Variance of residual errors in the $58$ LME models for males and females across the $29$ countries.}
    \label{fig25}
\end{figure}

In our analysis of the 58 datasets using the LME model, we consistently observe that the means of the random effects for both the intercept and slope are nearly zero. This indicates that, on average, the random effects associated with the intercept and slope do not significantly deviate from the overall baseline values within each dataset. Additionally, we find that the variances of the random effects for the intercept and slope are consistently small in the majority of the LME models, with values less than 1 observed in 44 out of the 58 models for the intercept and 57 out of the 58 models for the slope. These small variances suggest that the random effects contribute relatively less to the overall variability in the model compared to the fixed effects. Furthermore, it is consistently observed that the variance in each model is split between the residuals and the random effects of the intercept and slope. The estimated variances of the random effects for the intercept and slope capture the variability specific to each age group, while the variance of the residuals represents the within-age group variability. This partitioning of variance underscores the importance of accounting for both within-age group and between-age group variability to obtain a comprehensive understanding of the data. Taken together, our findings across all 58 datasets support the consistent observation of near-zero means and small variances for the random effects of the intercept and slope, indicating that the random effects have a relatively smaller influence on the overall variability compared to the fixed effects. However, it is important to acknowledge that the random effects still contribute to capturing the variability specific to each age group, see Figures \ref{fig24} and \ref{fig25}.


\begin{figure}[H]
    \centering
    \includegraphics[width=\textwidth]{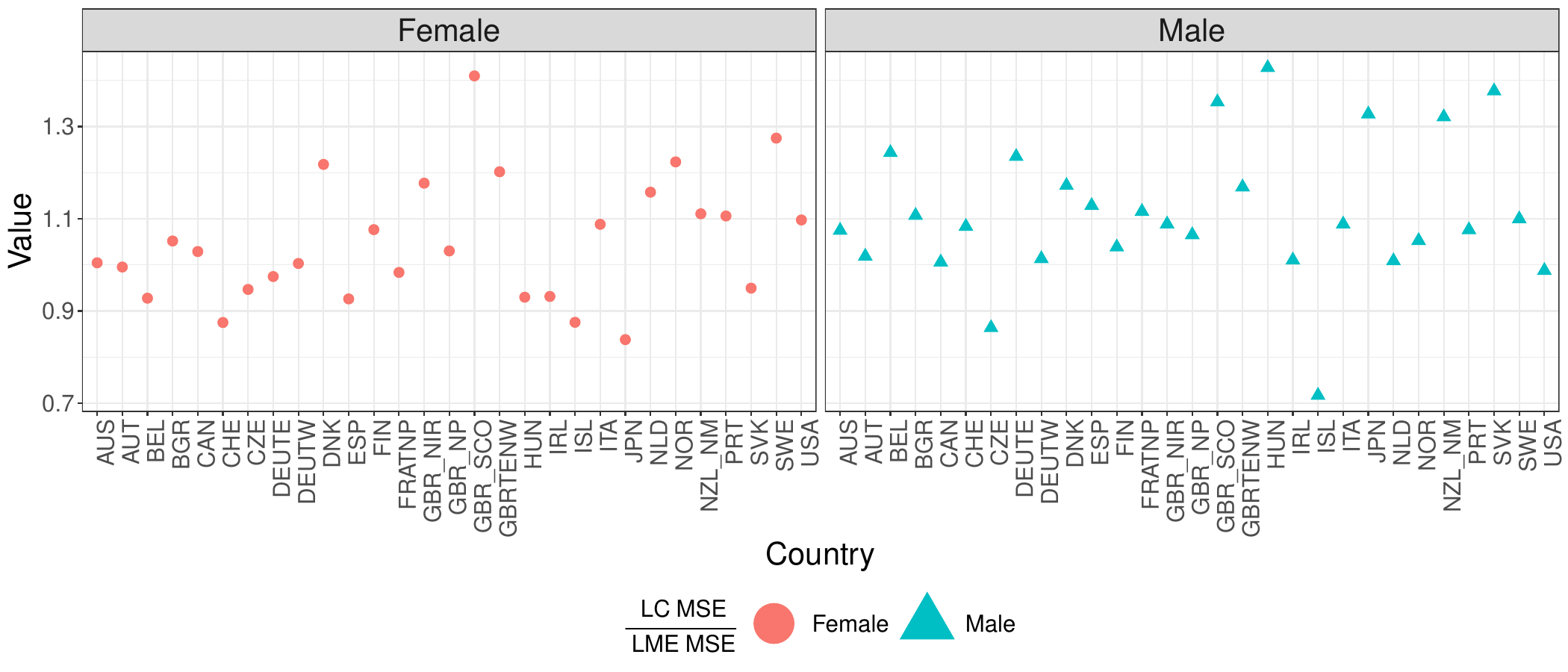}
    \caption{Ratio visualization of LC MSE divided by LME MSE (log ASDRs). Ratios greater than $1$ signify that the LC MSE is larger than the LME MSE, implying potential higher prediction errors in the LC model relative to the LME model.}
    \label{fig26}
\end{figure}

We use a random walk with drift to model and forecast the covariate $k_t$. However, the estimated fixed effects and the predicted random effects are assumed to be unchanged in forecasting. After computing the MSE of the models \eqref{Eq24} and LC for the 58 populations, we observed that model \eqref{Eq24} provides better forecast accuracy than the LC models for $43$ out of $58$ populations. These results are depicted in Figure \ref{fig26}, which visually demonstrates the superior forecasting performance of model \eqref{Eq24} compared to the LC models.

\section{Exploring the Practical Applications of the LME Model in Life Insurance}\label{sec5}

Our analysis revolves around an imaginary portfolio consisting of 20 individuals, with Year of Birth (YoB) and premiums denominated in CZK, see Tables \ref{tab10} and \ref{tab11} in \ref{C}. To enhance our modelling, we incorporate hypothetical experience factors that are specific to different age groups within the Czech Republic. These experience factors are used to adjust the observed mortality rates for each age group, reflecting their specific mortality experience compared to the overall population. By incorporating these age-specific experience factors into our models, we can better capture the unique characteristics and risk profiles of different age groups, leading to more accurate projections and insights in the context of life insurance. Experience factors adjust observed mortality rates to reflect the specific mortality experience of a group. When close to 1, mortality rates align with expectations, enabling confident risk assessment and pricing. Values below 1 indicate lower mortality rates, indicating favorable experience. Conversely, values above 1 imply higher mortality rates, indicating adverse experience. Insurers can use experience factors to refine risk assessment and pricing, improving liability estimation and financial stability.

The Best Estimate Liability (BEL) is the unbiased estimate of
the present value of expected future cash flows. In essence, it involves valuing these cash flows by employing the most accurate assumptions available, without including any explicit adjustments or safety margins. In simpler terms, it assumes that the estimated future cash flows will unfold as expected, without incorporating specific provisions for uncertainties or risks.

Our main goal is to calculate the BEL for both the LME model \eqref{Eq23} and the LC model \eqref{Eq2} \citep[see][for an overview]{dickson2019actuarial,wuthrich2013financial,gerber2013life,promislow2014fundamentals}. To achieve this, we will fit the models using the Czech mortality dataset, which covers age groups $45$ to $110$ with $1$-year intervals, spanning from $1950$ to $2019$. Subsequently, we will forecast the models for a $120$-year period, enabling us to project the BEL estimates beyond the available data range. To compute the BEL, various factors such as mortality rates, the insured individual's age, maximum age ($110$ years), retirement age ($65$ in the Czech Republic), valuation year, interest rate, and an indicator for discounting premium or annuity (type) are taken into account. In this analysis, we will utilize a valuation year of $2023$ and a interest rate of $0.1\%$ . These inputs will be utilized in the calculations to estimate the BEL for the specified scenario.

The non-deterministic nature of insurance cash flows means that solely considering the BEL as the fair value of the liability would lead to an underestimation. The BEL represents the average outstanding liability, but there is a potential risk of actual experience being more adverse than expected or the occurrence of catastrophic events, leading to significantly larger cash flows than anticipated. Therefore, it is essential to incorporate an additional risk margin component alongside the BEL to effectively address these uncertainties and ensure a comprehensive assessment of the liability.

\begin{figure}[H]
    \centering
    \includegraphics[width=\textwidth]{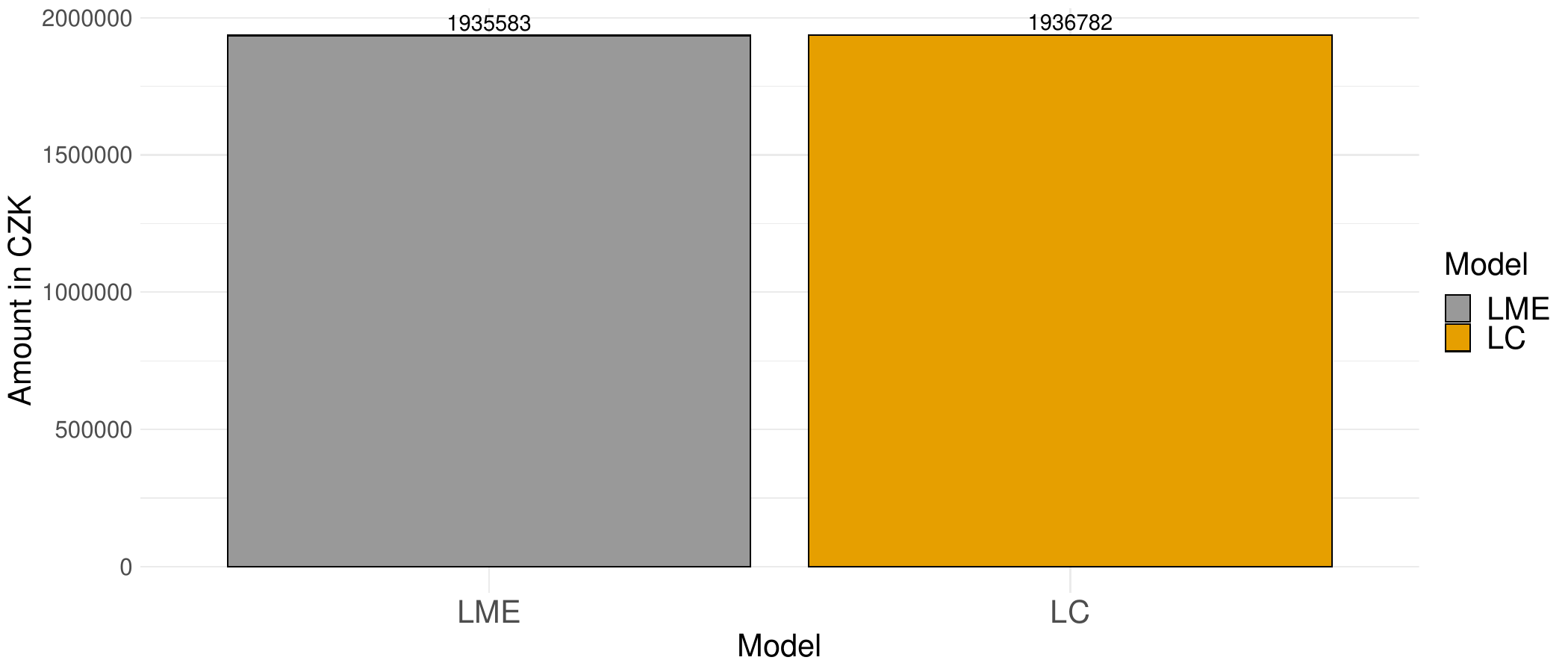}
    \caption{Comparing estimated values of BEL: LME vs. LC models.}
    \label{fig27}
\end{figure}

Beyond computing the BEL, we also conduct simulations to determine the Solvency Capital Requirement (SCR), which represents the capital amount an insurance company must hold to survive a 1 in 200 event \citep[see][for an overview]{wuthrich2013financial,promislow2014fundamentals,dickson2019actuarial}. Our analysis produces four barplots. The initial two barplots display the BEL values for the LME model and the LC model, yielding $1,935,583$ CZK and $1,936,782$ CZK, respectively, see Figure \ref{fig27}.

The barplots of BELs for the LME and LC models stand shoulder to shoulder, exhibiting nearly identical values.

\begin{figure}[H]
    \centering
    \includegraphics[width=\textwidth]{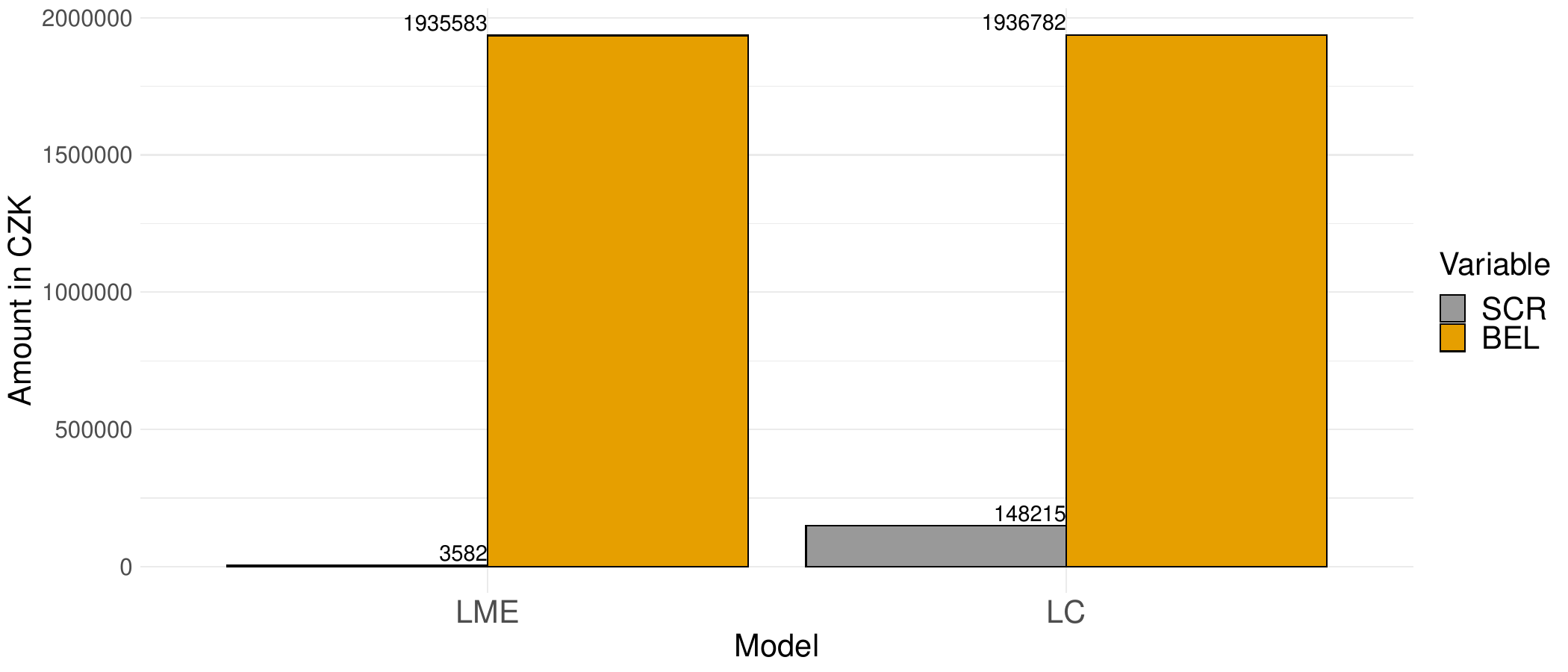}
    \caption{Comparing estimated values of BEL and SCR: LME vs. LC models.}
    \label{fig28}
\end{figure}

The subsequent two barplots present SCR values of $3,582$ CZK and $148,215$ CZK for the LME and LC models, respectively. These values are derived under the assumption of a Czech individual, with a retirement age of 65 and an annual pension of $23,700$ CZK ($\sim 1000$ Euros), see Figure \ref{fig28}.

\begin{figure}[H]
    \centering
    \includegraphics[width=\textwidth]{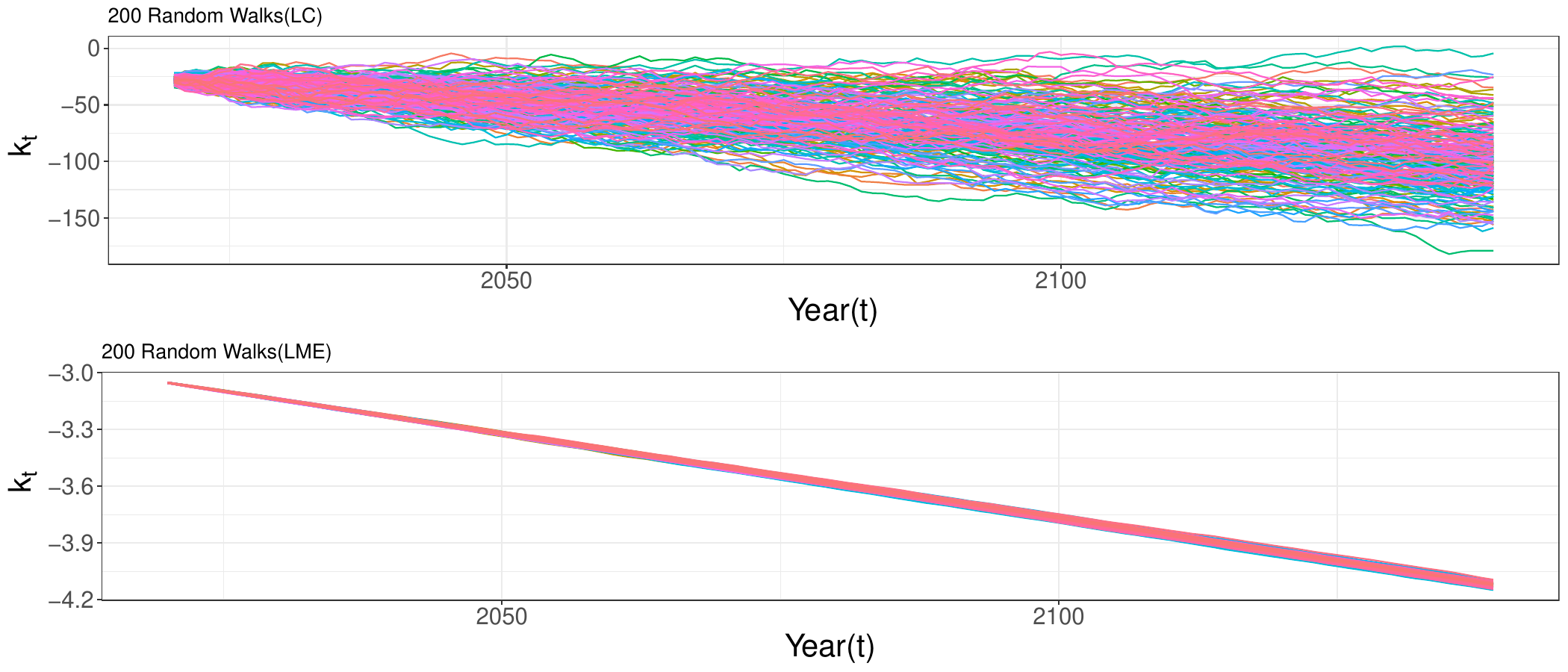}
    \caption{Simulated stochastic processes for $k_t$: LC vs. LME models.}
    \label{fig29}
\end{figure}

It is notable that both the LME and LC models provide similar estimates for the BEL, indicating their comparable effectiveness in forecasting future liabilities. However, the distinction in SCR values arises from the variance of the random walks component within each model. Specifically, the LC model exhibits a higher random walk variance of $2.40221$ compared to the significantly lower variance of $0.00089$ observed in the LME model. This discrepancy has implications for the simulation of the stochastic process governing the variable $k_t$, resulting in the LME model generating simulations that are more closely clustered together, in contrast to the wider spread of outcomes observed in the LC model. This characteristic of the LME model highlights its ability to produce more precise and concentrated distributions of potential outcomes, offering valuable insights for risk assessment and decision-making, see Figure \ref{fig29}.

\section{Conclusion}\label{sec6}

Mixed-effects models are a broad and active area of research in applied statistics. We propose an identifiable linear mixed-effects model for forecasting ASDRs. A crucial component of our model is the covariate $k_t$, which captures the common time effect affecting all age groups. It describes the common level of log ASDRs decline/growth. Additionally, $k_{ct}$ captures the impact of time on mortality rates for individuals in the younger age groups. The LME model also incorporates cohort effects, which vary across age, gender, and country. We further refine the LME model to incorporate GDP per capita as an economic growth factor, revealing that the random GDP effect varies across countries. However, our approach does have two primary limitations. Firstly, the meaningful selection of  populations, especially in the context of mortality projections for multiple populations, is a critical consideration. Expanding our exploration of common effects across different populations can enhance the reliability of parameter estimation. Therefore, the careful selection of populations is essential for robust results. Secondly, it is important to acknowledge that we rely on time series methods to predict future values of $k_{ct}$ and $k_t$. These methods introduce a level of uncertainty into our predictions. Nevertheless, our innovative approach, which combines linear mixed-effects models with time series forecasting of covariates, holds significant promise. Pursuing and implementing this approach could yield valuable insights and improvements in mortality rate projections.

As we conclude this study, it is worth acknowledging an intriguing open problem, one that bears the potential to significantly advance our understanding of longevity risk and mortality forecasting. This challenge centers on the development of a predictive model for the introduction of future vaccines, particularly cancer vaccines, and their potential impact on mortality rates. Historical evidence, such as the steep mortality improvement experienced between 1940 and 1960, attributed in large part to the introduction of Penicillin, underscores the profound influence of medical breakthroughs on mortality trends. If future vaccines, like cancer vaccines, exhibit mortality rate improvements akin to Penicillin, it becomes paramount to consider them from a longevity risk perspective. The central question is whether we can construct a model capable of anticipating the introduction and impact of such transformative medical interventions. By doing so, we could not only enhance our mortality forecasting capabilities but also proactively integrate these advancements into models like the LME model, thereby significantly refining our ability to project mortality rates into the future. This open problem invites researchers to explore the dynamics of medical innovation and its cascading effects on human longevity. 

\bibliographystyle{chicago}
\bibliography{mybib}

\appendix
\setcounter{table}{0}
\section{Random effects tables}\label{A}

\begin{table}[H]
\Huge
    \centering
\resizebox{15 cm}{!}{
\setlength{\tabcolsep}{14pt}
\renewcommand{\arraystretch}{1}
    \centering
\begin{tabular}{|c|c|c|c|c|c|c|c|c|c|c|c|c|}
%\toprule
\multicolumn{13}{c}{} \\\cmidrule{1-13}
%\midrule
 \multicolumn{1}{|c}{}&\multicolumn{2}{|c}{AUT} & \multicolumn{2}{|c}{BEL} & \multicolumn{2}{|c}{CHE} & \multicolumn{2}{|c}{CZE} & \multicolumn{2}{|c}{DNK} & \multicolumn{2}{|c|}{SWE} \\
 
\multicolumn{1}{|c}{$x$}&\multicolumn{1}{|c}{Female} & \multicolumn{1}{c}{Male} &\multicolumn{1}{|c}{Female} & \multicolumn{1}{c}{Male} &\multicolumn{1}{|c}{Female} & \multicolumn{1}{c}{Male} &\multicolumn{1}{|c}{Female} & \multicolumn{1}{c}{Male} &\multicolumn{1}{|c}{Female} & \multicolumn{1}{c}{Male} &\multicolumn{1}{|c}{Female} & \multicolumn{1}{c|}{Male}  \\
 \cmidrule{2-13}
 0& -0.01151 & 0.00366 & 0.01611 & -0.00345 & -0.00818 & 0.01209 & 0.02857 & 0.02340 & 0.01339 & 0.00498 & 0.11389 & 0.12112 \\ 
  1 & 0.00276 & 0.02327 & 0.01182 & -0.01832 & -0.05493 & -0.06192 & 0.07966 & 0.06070 & -0.03965 & -0.00565 & -0.01744 & 0.04554 \\ 
  5 & 0.08846 & 0.11427 & 0.00610 & 0.07612 & -0.02238 & -0.04949 & -0.04129 & -0.04754 & 0.04919 & -0.00976 & -0.04850 & -0.00224 \\ 
  10 & 0.07349 & 0.10511 & -0.01756 & 0.05748 & -0.00839 & -0.05696 & 0.01638 & -0.06191 & -0.00521 & -0.01750 & -0.05317 & -0.03556 \\ 
  15 & -0.08016 & -0.24013 & 0.00304 & 0.02741 & -0.03671 & -0.03810 & 0.04144 & 0.05331 & 0.09103 & 0.09649 & -0.04692 & 0.09770 \\ 
  20 & -0.00095 & -0.12063 & -0.04838 & -0.02360 & -0.06819 & -0.13067 & 0.06376 & 0.03280 & 0.06956 & 0.15081 & -0.07319 & 0.01421 \\ 
  25 & 0.07375 & -0.04847 & -0.05554 & -0.01523 & -0.05993 & -0.07102 & 0.11101 & 0.00252 & -0.01894 & 0.07620 & -0.09129 & -0.04950 \\ 
  30 & 0.04930 & -0.02965 & -0.04175 & 0.01076 & 0.02124 & 0.03061 & 0.06752 & -0.06929 & -0.06805 & 0.01659 & -0.03076 & -0.03196 \\ 
  35 & 0.03557 & -0.05722 & -0.03700 & 0.02955 & 0.12217 & 0.12467 & 0.02485 & -0.14827 & -0.13414 & -0.03685 & 0.00670 & 0.02398 \\ 
  40 & 0.03540 &\cellcolor{blue!25}  -0.07183 & -0.04041 & 0.03634 & 0.18222 & 0.17952 & -0.03403 & -0.22655 & -0.16926 & -0.02251 & 0.07013 & 0.10067 \\ 
  45 & 0.02931 & -0.06576 & -0.04294 & -0.03330 & 0.07467 & -0.04630 & 0.08589 & -0.02307 & -0.15862 & 0.03781 & 0.06000 & 0.04872 \\ 
  50 &\cellcolor{blue!25}  0.04290 & -0.05007 & -0.03251 & -0.03540 & 0.09585 & -0.02998 & 0.05529 & -0.03619 & -0.17932 & 0.00807 & 0.06528 & 0.06228 \\ 
  55 & 0.05385 & -0.02879 & -0.00184 & -0.03129 & 0.06547 & -0.01710 & 0.05550 & -0.06020 & -0.17671 & 0.02095 & 0.01545 & 0.07702 \\ 
  60 & 0.06294 & -0.00705 & 0.02504 & -0.01825 & 0.09300 & 0.02428 & 0.02706 & -0.10491 & -0.17809 & 0.01699 & 0.00606 & 0.10145 \\ 
  65 & 0.06907 & 0.01596 & 0.04052 & -0.01271 & 0.07606 & -0.00612 & 0.02162 & -0.08172 & -0.14171 & 0.03386 & -0.01427 & 0.05931 \\ 
  70 & 0.05242 & 0.01983 & 0.04152 & -0.01586 & 0.07437 & -0.01178 & 0.00400 & -0.06261 & -0.08214 & 0.04230 & -0.01176 & 0.02628 \\ 
  75 & 0.02610 & 0.01080 & 0.02939 & -0.01602 & 0.06513 & -0.04020 & -0.01376 & -0.02191 & -0.00213 & 0.05641 & -0.01331 & -0.02934 \\ 
  80 & -0.00939 & -0.00381 & 0.01641 & -0.00998 & 0.02992 & -0.05383 & -0.02430 & -0.00419 & 0.07197 & 0.07613 & -0.00859 & -0.04831 \\ 
  85 & -0.02261 & -0.00526 & 0.01466 & -0.00654 & -0.00501 & -0.04486 & -0.00429 & 0.00453 & 0.08212 & 0.07322 & -0.01954 & -0.05378 \\ 
  90 & -0.02555 & -0.01128 & 0.01488 & 0.00280 & -0.04207 & -0.04534 & 0.01216 & 0.01413 & 0.06708 & 0.06156 & -0.02998 & -0.04930 \\ 
  95& -0.02906 & -0.00833 & 0.01421 & 0.00748 & -0.05964 & -0.03863 & 0.02204 & 0.02456 & 0.06393 & 0.04768 & -0.03685 & -0.05374 \\ 
  100 & -0.02807 & -0.00762 & 0.01104 & 0.01023 & -0.06634 & -0.03321 & 0.02510 & 0.02823 & 0.05368 & 0.03458 & -0.03649 & -0.04877 \\ 
  105 & -0.02377 & -0.00632 & 0.00767 & 0.01094 & -0.06163 & -0.02492 & 0.02341 & 0.02386 & 0.04083 & 0.02602 & -0.03185 & -0.03908 \\ 
  $110^{+} $& -0.01913 & -0.00526 & 0.00516 & 0.01031 & -0.05291 & -0.02066 & 0.01993 & 0.02146 & 0.03042 & 0.01891 & -0.02653 & -0.03241 \\ 
  
 \hline
 \end{tabular}
    }
    \caption{Predicted random effects $\hat{\eta}_{0cgx}$.}
    \label{tab5}
\end{table}
\begin{table}[H]
\Huge
    \centering
\resizebox{15 cm}{!}{
\setlength{\tabcolsep}{10pt}
\renewcommand{\arraystretch}{1}
    \centering
\begin{tabular}{|c|c|c|c|c|c|c|c|c|c|c|c|c|}
%\toprule
\multicolumn{13}{c}{} \\\cmidrule{1-13}
%\midrule
 \multicolumn{1}{|c}{}&\multicolumn{2}{|c}{AUT} & \multicolumn{2}{|c}{BEL} & \multicolumn{2}{|c}{CHE} & \multicolumn{2}{|c}{CZE} & \multicolumn{2}{|c}{DNK} & \multicolumn{2}{|c|}{SWE} \\
 
\multicolumn{1}{|c}{$x$}&\multicolumn{1}{|c}{Female} & \multicolumn{1}{c}{Male} &\multicolumn{1}{|c}{Female} & \multicolumn{1}{c}{Male} &\multicolumn{1}{|c}{Female} & \multicolumn{1}{c}{Male} &\multicolumn{1}{|c}{Female} & \multicolumn{1}{c}{Male} &\multicolumn{1}{|c}{Female} & \multicolumn{1}{c}{Male} &\multicolumn{1}{|c}{Female} & \multicolumn{1}{c|}{Male}  \\
 \cmidrule{2-13}

0 & -0.06504 & -0.07022 & -0.09558 & -0.09695 & -0.03854 & -0.05150 & -0.12417 & -0.12234 & -0.05445 & -0.07802 & -0.11349 & -0.11925 \\ 
 1 & 0.05462 & 0.01272 & 0.01129 & -0.00540 & -0.01596 & -0.05953 & 0.03161 & -0.00696 & -0.01911 & -0.04323 & 0.02104 & -0.02634 \\ 
  5 & 0.04329 & 0.00206 & -0.00246 & -0.02718 & -0.05120 & -0.06943 & 0.02845 & -0.01822 & -0.07170 & -0.08968 & -0.04788 & -0.08159 \\ 
 10 & 0.01157 & 0.02360 & -0.00639 & -0.00461 & 0.00199 & -0.01564 & 0.02585 & 0.03044 & -0.02109 & -0.01886 & -0.02665 & 0.00751 \\ 
  15 & 0.02871 & 0.00104 & 0.01483 & -0.00229 & 0.01957 & -0.00909 & -0.00334 & -0.00448 & -0.00083 & 0.01197 & 0.01539 & -0.00129 \\ 
  20 & 0.03594 & 0.02347 & 0.00808 & 0.03722 & 0.03556 & 0.01610 & 0.01859 & 0.02774 & 0.02421 & 0.06468 & 0.05188 & 0.09314 \\ 
  25 & 0.01655 & 0.01832 & 0.02623 & 0.06152 & 0.04479 & 0.06070 & -0.00386 & 0.03160 & 0.02849 & 0.08744 & 0.01980 & 0.09492 \\ 
 30 & -0.02311 & -0.01462 & 0.00542 & 0.04589 & 0.03144 & 0.05215 & 0.00407 & 0.02747 & -0.00901 & 0.09048 & -0.00147 & 0.04893 \\ 
  35 & -0.03591 & -0.01726 & 0.00222 & 0.02911 & 0.00437 & 0.03659 & -0.00167 & 0.02275 & -0.00543 & 0.07388 & -0.00595 & 0.01975 \\ 
  40 & -0.04848 & \cellcolor{blue!25} -0.05167 & -0.01843 & -0.01064 & -0.02116 & -0.00833 & -0.01222 & -0.00188 & -0.02322 & 0.04900 & -0.02191 & 0.00065 \\ 
  45 & -0.03687 & 0.04134 & -0.00664 & 0.04925 & -0.01416 & 0.06726 & -0.02782 & 0.01152 & -0.03851 & 0.02923 & -0.03231 & 0.02422 \\ 
  50 &\cellcolor{blue!25}  -0.02821 & 0.05138 & -0.00548 & 0.03705 & -0.03434 & 0.04539 & -0.01774 & 0.02865 & -0.02312 & 0.03222 & -0.03268 & 0.02380 \\ 
  55 & -0.00629 & 0.02995 & 0.00101 & 0.00521 & -0.01347 & 0.01662 & 0.00139 & 0.02191 & 0.00309 & 0.00067 & -0.00704 & 0.00009 \\ 
  60 & -0.02301 & -0.01731 & -0.01875 & -0.03227 & -0.03763 & -0.01268 & -0.00293 & -0.01097 & 0.01243 & -0.02289 & -0.01786 & -0.01219 \\ 
  6 & -0.02886 & -0.02798 & -0.03048 & -0.03365 & -0.03906 & -0.00114 & -0.00959 & -0.03658 & 0.01522 & -0.02129 & -0.01807 & -0.00027 \\ 
  70 & -0.04125 & -0.02755 & -0.04216 & -0.02890 & -0.05086 & 0.00912 & -0.01945 & -0.04937 & 0.00707 & -0.01141 & -0.03634 & 0.00710 \\ 
  75 & -0.03995 & -0.00670 & -0.04414 & 0.00175 & -0.06280 & 0.03147 & -0.01847 & -0.04019 & -0.00308 & 0.00967 & -0.05160 & 0.02272 \\ 
  80 & -0.02972 & 0.00209 & -0.03235 & 0.01511 & -0.04799 & 0.03584 & -0.01465 & -0.02460 & -0.01600 & 0.01260 & -0.04247 & 0.02801 \\ 
  85 & -0.00426 & 0.00984 & -0.01136 & 0.01914 & -0.02214 & 0.02692 & 0.00189 & -0.00620 & -0.02178 & 0.00913 & -0.02113 & 0.02498 \\ 
  90 & 0.01611 & 0.01452 & 0.00742 & 0.02185 & 0.00987 & 0.02256 & 0.01664 & 0.00927 & -0.00896 & 0.01181 & 0.00733 & 0.02296 \\ 
  95 & 0.02811 & 0.02046 & 0.01723 & 0.02470 & 0.02224 & 0.01852 & 0.02341 & 0.01998 & -0.00271 & 0.00964 & 0.01955 & 0.02112 \\ 
  100 & 0.03335 & 0.02162 & 0.02300 & 0.02380 & 0.03120 & 0.01457 & 0.02562 & 0.02461 & 0.00206 & 0.00867 & 0.02774 & 0.01804 \\ 
  105& 0.03210 & 0.01891 & 0.02356 & 0.01980 & 0.03251 & 0.00916 & 0.02338 & 0.02289 & 0.00472 & 0.00680 & 0.02916 & 0.01297 \\ 
  $110^{+}$ & 0.02817 & 0.01669 & 0.02152 & 0.01682 & 0.02989 & 0.00702 & 0.01969 & 0.02070 & 0.00565 & 0.00589 & 0.02702 & 0.01027 \\ 
   \hline
 \end{tabular}
    }
    \caption{Predicted random effects $\hat{\eta}_{2cgx}$.}
    \label{tab6}
\end{table}
\begin{table}[H]
\Huge
    \centering
    \resizebox{15 cm}{!}{
\setlength{\tabcolsep}{10pt}
\renewcommand{\arraystretch}{1}
    \centering
\begin{tabular}{|c|c|c|c|c|c|c|c|c|c|c|c|c|}
%\toprule
\multicolumn{13}{c}{} \\\cmidrule{1-13}
%\midrule
 \multicolumn{1}{|c}{}&\multicolumn{2}{|c}{AUT} & \multicolumn{2}{|c}{BEL} & \multicolumn{2}{|c}{CHE} & \multicolumn{2}{|c}{CZE} & \multicolumn{2}{|c}{DNK} & \multicolumn{2}{|c|}{SWE} \\
 
\multicolumn{1}{|c}{$x$}&\multicolumn{1}{|c}{Female} & \multicolumn{1}{c}{Male} &\multicolumn{1}{|c}{Female} & \multicolumn{1}{c}{Male} &\multicolumn{1}{|c}{Female} & \multicolumn{1}{c}{Male} &\multicolumn{1}{|c}{Female} & \multicolumn{1}{c}{Male} &\multicolumn{1}{|c}{Female} & \multicolumn{1}{c}{Male} &\multicolumn{1}{|c}{Female} & \multicolumn{1}{c|}{Male}  \\
 \cmidrule{2-13}
0 & 0.000722 & 0.000763 & 0.001026 & 0.001062 & 0.000429 & 0.000549 & 0.001325 & 0.001311 & 0.000580 & 0.000846 & 0.001119 & 0.001175 \\ 
  1& -0.000599 & -0.000163 & -0.000136 & 0.000078 & 0.000232 & 0.000715 & -0.000429 & 0.000012 & 0.000250 & 0.000478 & -0.000211 & 0.000240 \\ 
  5 & -0.000565 & -0.000142 & 0.000020 & 0.000217 & 0.000582 & 0.000810 & -0.000267 & 0.000249 & 0.000731 & 0.000989 & 0.000573 & 0.000893 \\ 
  10 & -0.000203 & -0.000368 & 0.000088 & -0.000010 & -0.000013 & 0.000230 & -0.000299 & -0.000267 & 0.000236 & 0.000224 & 0.000347 & -0.000045 \\ 
  15 & -0.000229 & 0.000240 & -0.000165 & -0.000004 & -0.000175 & 0.000139 & -0.000007 & -0.000007 & -0.000086 & -0.000232 & -0.000119 & -0.000088 \\ 
  20 & -0.000391 & -0.000130 & -0.000037 & -0.000382 & -0.000317 & -0.000039 & -0.000270 & -0.000337 & -0.000337 & -0.000864 & -0.000490 & -0.001032 \\ 
  25 & -0.000258 & -0.000149 & -0.000228 & -0.000656 & -0.000426 & -0.000588 & -0.000074 & -0.000347 & -0.000291 & -0.001034 & -0.000120 & -0.000984 \\ 
  30 & 0.000201 & 0.000191 & -0.000015 & -0.000512 & -0.000365 & -0.000601 & -0.000115 & -0.000227 & 0.000170 & -0.001005 & 0.000048 & -0.000501 \\ 
  35 & 0.000355 & 0.000248 & 0.000015 & -0.000349 & -0.000176 & -0.000530 & -0.000008 & -0.000093 & 0.000200 & -0.000768 & 0.000058 & -0.000241 \\ 
  40 & 0.000492 & \cellcolor{blue!25} 0.000639 & 0.000243 & 0.000078 & 0.000040 & -0.000097 & 0.000169 & 0.000258 & 0.000431 & -0.000511 & 0.000166 & -0.000113 \\ 
  45 & 0.000372 & -0.000382 & 0.000117 & -0.000503 & 0.000076 & -0.000686 & 0.000214 & -0.000102 & 0.000587 & -0.000359 & 0.000290 & -0.000315 \\ 
  50 &\cellcolor{blue!25}  0.000263 & -0.000508 & 0.000094 & -0.000367 & 0.000274 & -0.000464 & 0.000136 & -0.000275 & 0.000440 & -0.000360 & 0.000288 & -0.000325 \\ 
  55& 0.000012 & -0.000297 & -0.000009 & -0.000024 & 0.000078 & -0.000163 & -0.000073 & -0.000176 & 0.000152 & -0.000029 & 0.000061 & -0.000082 \\ 
  60 & 0.000185 & 0.000196 & 0.000178 & 0.000371 & 0.000313 & 0.000113 & 0.000004 & 0.000230 & 0.000051 & 0.000232 & 0.000189 & 0.000027 \\ 
  65 & 0.000243 & 0.000289 & 0.000290 & 0.000381 & 0.000347 & 0.000019 & 0.000082 & 0.000485 & -0.000018 & 0.000197 & 0.000212 & -0.000059 \\ 
  70 & 0.000395 & 0.000280 & 0.000417 & 0.000332 & 0.000477 & -0.000087 & 0.000208 & 0.000604 & 0.000009 & 0.000080 & 0.000409 & -0.000105 \\ 
  75 & 0.000409 & 0.000062 & 0.000451 & -0.000002 & 0.000617 & -0.000301 & 0.000216 & 0.000462 & 0.000036 & -0.000165 & 0.000577 & -0.000217 \\ 
  80 & 0.000334 & -0.000019 & 0.000336 & -0.000154 & 0.000492 & -0.000335 & 0.000185 & 0.000273 & 0.000099 & -0.000217 & 0.000473 & -0.000255 \\ 
  85& 0.000070 & -0.000102 & 0.000109 & -0.000202 & 0.000247 & -0.000247 & -0.000016 & 0.000063 & 0.000152 & -0.000176 & 0.000251 & -0.000216 \\ 
  90 & -0.000149 & -0.000147 & -0.000097 & -0.000241 & -0.000064 & -0.000199 & -0.000194 & -0.000116 & 0.000027 & -0.000193 & -0.000049 & -0.000199 \\ 
  95 & -0.000276 & -0.000215 & -0.000203 & -0.000277 & -0.000180 & -0.000162 & -0.000279 & -0.000244 & -0.000037 & -0.000155 & -0.000175 & -0.000174 \\ 
  100 & -0.000335 & -0.000228 & -0.000263 & -0.000271 & -0.000271 & -0.000124 & -0.000306 & -0.000298 & -0.000079 & -0.000131 & -0.000264 & -0.000146 \\ 
  105 & -0.000325 & -0.000200 & -0.000265 & -0.000228 & -0.000290 & -0.000074 & -0.000280 & -0.000275 & -0.000094 & -0.000102 & -0.000285 & -0.000101 \\ 
 $ 110^{+}$& -0.000287 & -0.000177 & -0.000240 & -0.000194 & -0.000271 & -0.000055 & -0.000236 & -0.000248 & -0.000094 & -0.000084 & -0.000267 & -0.000078 \\
  \hline
 \end{tabular}
    }
    \caption{Predicted random effects $\hat{\eta}_{{\rm{cohort}}_{cgx}}$.}
    \label{tab7}
\end{table}
\setcounter{table}{0}
\section{Random walks tables}\label{B}

\begin{table}[H]
\Huge
\centering
\scalebox{.27}{
\setlength{\tabcolsep}{10pt}
\renewcommand{\arraystretch}{1}
    \centering
\begin{tabular}{|c|c|c|c|c|c|c|c|c|c|c|c|c|}
%\toprule
\multicolumn{13}{c}{} \\\cmidrule{1-13}
%\midrule
 \multicolumn{1}{|c}{}&\multicolumn{2}{|c}{AUT} & \multicolumn{2}{|c}{BEL} & \multicolumn{2}{|c}{CHE} & \multicolumn{2}{|c}{CZE} & \multicolumn{2}{|c}{DNK} & \multicolumn{2}{|c|}{SWE} \\
 
\multicolumn{1}{|c}{t}&\multicolumn{1}{|c}{Age 0-40} & \multicolumn{1}{c}{Age 45-110} &\multicolumn{1}{|c}{Age 0-40} & \multicolumn{1}{c}{Age 45-110} &\multicolumn{1}{|c}{Age 0-40} & \multicolumn{1}{c}{Age 45-110} &\multicolumn{1}{|c}{Age 0-40} & \multicolumn{1}{c}{Age 45-110} &\multicolumn{1}{|c}{Age 0-40} & \multicolumn{1}{c}{Age 45-110} &\multicolumn{1}{|c}{Age 0-40} & \multicolumn{1}{c|}{Age 45-110}  \\
 \cmidrule{2-13}

  1961 & -6.466 & -2.468 & -6.622 & -2.479 & -6.641 & -2.534 & -6.645 & -2.419 & -6.810 & -2.560 & -6.873 & -2.600 \\ 
  1962 & -6.497 & -2.404 & -6.609 & -2.431 & -6.626 & -2.479 & -6.618 & -2.342 & -6.819 & -2.526 & -6.875 & -2.580 \\ 
   1963 & -6.502 & -2.413 & -6.628 & -2.412 & -6.671 & -2.461 & -6.632 & -2.417 & -6.809 & -2.539 & -6.878 & -2.589 \\ 
  1964 & -6.506 & -2.466 & -6.636 & -2.482 & -6.729 & -2.541 & -6.676 & -2.438 & -6.844 & -2.532 & -6.876 & -2.622 \\ 
  1965 & -6.548 & -2.417 & -6.636 & -2.456 & -6.786 & -2.519 & -6.648 & -2.419 & -6.804 & -2.544 & -6.890 & -2.617 \\ 
   1966 & -6.526 & -2.464 & -6.650 & -2.462 & -6.766 & -2.533 & -6.654 & -2.434 & -6.845 & -2.517 & -6.927 & -2.634 \\ 
  1967 & -6.534 & -2.442 & -6.703 & -2.474 & -6.764 & -2.582 & -6.667 & -2.411 & -6.888 & -2.567 & -6.962 & -2.636 \\ 
   1968 & -6.575 & -2.442 & -6.674 & -2.421 & -6.790 & -2.538 & -6.630 & -2.385 & -6.832 & -2.597 & -6.913 & -2.629 \\ 
   1969 & -6.548 & -2.420 & -6.670 & -2.447 & -6.770 & -2.553 & -6.591 & -2.371 & -6.845 & -2.596 & -6.951 & -2.633 \\ 
  1970 & -6.546 & -2.425 & -6.685 & -2.471 & -6.803 & -2.588 & -6.629 & -2.352 & -6.857 & -2.605 & -6.950 & -2.690 \\ 
   1971 & -6.531 & -2.439 & -6.699 & -2.471 & -6.762 & -2.589 & -6.656 & -2.373 & -6.835 & -2.608 & -6.980 & -2.690 \\ 
  1972 & -6.561 & -2.473 & -6.715 & -2.491 & -6.807 & -2.630 & -6.719 & -2.401 & -6.894 & -2.614 & -6.996 & -2.682 \\ 
  1973 & -6.611 & -2.498 & -6.753 & -2.490 & -6.903 & -2.633 & -6.746 & -2.376 & -6.890 & -2.619 & -7.032 & -2.689 \\ 
   1974 & -6.645 & -2.492 & -6.774 & -2.511 & -6.904 & -2.665 & -6.784 & -2.366 & -7.006 & -2.609 & -7.051 & -2.696 \\ 
   1975 & -6.634 & -2.472 & -6.817 & -2.495 & -6.926 & -2.688 & -6.796 & -2.394 & -6.949 & -2.640 & -7.029 & -2.692 \\ 
  1976 & -6.727 & -2.473 & -6.803 & -2.500 & -6.999 & -2.668 & -6.833 & -2.382 & -6.995 & -2.587 & -7.043 & -2.679 \\ 
 1977 & -6.748 & -2.518 & -6.839 & -2.573 & -7.007 & -2.739 & -6.830 & -2.390 & -7.029 & -2.668 & -7.126 & -2.720 \\ 
  1978 & -6.769 & -2.501 & -6.812 & -2.554 & -6.985 & -2.715 & -6.843 & -2.401 & -7.045 & -2.663 & -7.127 & -2.731 \\ 
  1979 & -6.776 & -2.537 & -6.846 & -2.600 & -6.993 & -2.732 & -6.866 & -2.400 & -7.012 & -2.634 & -7.133 & -2.727 \\ 
   1980 & -6.789 & -2.538 & -6.867 & -2.592 & -7.026 & -2.729 & -6.852 & -2.349 & -7.025 & -2.625 & -7.203 & -2.725 \\ 
  1981 & -6.851 & -2.540 & -6.937 & -2.611 & -7.030 & -2.748 & -6.889 & -2.390 & -7.021 & -2.637 & -7.302 & -2.737 \\ 
  1982 & -6.843 & -2.570 & -6.968 & -2.629 & -7.099 & -2.768 & -6.920 & -2.389 & -7.092 & -2.677 & -7.329 & -2.768 \\ 
   1983 & -6.882 & -2.553 & -6.966 & -2.611 & -7.075 & -2.761 & -6.903 & -2.363 & -7.087 & -2.644 & -7.340 & -2.775 \\ 
  1984 & -6.915 & -2.609 & -7.028 & -2.656 & -7.132 & -2.810 & -6.957 & -2.380 & -7.125 & -2.659 & -7.362 & -2.804 \\ 
   1985 & -6.985 & -2.593 & -7.043 & -2.654 & -7.129 & -2.813 & -6.984 & -2.393 & -7.062 & -2.646 & -7.346 & -2.779 \\ 
  1986 & -7.027 & -2.626 & -7.072 & -2.667 & -7.137 & -2.826 & -6.973 & -2.377 & -7.130 & -2.676 & -7.344 & -2.800 \\ 
  1987 & -7.086 & -2.657 & -7.083 & -2.733 & -7.202 & -2.853 & -7.030 & -2.423 & -7.157 & -2.678 & -7.368 & -2.808 \\ 
   1988 & -7.146 & -2.683 & -7.125 & -2.748 & -7.140 & -2.854 & -7.042 & -2.441 & -7.145 & -2.683 & -7.345 & -2.793 \\ 
  1989 & -7.134 & -2.680 & -7.126 & -2.723 & -7.180 & -2.874 & -7.067 & -2.412 & -7.153 & -2.676 & -7.392 & -2.847 \\ 
  1990 & -7.189 & -2.710 & -7.118 & -2.762 & -7.178 & -2.841 & -7.019 & -2.415 & -7.174 & -2.664 & -7.407 & -2.839 \\ 
  1991 & -7.194 & -2.705 & -7.164 & -2.771 & -7.139 & -2.884 & -7.029 & -2.436 & -7.206 & -2.700 & -7.462 & -2.841 \\ 
   1992 & -7.239 & -2.712 & -7.171 & -2.779 & -7.202 & -2.904 & -7.038 & -2.473 & -7.255 & -2.678 & -7.532 & -2.854 \\ 
  1993 & -7.187 & -2.720 & -7.206 & -2.765 & -7.224 & -2.904 & -7.100 & -2.494 & -7.287 & -2.660 & -7.503 & -2.849 \\ 
  1994 & -7.217 & -2.753 & -7.219 & -2.795 & -7.230 & -2.928 & -7.082 & -2.505 & -7.306 & -2.694 & -7.626 & -2.907 \\ 
   1995 & -7.323 & -2.757 & -7.228 & -2.794 & -7.300 & -2.913 & -7.117 & -2.510 & -7.286 & -2.669 & -7.666 & -2.901 \\ 
  1996 & -7.381 & -2.763 & -7.292 & -2.808 & -7.385 & -2.943 & -7.245 & -2.555 & -7.325 & -2.703 & -7.709 & -2.906 \\ 
   1997 & -7.435 & -2.791 & -7.344 & -2.823 & -7.438 & -2.936 & -7.225 & -2.561 & -7.424 & -2.725 & -7.705 & -2.930 \\ 
   1998 & -7.472 & -2.813 & -7.330 & -2.833 & -7.496 & -2.964 & -7.339 & -2.596 & -7.462 & -2.762 & -7.694 & -2.942 \\ 
   1999 & -7.465 & -2.830 & -7.362 & -2.836 & -7.497 & -2.976 & -7.363 & -2.596 & -7.486 & -2.750 & -7.749 & -2.939 \\ 
   2000 & -7.465 & -2.854 & -7.362 & -2.847 & -7.537 & -2.994 & -7.367 & -2.611 & -7.525 & -2.775 & -7.788 & -2.952 \\ 
   2001 & -7.554 & -2.885 & -7.346 & -2.860 & -7.583 & -3.017 & -7.425 & -2.633 & -7.497 & -2.780 & -7.781 & -2.955 \\ 
  2002 & -7.596 & -2.885 & -7.433 & -2.849 & -7.629 & -3.023 & -7.420 & -2.645 & -7.549 & -2.781 & -7.817 & -2.960 \\ 
   2003 & -7.557 & -2.892 & -7.506 & -2.847 & -7.681 & -3.019 & -7.463 & -2.622 & -7.594 & -2.797 & -7.806 & -2.989 \\ 
  2004 & -7.610 & -2.938 & -7.592 & -2.905 & -7.650 & -3.072 & -7.523 & -2.673 & -7.594 & -2.834 & -7.869 & -3.015 \\ 
  2005 & -7.614 & -2.948 & -7.576 & -2.902 & -7.739 & -3.077 & -7.532 & -2.674 & -7.750 & -2.853 & -7.862 & -3.015 \\ 
   2006 & -7.684 & -2.975 & -7.597 & -2.939 & -7.767 & -3.105 & -7.626 & -2.714 & -7.727 & -2.856 & -7.873 & -3.036 \\ 
   2007 & -7.783 & -2.988 & -7.586 & -2.949 & -7.834 & -3.112 & -7.625 & -2.742 & -7.728 & -2.864 & -7.914 & -3.041 \\ 
   2008 & -7.791 & -2.994 & -7.658 & -2.943 & -7.895 & -3.128 & -7.664 & -2.755 & -7.761 & -2.886 & -7.930 & -3.058 \\ 
  2009 & -7.720 & -2.989 & -7.668 & -2.962 & -7.884 & -3.133 & -7.655 & -2.750 & -7.836 & -2.900 & -7.887 & -3.074 \\ 
  2010 & -7.791 & -3.008 & -7.694 & -2.982 & -8.024 & -3.147 & -7.702 & -2.778 & -7.944 & -2.920 & -7.981 & -3.092 \\ 
  2011 & -7.818 & -3.019 & -7.716 & -2.992 & -8.052 & -3.159 & -7.724 & -2.786 & -7.968 & -2.927 & -8.004 & -3.102 \\ 
  2012 & -7.845 & -3.030 & -7.738 & -3.002 & -8.080 & -3.172 & -7.745 & -2.793 & -7.991 & -2.935 & -8.027 & -3.112 \\ 
  2013 & -7.872 & -3.041 & -7.760 & -3.012 & -8.108 & -3.184 & -7.767 & -2.800 & -8.014 & -2.942 & -8.049 & -3.123 \\ 
   2014 & -7.899 & -3.052 & -7.782 & -3.023 & -8.137 & -3.197 & -7.788 & -2.808 & -8.037 & -2.949 & -8.072 & -3.133 \\ 
   2015 & -7.927 & -3.063 & -7.803 & -3.033 & -8.165 & -3.209 & -7.810 & -2.815 & -8.060 & -2.957 & -8.095 & -3.143 \\ 
  2016 & -7.954 & -3.074 & -7.825 & -3.043 & -8.193 & -3.222 & -7.832 & -2.822 & -8.083 & -2.964 & -8.117 & -3.153 \\ 
  2017 & -7.981 & -3.086 & -7.847 & -3.053 & -8.221 & -3.234 & -7.853 & -2.830 & -8.106 & -2.971 & -8.140 & -3.163 \\ 
   2018 & -8.008 & -3.097 & -7.869 & -3.064 & -8.249 & -3.247 & -7.875 & -2.837 & -8.130 & -2.979 & -8.162 & -3.173 \\ 
  2019 & \cellcolor{blue!25}-8.035 & \cellcolor{blue!25}-3.108 & -7.891 & -3.074 & -8.278 & -3.259 & -7.896 & -2.844 & -8.153 & -2.986 & -8.185 & -3.183 \\ 
   \hline
 \end{tabular}
    }
    \caption{Random walks and forecasted values for country-specific trends in the six European countries.}
    \label{tab8}
\end{table}

\begin{table}[H]
\Large
    \centering
\scalebox{.7}{
\begin{tabular}{|c|c|c|c|c|c|c|c|c|}
  \hline
t & $k_t$ & t & $k_t$  & t & $k_t$ & t & $k_t$   \\ 
  \hline
         1961 & -4.246 &       1976 & -4.361 &       1991 & -4.588 &       2006 & -4.927 \\ 
         1962 & -4.216 &       1977 & -4.405 &       1992 & -4.611 &       2007 & -4.948 \\ 
        1963 & -4.228 &       1978 & -4.401 &       1993 & -4.615 &       2008 & -4.970 \\ 
        1964 & -4.263 &       1979 & -4.410&       1994 & -4.646 &       2009 & -4.971 \\ 
        1965 & -4.256 &       1980 & -4.413 &       1995 & -4.659 &       2010 & -5.016 \\ 
        1966 & -4.266 &       1981 & -4.441 &       1996 & -4.700 &       2011 & -5.032 \\ 
         1967 & -4.283 &       1982 & -4.470 &       1997 & -4.725 &       2012 & -5.048 \\ 
        1968 & -4.266 &       1983 & -4.461 &       1998 & -4.755 &       2013 & -5.063 \\ 
         1969 & -4.264 &       1984 & -4.500 &       1999 & -4.765 &       2014 & -5.079\\ 
         1970 & -4.281 &       1985 & -4.499 &       2000 & -4.784 &       2015 & -5.095\\ 
        1971 & -4.285 &       1986 & -4.517 &       2001 & -4.803 &       2016 & -5.111 \\ 
         1972 & -4.312 &       1987 & -4.551 &       2002 & -4.823 &       2017 & -5.126 \\ 
       1973 & -4.331 &       1988 & -4.557 &       2003 & -4.836 &       2018 & -5.142\\ 
        1974 & -4.350 &       1989 & -4.566 &       2004 & -4.879 &       2019 & \cellcolor{blue!25}-5.158 \\ 
       1975 & -4.353 &       1990 & -4.570 &       2005 & -4.89    &&  \\ 
   \hline
\end{tabular}
}
\caption{Observations and forecasts of the random walk $k_t$.}
    \label{tab9}
\end{table}

\section{A hypothetical portfolio and experience factors}\label{C}
\setcounter{table}{0}

\begin{table}[H]
    \centering
\scalebox{.7}{
\begin{tabular}{|c|c|c|}
  \hline
  YoB & Premium\\ 
 \hline
1965 &  CZK 6,925.00  \\ 
  1958 &  CZK 5,540.00  \\ 
  1962 &  CZK 8,480.00  \\ 
  1943 &  CZK 7,632.00  \\ 
  1937 &  CZK 5,144.00  \\ 
  1951 &  CZK 4,598.00  \\ 
  1969 &  CZK 6,396.00  \\ 
  1925 &  CZK 5,984.00  \\ 
  1949 &  CZK 6,073.00  \\ 
  1956 &  CZK 7,481.00  \\ 
  1961 &  CZK 7,022.00  \\ 
  1928 &  CZK 6,245.00  \\ 
  1971 &  CZK 7,991.00  \\ 
  1934 &  CZK 7,475.00  \\ 
  1952 &  CZK 7,433.00  \\ 
  1930 &  CZK 8,640.00  \\ 
  1970 &  CZK 8,485.00  \\ 
  1941 &  CZK 8,133.00  \\ 
  1936 &  CZK 8,812.00  \\ 
  1955 &  CZK 9,307.00  \\ 
   \hline
\end{tabular}
}
\caption{Hypothetical portfolio: Year of birth (YoB) and premiums (in CZK) for 20 insured Individuals.}
    \label{tab10}
\end{table}

\begin{table}[H]
    \centering
\scalebox{.7}{
\begin{tabular}{|c|c|c|c|c|c|c|c|c|c|c|c|c|}
  \hline
  Age & ExpF& Age & ExpF& Age & ExpF&Age & ExpF\\ 
  \hline
 0 & 0.40 & 28 & 0.56 & 56 & 0.72 & 84 & 0.90 \\ 
   1 & 0.40 & 29 & 0.56 & 57 & 0.73 & 85 & 0.90 \\ 
   2 & 0.40 & 30 & 0.57 & 58 & 0.74 & 86 & 0.90 \\ 
   3 & 0.41 & 31 & 0.57 & 59 & 0.74 & 87 & 0.92 \\ 
   4 & 0.41 & 32 & 0.58 & 60 & 0.76 & 88 & 0.92 \\ 
   5 & 0.42 & 33 & 0.59 & 61 & 0.76 & 89 & 0.92 \\ 
   6 & 0.43 & 34 & 0.59 & 62 & 0.76 & 90 & 0.93 \\ 
   7 & 0.43 & 35 & 0.60 & 63 & 0.77 & 91 & 0.94 \\ 
   8 & 0.44 & 36 & 0.61 & 64 & 0.78 & 92 & 0.94 \\ 
   9 & 0.45 & 37 & 0.61 & 65 & 0.78 & 93 & 0.95 \\ 
   10 & 0.46 & 38 & 0.62 & 66 & 0.79 & 94 & 0.95 \\ 
   11 & 0.46 & 39 & 0.63 & 67 & 0.80 & 95 & 0.96 \\ 
   12 & 0.47 & 40 & 0.63 & 68 & 0.80 & 96 & 0.97 \\ 
   13 & 0.47 & 41 & 0.64 & 69 & 0.80 & 97 & 0.97 \\ 
   14 & 0.48 & 42 & 0.64 & 70 & 0.81 & 98 & 0.98 \\ 
   15 & 0.48 & 43 & 0.65 & 71 & 0.82 & 99 & 0.98 \\ 
   16 & 0.49 & 44 & 0.66 & 72 & 0.82 & 100 & 0.99 \\ 
   17 & 0.49 & 45 & 0.66 & 73 & 0.83 & 101 & 1.00 \\ 
  18 & 0.50 & 46 & 0.67 & 74 & 0.83 & 102 & 1.00 \\ 
   19 & 0.51 & 47 & 0.68 & 75 & 0.84 & 103 & 1.01 \\ 
   20 & 0.52 & 48 & 0.68 & 76 & 0.85 & 104 & 1.01 \\ 
   21 & 0.52 & 49 & 0.68 & 77 & 0.85 & 105 & 1.02 \\ 
   22 & 0.52 & 50 & 0.69 & 78 & 0.86 & 106 & 1.02 \\ 
   23 & 0.53 & 51 & 0.70 & 79 & 0.86 & 107 & 1.03 \\ 
   24 & 0.54 & 52 & 0.70 & 80 & 0.87 & 108 & 1.04 \\ 
  25 & 0.54 & 53 & 0.71 & 81 & 0.88 & 109 & 1.04 \\ 
   26 & 0.55 & 54 & 0.71 & 82 & 0.88 & 110 & 1.05 \\ 
  27 & 0.56 & 55 & 0.72 & 83 & 0.89 & &  \\ 
   \hline
\end{tabular}
}
\caption{Hypothetical experience factors: This table presents a hypothetical Experience Factors table for age groups ranging from $0$ to $110$, specifically designed for an insurance company operating in the Czech Republic.}
    \label{tab11}
\end{table}

\end{document}